\documentclass[pdftex,twocolumn,epjc3]{svjour3}          

\RequirePackage[T1]{fontenc}

\smartqed  

\RequirePackage{graphicx}
\RequirePackage{mathptmx}      
\RequirePackage{flushend}
\RequirePackage[numbers,sort&compress]{natbib}
\RequirePackage[colorlinks,citecolor=blue,urlcolor=blue,linkcolor=blue]{hyperref}
\RequirePackage{amssymb}

\journalname{Eur. Phys. J. C}

\begin{document}
\newcommand{\ti}[1]{\mbox{\tiny{#1}}}
\def\be{\begin{equation}}
\def\ee{\end{equation}}
\def\bea{\begin{eqnarray}}
\def\eea{\end{eqnarray}}
\newcommand{\bt}[1]{\mathbf{\mathtt{#1}}}
\newcommand{\tb}[1]{\textbf{\texttt{#1}}}
\newcommand{\btb}[1]{\textcolor[rgb]{0.00,0.00,1.00}{\tb{#1}}}
\newcommand{\il}{~}
\newcommand{\rtb}[1]{\textcolor[rgb]{1.00,0.00,0.00}{\tb{#1}}}
\newcommand{\ptb}[1]{\textcolor[rgb]{0.50,0.00,0.50}{\tb{#1}}}
\newcommand{\gtb}[1]{\textcolor[rgb]{0.00,0.80,0.50}{\tb{#1}}}
\newcommand{\otb}[1]{\textcolor[rgb]{1.00,0.50,0.25}{\textbf{#1}}}

\title{General classification of charged test particle circular orbits in Reissner--Nordstr\"om spacetime}

\author{
D. Pugliese\thanksref{e1,addr1}, H. Quevedo\thanksref{addr2,addr3,addr4,addr5} \and R. Ruffini\thanksref{addr2,addr3}
}

\thankstext{e1}{e-mail:d.pugliese.physics@gmail.com}

\institute{Institute of Physics, Faculty of Philosophy \& Science,
  Silesian University in Opava,
 Bezru\v{c}ovo n\'{a}m\v{e}st\'{i} 13, CZ-74601 Opava, Czech Republic\label{addr1}\\
 \and
          Dipartimento di Fisica and ICRA, Universit\`a di Roma ``La Sapienza",
                 Piazzale Aldo Moro 5, I-00185 Roma, Italy\label{addr2}
 \and
                 ICRANet, Piazzale della Repubblica 10, I-65122 Pescara, Italy\label{addr3}
\and Instituto de Ciencias Nucleares, Universidad Nacional Aut\'onoma de M\'exico,\\
AP 70543, M\'exico, DF 04510, Mexico\label{addr4}
\and
Department of Theoretical and Nuclear Physics,
Kazakh National University,
050040 Almaty, Kazakhstan\label{addr5}
}

\date{Received: date / Accepted: date}

\maketitle

\begin{abstract}
We investigate  charged  particles circular motion   in the gravitational field of a charged mass distribution described by the Reissner-Nordstr\"om spacetime. We introduce a set of independent parameters completely characterizing  the different spatial regions in  which circular motion is allowed.
We provide  most complete classification of circular orbits for different sets of particle and  source charge-to-mass ratios. We study both black holes and naked singularities and show that the behavior of charged   particles depend drastically on the
type of source.  Our analysis shows in an alternative manner that the behavior of circular orbits can   in principle be used to distinguish between black holes and naked singularities.  From this  analysis,  special  limiting  values for the dimensionless charge  of  black hole and naked singularity   emerge, namely, Q/M=1/2, $Q/M=\sqrt{13}/5$ and  $Q/M=\sqrt{2/3}$ for the black hole case and Q/M=1, $Q/M=5/ (2 \sqrt{6})$, $Q/M=3 \sqrt{6}/7$, and  finally $Q/M= \sqrt{9/8}$ for the naked singularity case. Similarly and surprisingly,  analogue   limits  emerge  for the orbiting  particles charge-to-mass ratio $\epsilon$,  for positive charges  $\epsilon=1$, $\epsilon=2$  and $\epsilon=M/Q$. These limits play an important role in the study of the coupled electromagnetic and gravitational interactions, and the investigation of the role of  the charge  in the gravitational collapse of compact objects.
\end{abstract}

\section{Introduction}

The dynamics of  massive or massless, charged or neutral test particles in the vicinity of compact objects is one of the most interesting problems of relativistic astrophysics. The motion of test particles depends explicitly on the properties of the source of gravity and, hence, the geometric and physical properties of the trajectories can be used to derive information about the compact object. Moreover, the structure of spacetime in the surrounding of astrophysical  compact objects can be explored in detail by using test particles
\cite{RuRR,sharp79,frolnov98}.

The study of particle motion   revealed to be  an essential and  useful method  for the determination of the geometric  and topological properties of the spacetime as  described by the  pseudo-Riemannian manifold of  general relativity. Some examples of this application are given in {\cite{Pugliese:2013zma,Zakharov:2014lqa,Kovar:2014tla,Toshmatov:2014qja,
Stuchlik:2014qja,Beheshti:2015bak,Stuchlik:2015nlt,Tursunov:2016dss,Kovar:2016kqh}}.

The utility of this approach of analysis resides in the possibility of proposing several  different methods for modeling the matter  behavior  even in those situations where the approximation of point-like objects (i.e.  without structure) is no longer valid, in particular,  in   models of   extended matter  as described by accretion disks. Indeed, accretion disks are predominantly toroidal structures that may
be  subjected to  many different factors  such as   hydrostatic  or radiative pressure, viscosity, resistance and  magnetic field
\cite{Pugliese:2012ub,Pugliese:2015zla,abra}.
 The basic study of  test particles motion is,   however, always essential to set up more sophisticated and rich studies of accretion disks configurations, especially  to characterize  the equilibrium, where the gravitational effects are so relevant to require a general relativistic treatment  or during the  interaction with the source as in the accretion, or in the  jet emission \cite{abra}.

 In this work,  we focus on the still intriguing and veiled challenge of  coupling between gravitation in Einstein's general relativistic formulation and the electromagnetic field in Maxwell's theory. Our analysis can be classified within the class of electromagnetic effects in curved spacetimes; in particular, we study the  circular motion of charged test (point-like) particles in a
 Reissner--Nordstr\"om  (\textbf{RN})  black hole (\textbf{BH})  and  naked singularity  (\textbf{NS}) family of geometries.

 It is well known that in the \textbf{RN}  spacetime,  the electromagnetic field contributes to the make up of the  geometry
\cite{Pugliese:2011aa,Pugliese:2014yla,Pugliese:2013gsa,Belvedere:2012uc}.
On the other hand, some of the fundamental problems to be addressed is to correctly combine the   curvature and electromagnetic effects in the description of extended bodies, for  example,  for the determination (definition and limits) of the  charge-to-mass ratio.
Here the charge-to-mass ratio determines the occurrence of a horizon or, on the contrary, the impossibility that it is formed.
Here, in order  to reduce other interaction effects such as   the presence of spin for the particle and the attractor, which  would imply
  a spin-spin  coupling and a spin-orbit coupling \cite{Pugliese:2013zma}, we consider  a test particle with charge in a spacetime with spherical symmetry described by the (static) \textbf{RN} solution; we focus on the the exterior  regions outside the exterior horizon.

The motion of charged material around singularities has been investigated extensively, for instance, in
\cite{PuQueRu-charge,Kovar:2016kqh,Tursunov:2016dss,Stuchlik:2015nlt,Kovar:2014tla,Pradhan:2010ws,Gladush:2011cz,Joshi,PatJosKimNa-2011,
Bib-Stuc,
Grg}.
Essentially,  the possibility has been analyzed that there could be a significant change in the behavior of matter under the effect of electromagnetic interaction in the presence  or absence of an event horizon. This may affect the evolutionary processes involving the event horizon and its interaction with the environment; see, for example, \cite{Rezzolla}.  It has been found, among other things, that
under certain circumstances even an arbitrarily small charge may have an essential role in the destruction of the horizon,  leading to very peculiar
stability properties of charged particles around  a \textbf{NS}. This could have a role in the events of horizon formation. A question  which starts to be predominantly timely in facing the problematic of merging of black holes or even the merging of stars, leading to an one-horizon structure.
This phenomenon, however, disappears for  particles with charge-to-mass ratios ($\epsilon$) sufficiently small, independently of the  attractive or repulsive nature of the electromagnetic interaction. The peculiar structure of the stability  is regulated by a set of special  charges  which we single out for the  attractors (\textbf{NS} or  \textbf{BH}). However, intriguingly,  the dimensionless  charge of the particle  have also a particular influence on the interaction and horizon evolution, depending on some special values we single out.

 We deep the analysis performed in \cite{PuQueRu-charge} in order to obtain the complete classification of  circular motion around
\textbf{RN} black holes  and naked singularities.
From a methodological view point, this is a new systematic study of circular motion which makes use of a  double classification procedure introduced  in  the double contest  of a \textbf{BH} and a \textbf{NS}.  This  allows us to reveal the motion characteristics on the basis of a dimensionless parameter related to the   attractor  and the particle; this type of analysis has not been considered in previous related works.
This kind of approach reveals to be  interesting  also to be implemented in other contexts where the dynamics could be equally considered in the geometries of black holes  or naked singularities.

This article is structured as follows. After reviewing the main properties of the \textbf{RN} spacetime in
Sec.\il(\ref{xanes}), the  circularly orbiting charged  particles are considered in Sec.\il\ref{sec:veff}.
In Sec.\il\ref{Sec:classBKRNAp}, the circular motion around a \textbf{RN} black hole is addressed, while
 Sec.\il\ref{NSNSTRE} focuses on the  \textbf{RN} naked singularity case. Particularly, we explore the case of a ``weak'' repulsive interaction,  $\epsilon\in]0,1[$, in Sec.\il \ref{sec:pos}, and a ``weak''  attraction, $\epsilon\in]-1,0[$, in Sec.\il \ref{sec:neg}. Conclusions close the paper in section\il(\ref{Sec:m.l-ka}). In the appendices, we present exact details about different circular orbits.

\section{Reissner--Nordstr\"om  spacetime and circularly orbiting charged  particles}
\label{xanes}
The Reissner--Nordstr\"om (\textbf{RN}) line element
\begin{equation}\label{11metrica}
ds^2=-\frac{\Delta}{r^2}dt^2+\frac{r^2}{\Delta}dr^2
+r^2\left(d\theta^2+\sin^2\theta d\phi^2\right)\ ,
\end{equation}
in
standard spherical coordinates, describes the background of a static gravitational source of mass $M$ and
charge $Q$,
where
\begin{equation}
\Delta = (r-r_+)(r-r_-)\ , \quad r_\pm = M\pm\sqrt{M^2 -Q^2}
\end{equation}
and $r_\pm$ are the radii of the outer and inner horizon, respectively.
We consider the motion of a test particle of charge $q$ and mass $\mu$ moving in a
\textbf{RN}   background (\ref{11metrica}) as described by  the specific charge of the test particle  $\epsilon = q/\mu$ (see
\cite{Bicak:2000ea}).
Due to the existence of spacetime symmetries, the
following two conserved quantities exist
\begin{eqnarray}
\label{Eq:10000} p_t&=& -\left(\frac{\Delta}{r^2}\dot{t}+\frac{\epsilon Q}{r}\right)=-\frac{E}{\mu},\quad   p_{\phi}=r^2\sin^2\theta \dot{\phi}=\frac{L}{\mu},
\end{eqnarray}
where $L$ and  $E$ are respectively the angular momentum and energy
of the particle as measured by an observer at rest at infinity. A  dot
represents differentiation with respect to the proper
time.
On the equatorial plane $\theta=\pi/2$, the motion equations can be reduced to the form $\dot r^2 + V^2 = E^2/\mu^2$ which
describes the motion of a test particle inside an effective potential $V$. Then, it is convenient to define the potential
\begin{equation}\label{9}
V_{\pm}=\frac{E_{\pm}}{\mu}=\frac{\epsilon Q}{r}\pm
\sqrt{\left(1+\frac{L^2}{\mu^2r^2}\right)\left(1-\frac{2M}{r}+\frac{Q^2}{r^2}\right)}
\end{equation}
which corresponds to the value of $E/\mu$ at which the
(radial) kinetic energy of the particle vanishes \cite{RuRR,Chandra,Levin:2008mq,Bilic:2006bh}, i.e.
it is the value at which $r$ is a ``turning point''
$(V=E/\mu)$. The effective potential with positive (negative) sign corresponds to
the solution with
$
\lim_{r\rightarrow\infty} E_{\pm}=\pm\mu.
$
Notice that in general\footnote{{A thorough discussion of the meaning of the double solutions  $V_{\pm}$ has been provided in
\cite{Bib-Stuc2} for the case of the Kerr-Newman solution.
 The positive root states are associated to the positive particle energy, as measured by local observers with future-oriented
time
component of the 4-velocity. Moreover, $V_-$  determines
 the  negative-root states with  negative locally measured energy and past-oriented
time component of the 4-velocity. In  the Kerr-Newman geometries,
for the case of non-extreme black holes, both  solutions $V_{\pm}$ determine  the particle motion.
In the case of extreme black holes and naked singularities,   
 the negative-energy particles  (with $V<V_-)$ have been interpreted in terms of Dirac negative energy sea and holes.}}
$E_+\geq E_-$ and $ E_+(L,\epsilon,r)=-E_{-}(L,-\epsilon,r)$.
In the limiting case of vanishing test charge, the effective potential reduces to
\begin{equation}
V = + \sqrt{\left(1+\frac{L^2}{\mu^2r^2}\right)\left(1-\frac{2M}{r}+\frac{Q^2}{r^2}\right)}\ .
\end{equation}
This case was analyzed previously in
\cite{Pugliese:2010ps} and \cite{Pugliese:2010he}, where we  found that the stability properties of neutral test particles strongly depend
on the nature of the central source. Indeed, in the case of a \textbf{\textbf{\textbf{RN}}} black hole there exists a minimum  radius at which the orbit is stable, and outside this radius all the orbits are stable so that there exists only one region of stability. In the case of a naked singularity, the situation is completely different; the region of stability splits into two non-connected regions so that a zone appears inside which no stable circular orbit can exist. This means that the stability properties of circular orbits could, in principle, be used to differentiate between a black hole and a naked singularity.

In the following section we face the study of the circular motion for charged particles. We will find the conditions for the existence of circular orbits, and will analyze their  properties in different \textbf{BH} and \textbf{NS} spacetimes. We perform a classification in accordance with the value of the specific charge $\epsilon$.  Different  ranges of the $\epsilon$ parameter  evidently serve to investigate  the   joint action  and balance  of the attractor bending effects    and the electromagnetic attraction or repulsion, when any backreaction effects on the spacetime are negligible. We then  investigate the range $\epsilon\in[-1,1]$, representing a ``small''
charge-to-mass ratio  and the (more realistic) cases  with   $\epsilon<-1$ and, particularly,  $\epsilon>1$ for the repulsive case.
{We will use an alternative approach in which the orbit properties are parametrized according to the values of $ Q / M $ and $\epsilon$.
Depending on the orbital properties and the electromagnetic interaction we  analyze, different sets of parameters are used. This alternative treatment, adapted to the individual cases under investigation, will allow us to derive limiting values for the dimensionless charges of the attractor and noticeably also of the particles, a fact that has not been revealed in former  studies.
Therefore, the numerical values of the  reference parameters  $Q/M$ and $\epsilon$   will be accompanied by the reference orbital radii  associated to the limiting photon  orbit or zero angular momentum orbit. }
\subsection{Circular motion}
\label{sec:veff}
The effective potential (\ref{9}) regulates the circular motion of charged test particles.
We will see that due to the presence of a test charge, many different possibilities appear which
require a detailed investigation. We therefore limit ourselves here to the  special case of the  positive solution $V_+$.
Also, in this case it is possible to compare our results with those obtained in the case of neutral test particles
 in \cite{Pugliese:2010ps,Pugliese:2010he}.

The extrema of the  function $V_{+}$, defined by  the relations
\begin{equation}\label{fg2}
\frac{d V_{+}}{d r}=0, \quad V_+=\frac{E_+}{\mu},
\end{equation}
determine  the radius of circular orbits and the corresponding values of the energy  $E$
and the angular momentum $L$. In the following analysis, we  drop the subindex $+$.
Usually, the properties of the circular motion are investigated by analyzing  the behavior of the effective potential
$V$ in terms of the parameters of the test particle. In this work, we follow a different approach in which
the physical behavior of the parameters is derived from the circular motion conditions. Indeed,
solving (\ref{fg2}) with respect to $L$,  we find
\be
\frac{(L_{\pm})^2}{\mu^2} = \frac{r^2}{2\Sigma^2}\left[ 2(Mr-Q^2)\Sigma
+ \epsilon^2 Q^2 \Delta\pm Q\Delta\sqrt{\epsilon^2 \left(4\Sigma+\epsilon^2  Q^2\right)}\right],
\label{ECHEEUNAL}
\ee
where
\be
\Sigma\equiv r^2-3Mr+2Q^2
\ee
%
which represents the specific angular momentum of the test particle on a circular orbit of radius $r$.
The corresponding energy reads
\begin{equation}
\label{Lagesp}
\frac{E_{\pm}}{\mu}=\frac{\epsilon Q}{r}
+\frac{\Delta\sqrt{2\Sigma+\epsilon^2Q^2\pm Q\sqrt{\epsilon^2(4\Sigma+\epsilon^2Q^2)}}}
{\sqrt{2}r|\Sigma|}\  .
\end{equation}

We see that in the general case of charged test particles, the presence of the additional term  $\frac{\epsilon Q}{r}$ changes completely the physical properties  of test particles moving along
circular orbits, and leads to several possibilities which must be analyzed separately for black holes and naked singularities.
%
%
%
Section\il\ref{Sec:classBKRNAp} focuses on the case of a \textbf{RN} black hole
while  the case of a \textbf{RN} naked singularity is addressed in
Sec.\il\ref{NSNSTRE}.
%


\section{Circular motion around a \textbf{RN} black hole}
\label{Sec:classBKRNAp}
We are interested in investigating  all the regions outside the outer horizon $r_+$ in which circular motion is allowed.
In this section, we present a classification of the circular orbits around a \textbf{RN} black hole ($Q\leq M$) by using certain values of the parameters $L$ and $\epsilon$
which follow from the conditions of circular motion.
It can be considered as an alternative study to that presented in \cite{PuQueRu-charge},
in which the value of the ratio $Q/M$ plays a central role.
An analysis of the conditions for circular motion shows that it is convenient to introduce the  parameters
\bea\label{TTTilde}
\widetilde{\epsilon}&\equiv&
\frac{1}{\sqrt{2}Q}\sqrt{5M^{2}-4Q^{2}+\sqrt{25M^{2}-24Q^{2}}},
\\\label{TTTilden}
\epsilon_n&\equiv&\sqrt{\frac{3M^2}{2 Q^2}-1+\frac{\sqrt{(9M^2-8 Q^2)M^2}}{2 Q^2}},
\eea
\be\label{Eq:Q-Q-def}
\widetilde{Q}\equiv M\frac{\sqrt{4+5 \epsilon ^2}}{ 2+\epsilon ^2},
\quad
Q_n\equiv M\frac{\sqrt{1+3 \epsilon ^2}}{ 1+\epsilon ^2},
\ee
which represent upper and minimum boundaries for the particle charge-to-mass ratio with respect to the \textbf{BH} charge-to-mass ratio
{\footnote{{The charges $\widetilde{\epsilon}$ and $\epsilon_n$  introduced in Eq.\il(\ref{TTTilde}) and Eq.\il(\ref{TTTilden}), respectively, have been found  by parameterizing the particle motion  with charge  $\epsilon>1$
for the dimensionless charge  $ Q/M\in[0,1]$ of the central black hole attractor.  Consequently, we define  the ranges of
\textbf{BH} charges  in
Table\il\ref{Tabdffaubis}.  The parametrization according to the  particle charge-to-mass ratio $\epsilon>1$,  as given in  Table\il\ref{Tabdffautris}, leads to the values $\widetilde{Q}$ and $Q_n$ given in Eq.\il(\ref{Eq:Q-Q-def}). }}}.
\begin{figure}
\centering
\begin{tabular}{c}
\includegraphics[scale=0.3]{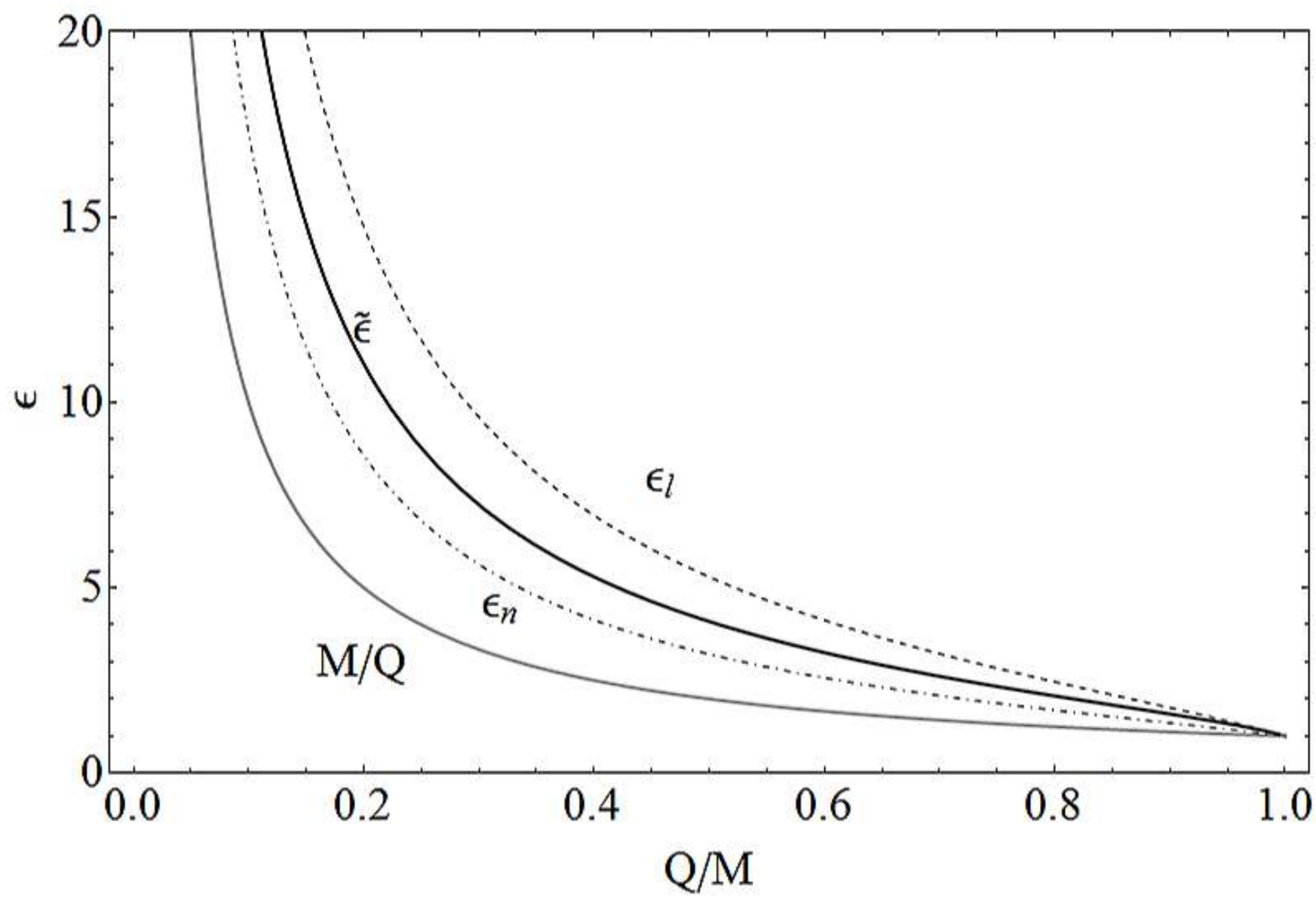}
\end{tabular}
\caption[font={footnotesize,it}]{\footnotesize{The charge parameters
$\epsilon_{l}$ (dashed curve) --defined in Eq.(\ref{20})--, ${\widetilde{\epsilon}}$ (black solid curve), $\epsilon_{n}$ (dotdashed curve), and $M/Q$ (gray curve)
as functions of the charge--to--mass ratio of the \textbf{RN} black hole.
 }}
\label{PETILQE}
\end{figure}
\begin{figure}
\centering
\begin{tabular}{cc}
\includegraphics[scale=0.3]{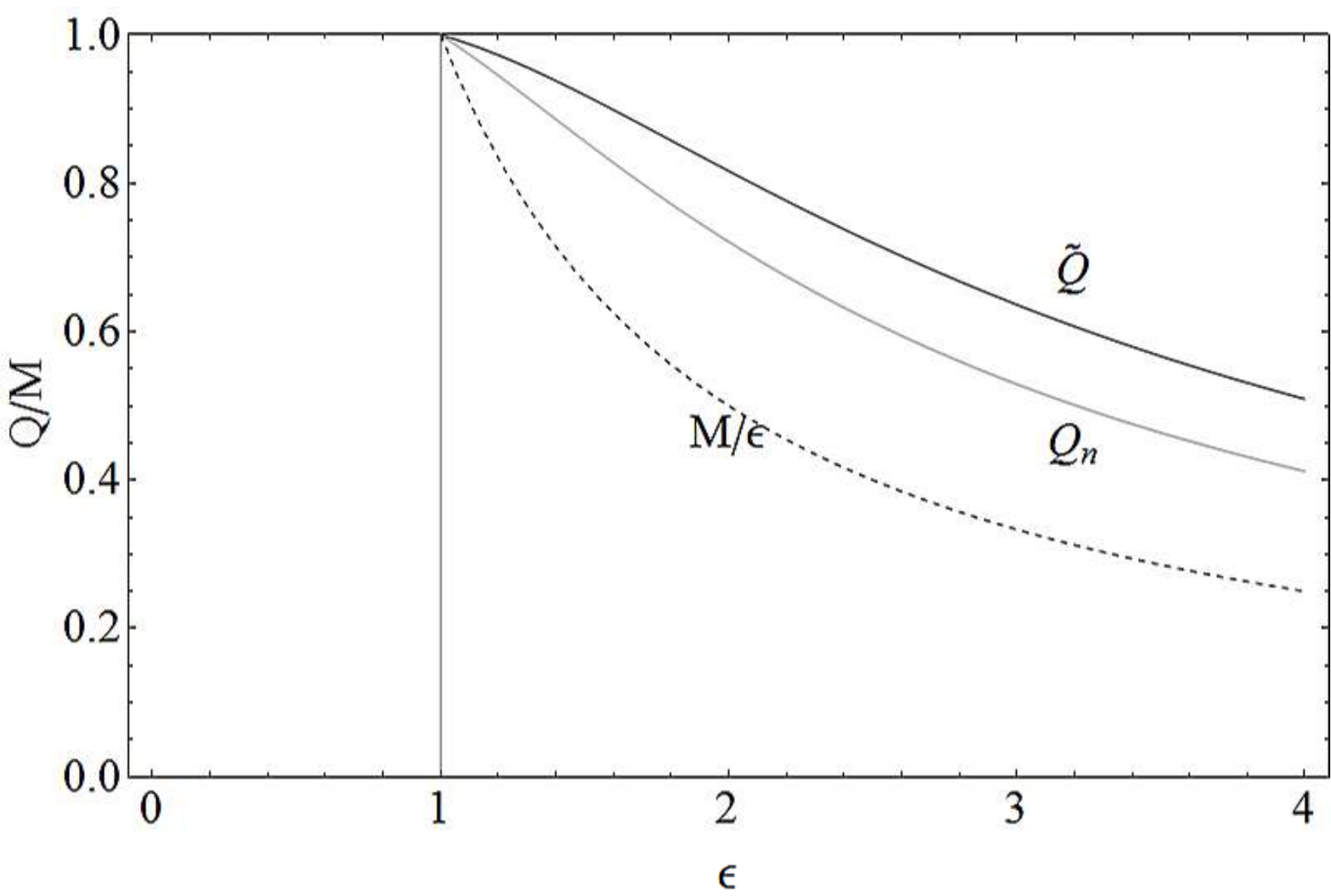}
\end{tabular}
\caption[font={footnotesize,it}]{\footnotesize{The charge parameters
$Q_{n}$ (gray curve), $\widetilde{Q}$ (black solid curve), and $M/\epsilon$ (dashed curve)
as functions of the charge--to--mass ratio $\epsilon$ of the test particle. }}
\label{PETILQQ}
\end{figure}
The behavior of these parameters is depicted in figures \ref{PETILQE} and \ref{PETILQQ}.

Moreover, let us introduce the radii
\be
r_{\gamma}^{\pm}\equiv\frac{3 M}{2}\pm\frac{1}{2} \sqrt{9 M^{2}-8 Q^{2}},
\ee
which represents the limiting radius at which neutral particles (photons) can be in circular motion around a \textbf{RN} black hole
\cite{Pugliese:2010ps},
\bea\label{Eq:lim-r-fot}
r^{\pm}_l&\equiv&\frac{3 M}{2}\pm\frac{1}{2} \sqrt{9 M^{2}-8 Q^{2}-Q^{2} \epsilon ^{2}} \ ,
\eea
where\footnote{{ The radii $r^{\pm}_l$, introduced  in Eq.\il(\ref{Eq:lim-r-fot}), are found by considering  the particle motion parameterized by both the particle charge-to-mass ratio $\epsilon$ and the dimensionless charge of the central attractor $Q/M$.
Investigating the particle angular momentum in the different orbital regions, the radii $r_l^{\pm}$ appear as  orbital boundary.
We note also the the photon orbits $r_{\gamma}^{\pm}$, which depend on the background parameters $(Q, M)$ only,
are the limiting values  of $r_l^{\pm}$, respectively. Then, in the case of black holes,
the radius $r_l^-$ defines only the properties of the charged particle,  while in the case of naked singularities both  solutions
$r_l^-$, and $r_l^+$ define the limiting orbital ranges. }} $\lim_{\epsilon\rightarrow0}r_{l}^{\pm}=r_{\gamma}^{\pm}$,
and
\be
r_{s}^{\pm}\equiv\frac{\left(\epsilon ^2-1\right)Q^2M }{\epsilon ^2Q^2 -M^2}\pm\sqrt{\frac{\epsilon^2Q^4 \left(\epsilon ^2-1\right) \left(M^2-Q^2\right) }{\left(\epsilon ^2Q^2 -M^2\right)^2}}\label{rE}\ ,
\ee
which corresponds to a zero angular momentum circular orbit (\textbf{ZAMPs})
\be\label{Eq:No-Lia}
L=0,\quad \frac{d V}{dr}=0
\ee
as seen by an observer at infinity. This special radius generalizes the concept of the classical radius $r_* = Q^2/M$ which is the limiting value for neutral test particles.

Furthermore, the
  parameter
\bea\nonumber
L^2_n\equiv  \frac{9M^7}{2\epsilon^2 Q^2} \left(3M+\sqrt{9M^2-8 Q^2}\right)+\frac{2Q^2M^2}{\epsilon^2} \left(1+\epsilon^2\right)
\\
 -\frac{M^3}{2\epsilon^2 } \left[27M+5 \sqrt{9M^2-8 Q^2}+3 \left(3M+\sqrt{9M^2-8 Q^2}\right) \epsilon ^2\right],
\eea
represents the value of the angular momentum
which satisfies the relationship $(dV/dr)(L_n, r_{\gamma}^+)=0$.

A careful analysis of the circular orbits properties shows that it is convenient to split the classification problem into
two different groups. The first group contains all the negative test charges and positive test charges with $\epsilon\leq 1$.
The second group contains only positive test charges with $\epsilon>1$. For both groups we investigate the values of the allowed orbit's radius
and the corresponding angular momentum. The results are summarized  for $\epsilon\leq 1$ in Table \ref{Tabdffautrismeno},
\begin{table*}
\caption[font={footnotesize,it}]{\label{Tabdffautrismeno} \footnotesize{
Characteristics of the circular orbits for charged test particles with charge--to--mass ratio
$\epsilon \leq 1$ in a \textbf{RN} black hole.}}

\begin{tabular}{@{}ll|ll}
\hline
$\epsilon\in]0,1] $&&
&$\epsilon<0$
\\
\hline
\textbf{\textbf{Region}}&\textbf{Momentum}
&\textbf{Region}&\textbf{Momentum}
\\
\hline
$r=r_{l}^{+}$&$ L=\pm L_-$
&$r>r_{\gamma}^{+}$&$ L=\pm L_+$
\\
$r_{l}^{+}<r<r_{\gamma}^{+}$&$ L=\pm L_{\pm}$
&$$&$$
\\
$r=r_{\gamma}^{+}$&$ L=\pm L_n$
&$$&$$
\\
$r>r_{\gamma}^{+}$&$ L=\pm L_-$
&$$&$$
\\
\hline
\end{tabular}
\end{table*}
and for $\epsilon>1$ in Tables \ref{Tabdffaubis} and \ref{Tabdffautris}. Moreover, the characteristic parameters for all
the different cases are listed in Table \ref{Tabdffau}.
\begin{table*}
\caption[font={footnotesize,it}]{\footnotesize{\label{Tabdffau} Values of the angular momentum $L$ which are possible in different regions
of the radial coordinate $r$.}}
\begin{tabular}{@{}lll|ll|ll}
\hline
&\textbf{Class} \textbf{I} &
&\textbf{Class} {\textbf{II}} &
&\textbf{Class} \textbf{III} &
\\
\hline
&\textbf{Region}&\textbf{Momentum}
&\textbf{Region}&\textbf{Momentum}
&\textbf{Region}&\textbf{Momentum}
\\
\hline
&$r=r_{l}^+$&$ L=\pm L_-$
&$r=r_{l}^+$&$ L=\pm L_-$
&$r=r_{l}^+$&$ L=\pm L_-$
\\
&$(r_{l}^+,r_{\gamma}^+) $&$ L=\pm L_{\pm}$
&$(r_{l}^+,r_{\gamma}^+) $&$ L=\pm L_{\pm}$
&$(r_{l}^+,r_{\gamma}^+) $&$ L=\pm L_{\pm}$
\\
&$r=r_{\gamma}^+ $&$ L=\pm L_n$
&$r=r_{\gamma}^+ $&$ L=\pm L_n$
&$r=r_{\gamma}^+ $&$ L=0$
\\
&$(r_{\gamma}^+ ,\infty)\ \ \ \ $&$ L=\pm L_-$
&$(r_{\gamma}^+ ,r_{s}^+)$&$ L=\pm L_-$
&$$ &$$
\\
\hline
&\textbf{Class} \textbf{IV} &
&\textbf{Class} \textbf{V} &
& \textbf{Class} \textbf{\textbf{VI}}&
\\
\hline
&\textbf{Region}&\textbf{Momentum}
&\textbf{Region}&\textbf{Momentum}
&\textbf{Region}&\textbf{Momentum}
\\
\hline
&$r=r_{l}^+$&$ L=\pm L_-$
&$r=r_{l}^+ (r_{s}^+)$&$ L=0$
&$r=r_{l}^+$&$ L=\pm L_-$
\\
&$(r_{l}^+,r_{s}^+) $&$ L=\pm L_{\pm}$
&$(r_{l}^+,r_{\gamma}^+) ((r_{s}^+,r_{\gamma}^+))\ \ \ \ $&$ L=\pm L_{+}$
&$(r_{l}^+,r_{\gamma}^+) $&$ L=\pm L_{\pm}$
\\
&$r=r_{s}^+ $&$ L=\pm L_+;\ L=0$
&$$&$ $
&$r=r_{\gamma}^+ $&$ L=\pm L_n$
\\
&$(r_{s}^+,r_{\gamma}^+ )$&$ L=\pm L_+$
&$$ &$$
&$(r_{\gamma}^+,r_{s}^+ )$&$ L=\pm L_-$
\\
&$$ &$$
&$$ &$$
&$r=r_s^+$&$ L=0$
\\
\hline
&\textbf{Class} \textbf{VII} &
&\textbf{Class} \textbf{VIII} &
&\textbf{Class} \textbf{IX} &
\\
\hline
&\textbf{Region}&\textbf{Momentum}
&\textbf{Region}&\textbf{Momentum}
&\textbf{Region}&\textbf{Momentum}
\\
\hline
&$r=r_{l}^+ (r_{s}^+)$&$ \pm L_-$
&$r=r_{l}^+$&$ L=0$
&$r=r_{s}^+ $&$L=0$
\\
&$(r_{l}^+,r_{s}^+) $&$ L=\pm L_{\pm}$
&$(r_{l}^+,r_{\gamma}^+) \ \ \ \ $&$ L=\pm L_{+}$
&$(r_{s}^+,r_{\gamma}^+) $&$ L=\pm  L_{+}$
\\
&$r=r_s^+$&$ L=0$
&$r=r_{\gamma}^+ $&$ L=\pm L_n$
&$$&$$
\\
\hline
\end{tabular}
\end{table*}
\begin{table*}
\caption[font={footnotesize,it}]
{\footnotesize{\label{Tabdffaubis} Classification of the circular orbits of a charged test particle with charge--to--mass ratio
$\epsilon>1$ in terms of the value of the ratio $Q/M$ of a \textbf{RN} black hole.}}
%
\begin{tabular}{@{}ll|ll|ll|llll}
\hline
{\footnotesize $Q/M\in(0,1/2]\ \ $}&
&{\footnotesize$Q/M\in[1/2,\sqrt{13}/5)$}&
&{\footnotesize $Q/M\in[\sqrt{13}/5,\sqrt{2/3})$}&
&{\footnotesize $Q/M\in[\sqrt{2/3},1)$}&
\\
\hline
\textbf{Region}&\textbf{Class}
&\textbf{Region}&\textbf{Class}
&\textbf{Region}&\textbf{Class}
&\textbf{Region}&\textbf{Class}
\\
\hline
$\epsilon\in]1,2]$&$ \mathbf{I}$
&$\epsilon\in]1,M/Q] $&$ \mathbf{I}$
&$\epsilon\in]1,M/Q] $&$ \mathbf{I}$
&$\epsilon\in]1,M/Q] $&$ \mathbf{I}$
\\
$\epsilon\in]2,M/Q] $&$\mathbf{I}$
&$\epsilon\in]M/Q,2]$&$\mathbf{II}$
&$\epsilon\in]M/Q,\epsilon_n ] $&$\mathbf{II}$
&$\epsilon\in]M/Q,\epsilon_n ] $&$\mathbf{II}$
\\
$\epsilon\in]M/Q,\epsilon_n ] $&$\textbf{VI}$
&$\epsilon\in]2,\epsilon_n [$&$\textbf{VI}$
&$\epsilon=\epsilon_n$&$\mathbf{III}$
&$\epsilon=\epsilon_n$&$\mathbf{III}$
\\
$\epsilon=\epsilon_n$&$\mathbf{VII}$
&$\epsilon=\epsilon_n$&$\mathbf{VII}$
&$\epsilon\in]\epsilon_n,2 ]$ &$\mathbf{IV}$
&$\epsilon\in]\epsilon_n,\widetilde{\epsilon}[$&$\mathbf{IV}$
\\
$\epsilon\in]\epsilon_n,\widetilde{\epsilon}[$&$\textbf{IV}$
&$\epsilon\in]\epsilon_n,\widetilde{\epsilon}[$&$\textbf{IV}$
&$\epsilon\in]2,\widetilde{\epsilon}[$&$\mathbf{IV}$
&$\epsilon\in]\widetilde{\epsilon},2]$&$\mathbf{V}$
\\
$\epsilon=\widetilde{\epsilon}$&$\textbf{VIII}$
&$\epsilon=\widetilde{\epsilon}$&$\textbf{VIII}$
&$\epsilon=\widetilde{\epsilon}$&$\mathbf{VIII}$
&$\epsilon>2$&$\mathbf{\mathbf{IX}}$
\\
$\epsilon>\widetilde{\epsilon}$&$\mathbf{IX}$
&$\epsilon>\widetilde{\epsilon}$&$\mathbf{IX}$
&$\epsilon>\widetilde{\epsilon}$&$\mathbf{IX}$
&$$&$$
\\
\hline
\end{tabular}
\end{table*}
\begin{table*}
\caption[font={footnotesize,it}]{\label{Tabdffautris}\footnotesize{Classification in terms of the charge--to--mass ratio
$\epsilon>1$ of the circular orbits of a charged test particle moving in the field of a \textbf{RN} black hole with
mass $M$ and charge $Q$. }}
\begin{tabular}{@{}ll|ll}
\hline
$\epsilon\in]1,2]$&&$\epsilon>2$
\\
\hline
\textbf{Region}&\textbf{Class}&\textbf{Region}&\textbf{Class}
\\
\hline
$Q\in]0,M/\epsilon]$&$\textbf{I}$&$Q\in]0,M/\epsilon]$&$ \textbf{I}$
\\
$Q\in]M/\epsilon,Q_n[$&$\mathbf{II}$&$Q\in]M/\epsilon,Q_n[$&$\textbf{VI}$
\\
$Q=Q_n$&$\textbf{III}$&$Q=Q_n$&$ \textbf{VII}$
\\
$Q\in]Q_n,\widetilde{Q}[$&$\textbf{IV}$&$Q\in]Q_n,\widetilde{Q}[$&$ \textbf{IV}$
\\
$Q\in[\widetilde{Q},M[$&$\textbf{V}$&$Q=\widetilde{Q}$&$\textbf{VIII}$
\\
$$&$$&$Q\in]\widetilde{Q},M[$&$ \textbf{IX}$
\\
\hline
\end{tabular}
\end{table*}
We can summarize the results as follows. For $\epsilon\leq 1$ , Table\il\ref{Tabdffautrismeno},
the infimum circular orbital radius  is located at $r_{\gamma}^+$ ($r_{l}^+$)  for $\epsilon<0$ ($0<\epsilon\leq1$).
For negative values of $\epsilon$, circular motion can occur for any radius greater than the infimum value $r_\gamma^+$.
The situation is much more complicated  for positive values of $\epsilon$ within the interval $0<\epsilon\leq1$.
In this case, the electromagnetic interaction is repulsive, but it is  balanced by the attractive gravitational component.
The region in which circular orbits are allowed is split by the radius $r_{\gamma}^+$ and $r_{l}^+$, and in each sub-region
different values of the angular momentum from the set $\{\pm L_{\pm},\pm L_n\}$ are possible.

The case $\epsilon>1$ should be considered apart.
This  corresponds to the real case of  charged elementary particles, like electrons, protons and ions, orbiting around a charged source and hence several possibilities for realistic motion exist.
In the case $\epsilon Q>0$, with $\epsilon>1$,  the circular motion dynamics is determined by the classification given in
Table \ref{Tabdffaubis}, for a fixed source charge-to-mass ratio, or in Table \ref{Tabdffautris},
for a fixed particle  charge-to-mass ratio.

\begin{figure}
\centering
\begin{tabular}{cc}
\includegraphics[scale=0.25]{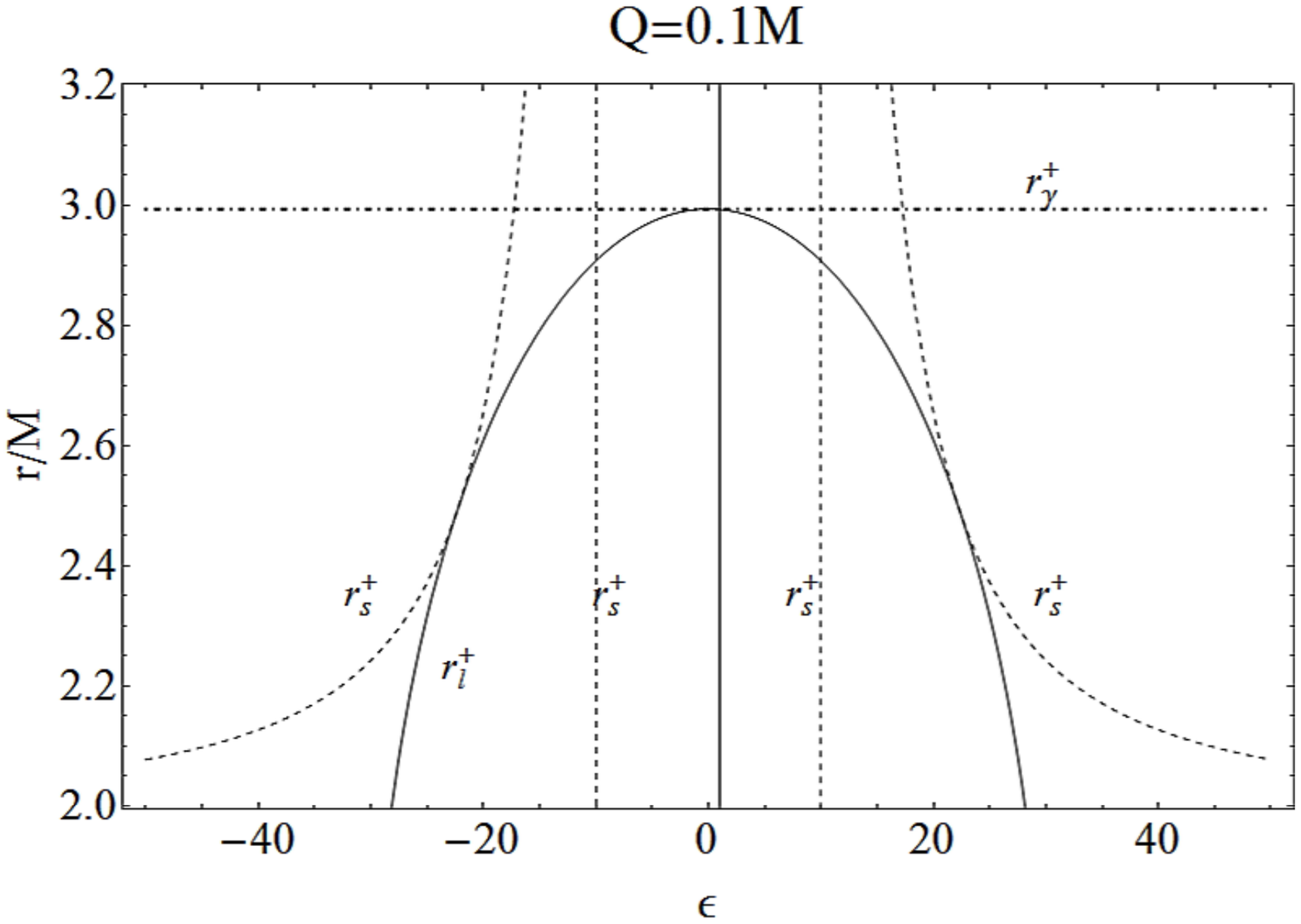}
\\
\includegraphics[scale=0.25]{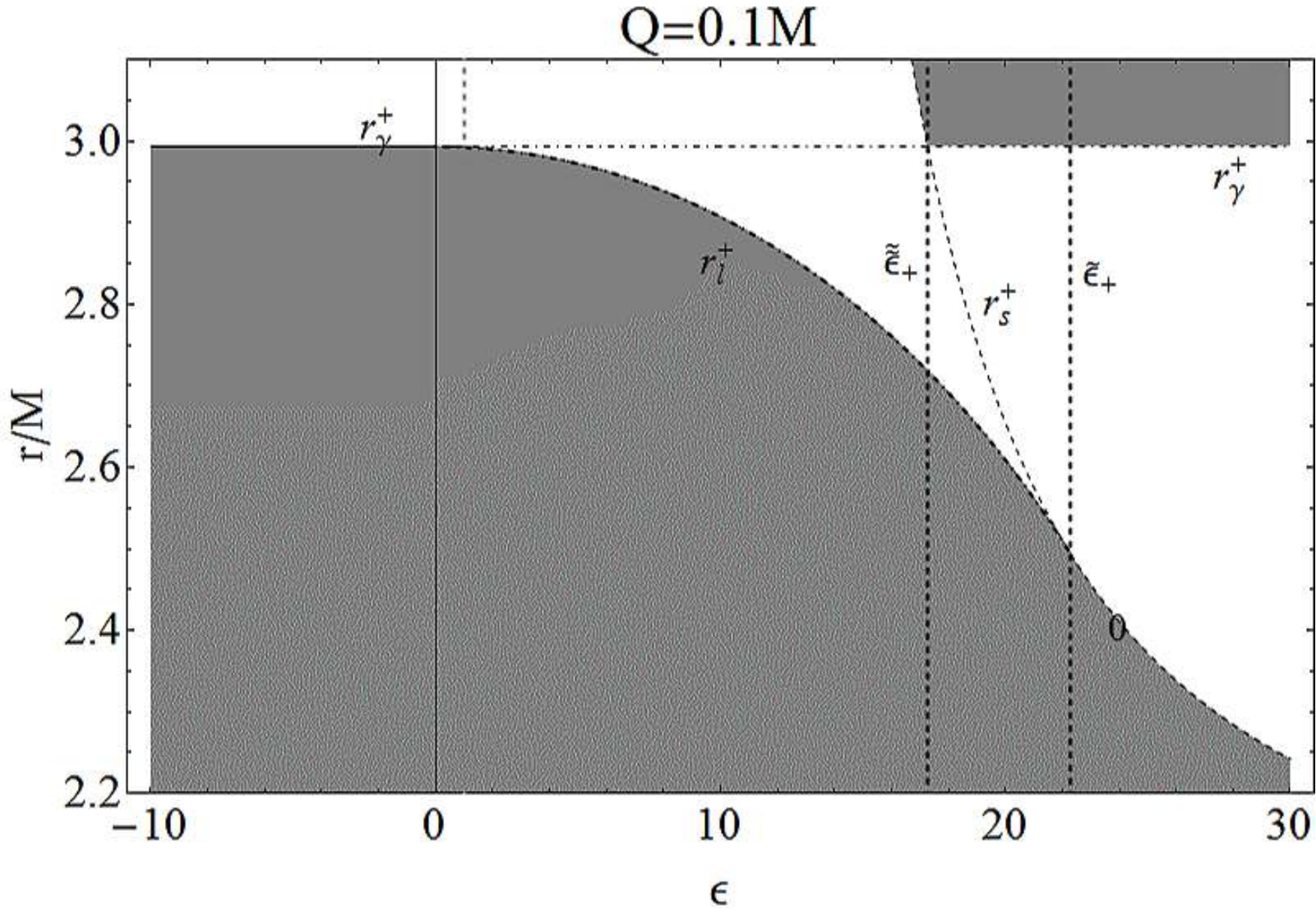}
\end{tabular}
\caption[font={footnotesize,it}]{\footnotesize{Black hole case with $Q=0.1M$. Left panel: The radii $r_s^+$, $r_l^+$ and $r_{\gamma}^+$ are plotted as functions of the test particle charge-to-mass ratio $\epsilon$. Shaded regions are forbidden (right panel).}}
 \label{01t}
\end{figure}

\begin{figure*}
\centering
\begin{tabular}{lccr}
\includegraphics[scale=0.162]{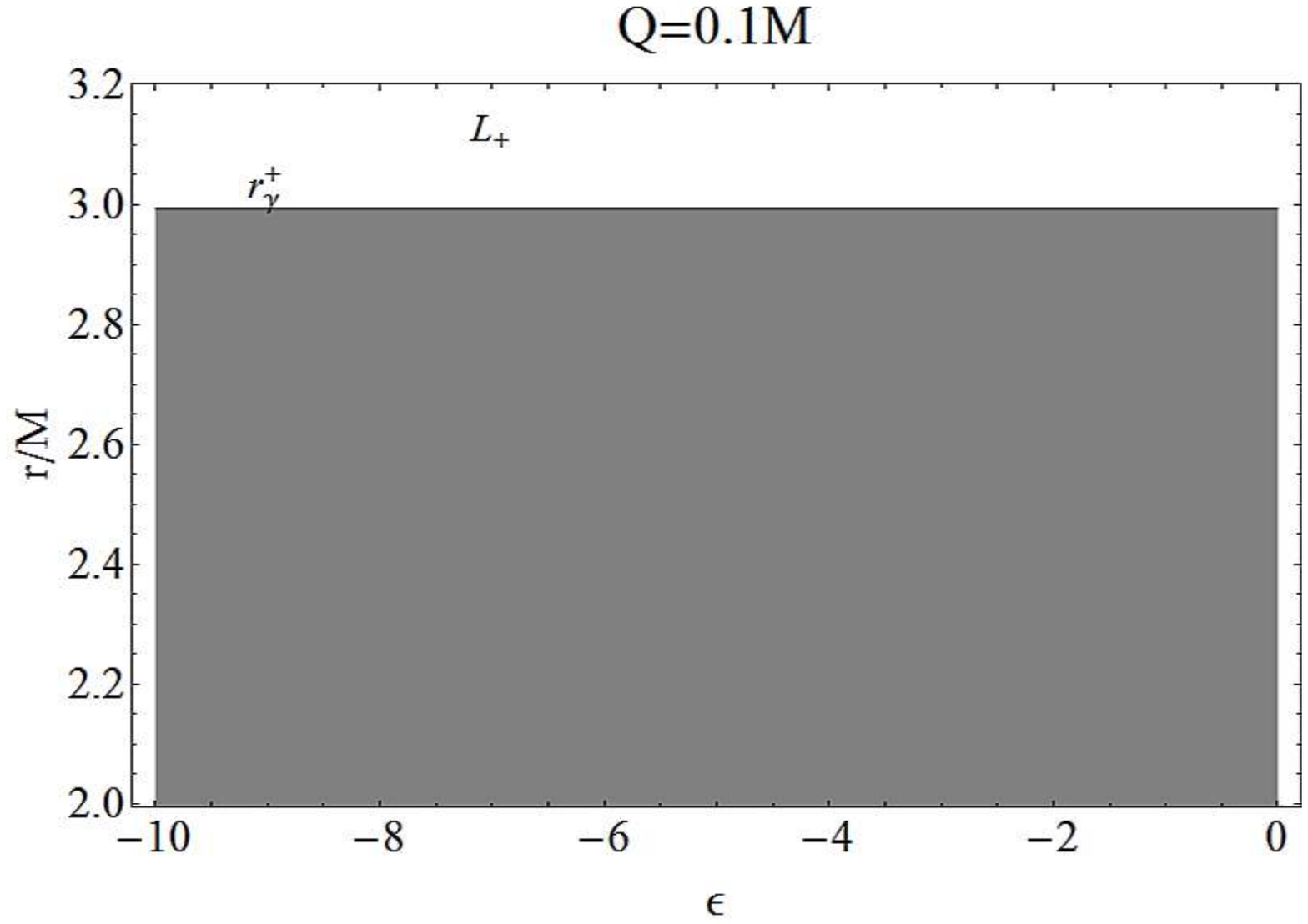}
\includegraphics[scale=0.162]{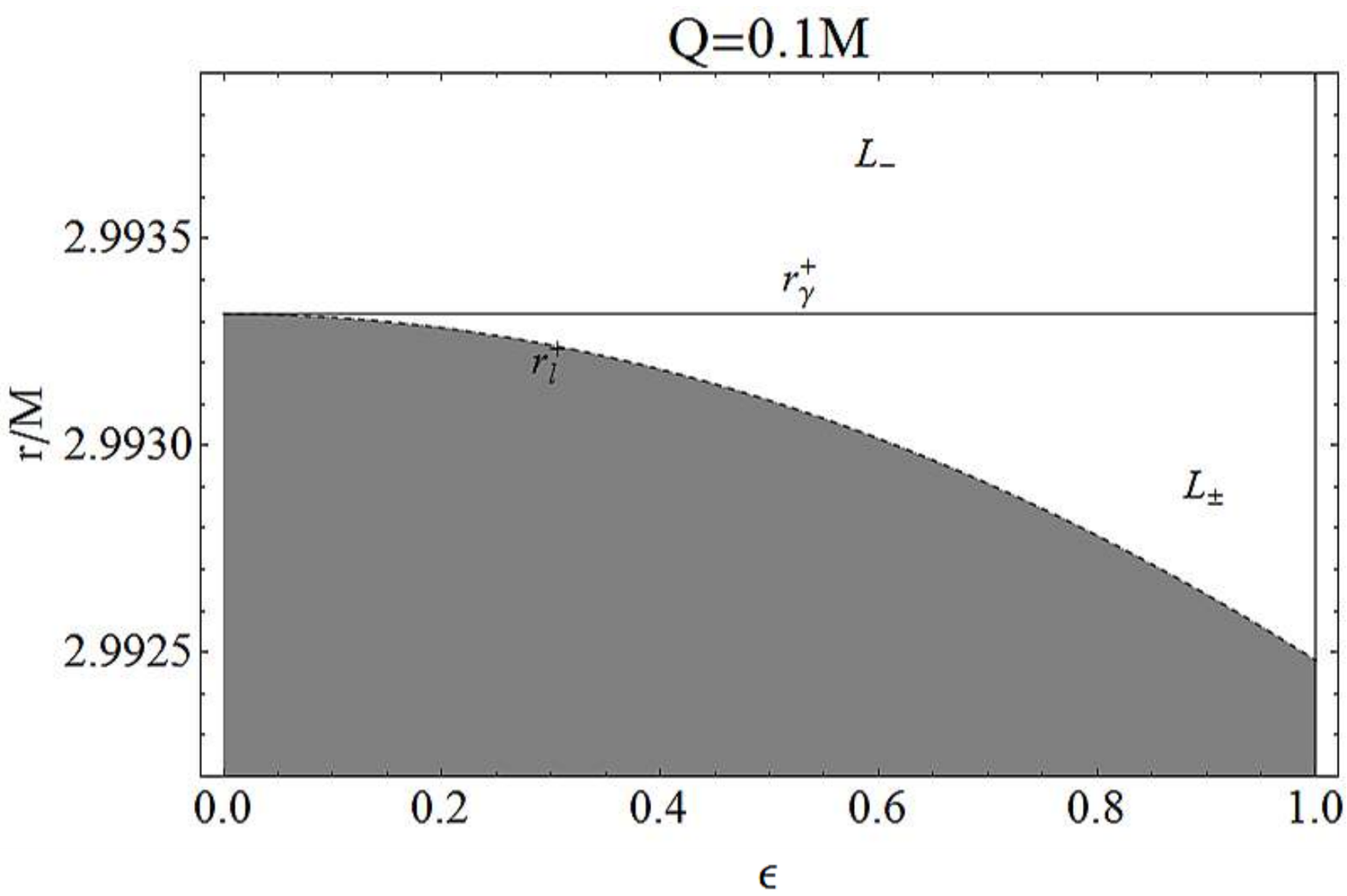}
\includegraphics[scale=0.162]{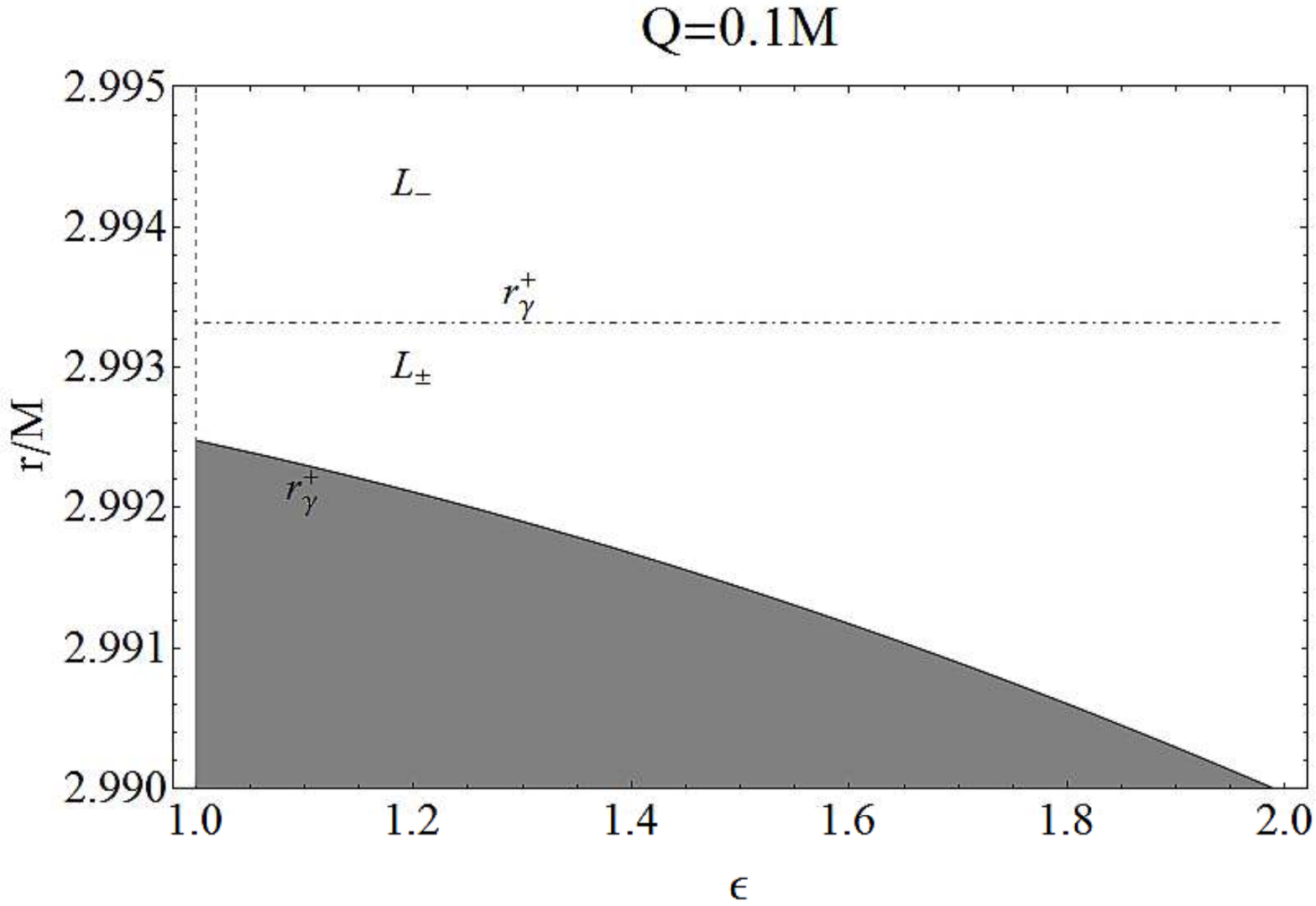}
\includegraphics[scale=0.162]{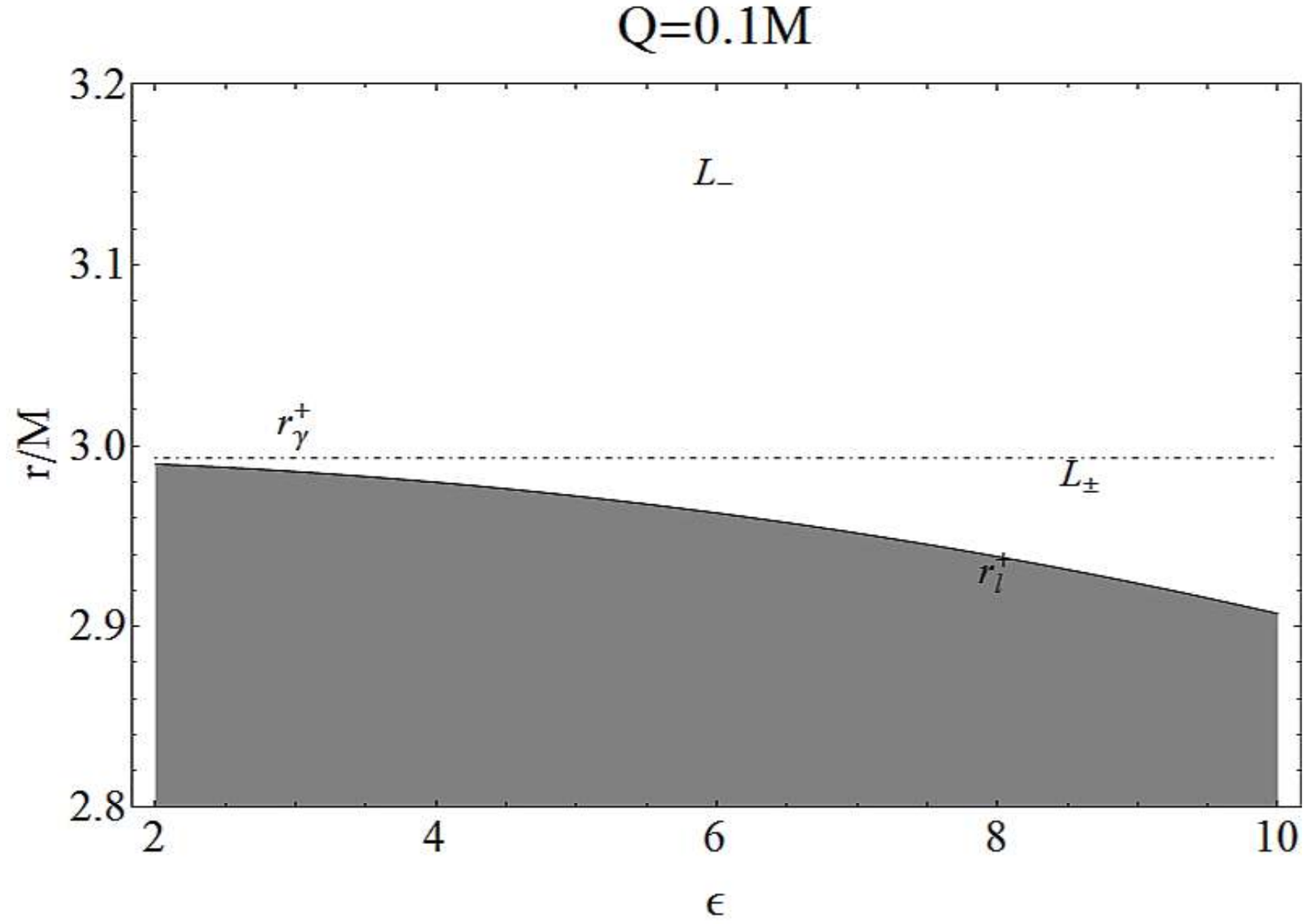}\\
\includegraphics[scale=0.172]{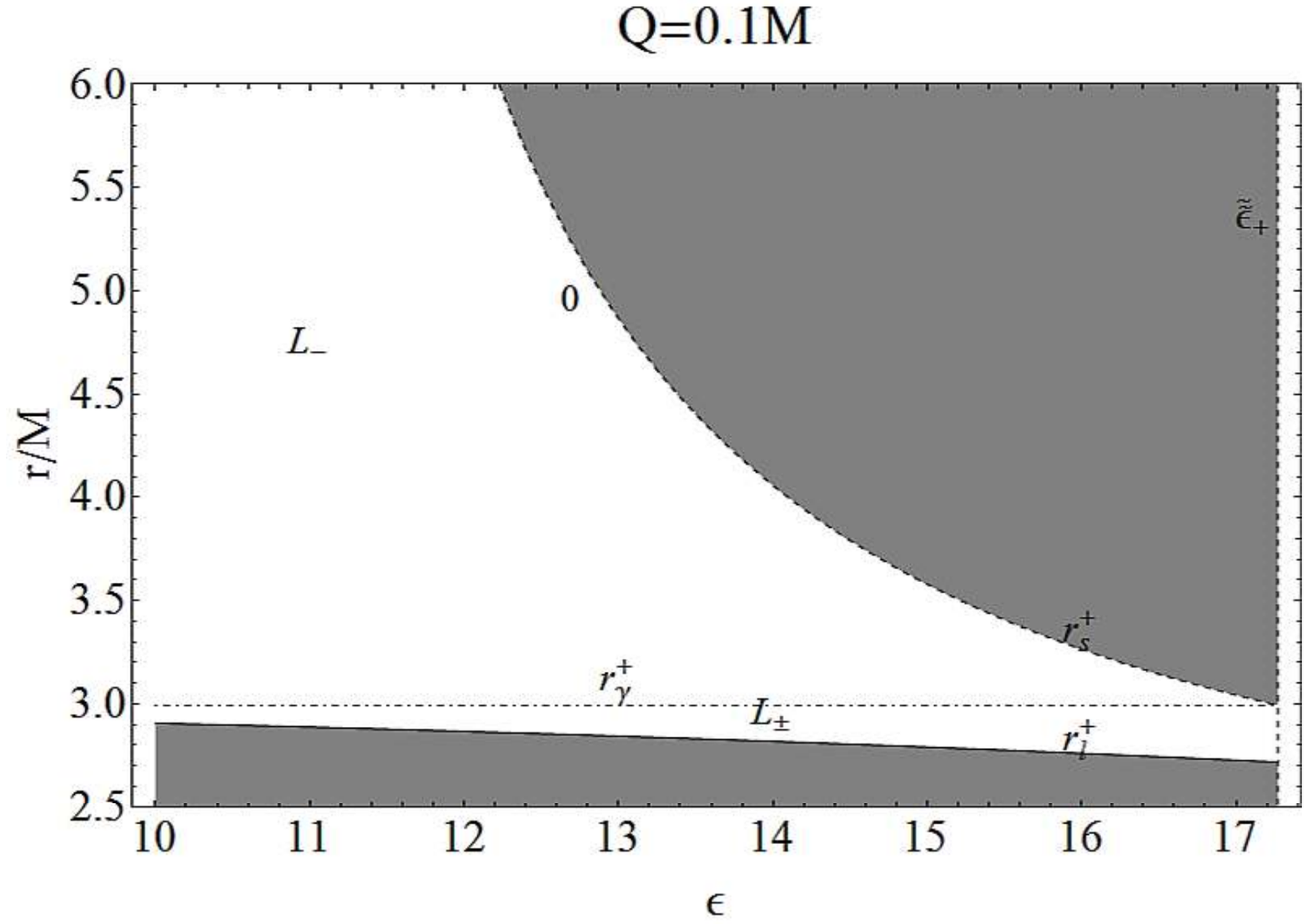}
\includegraphics[scale=0.172]{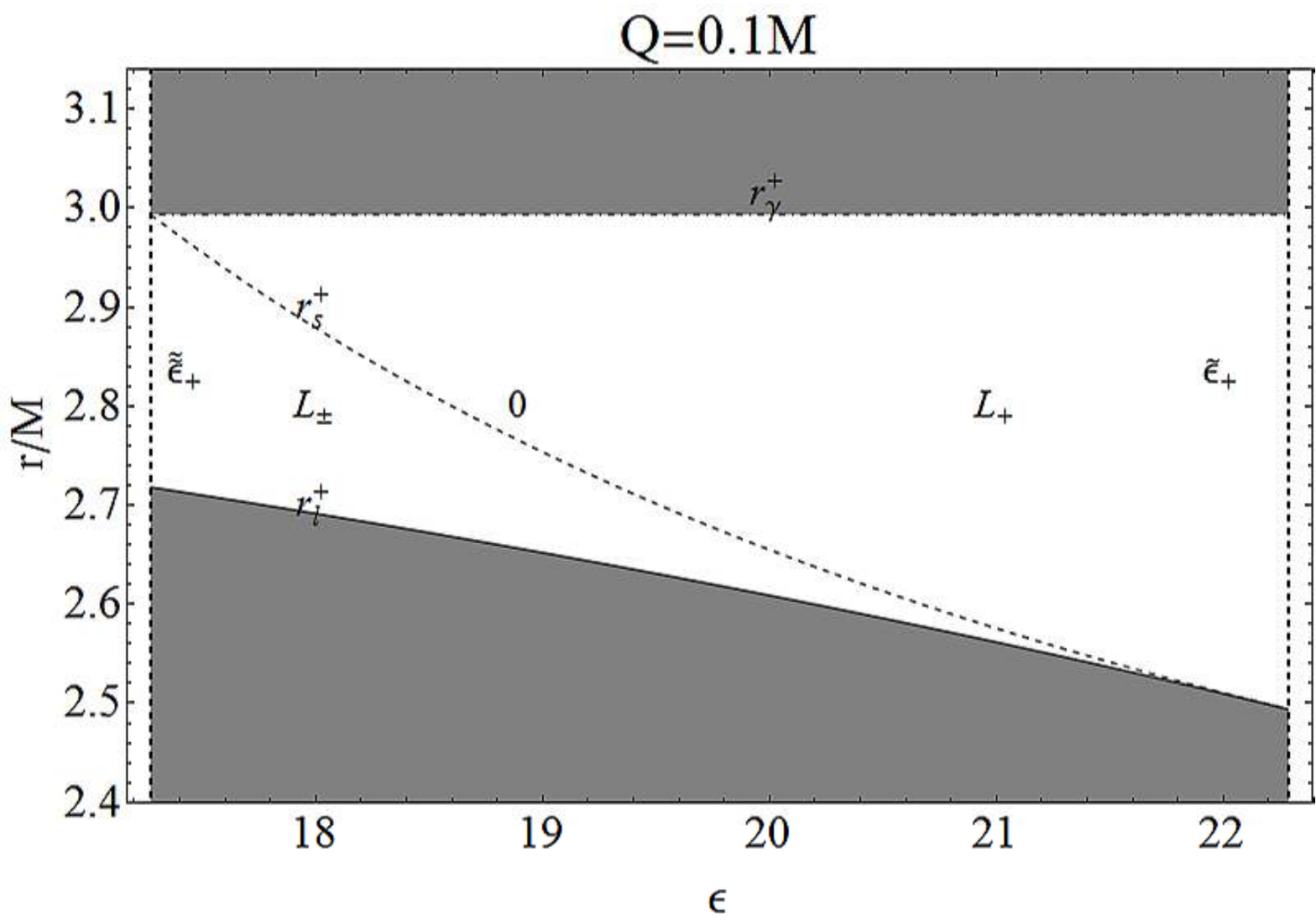}
\includegraphics[scale=0.172]{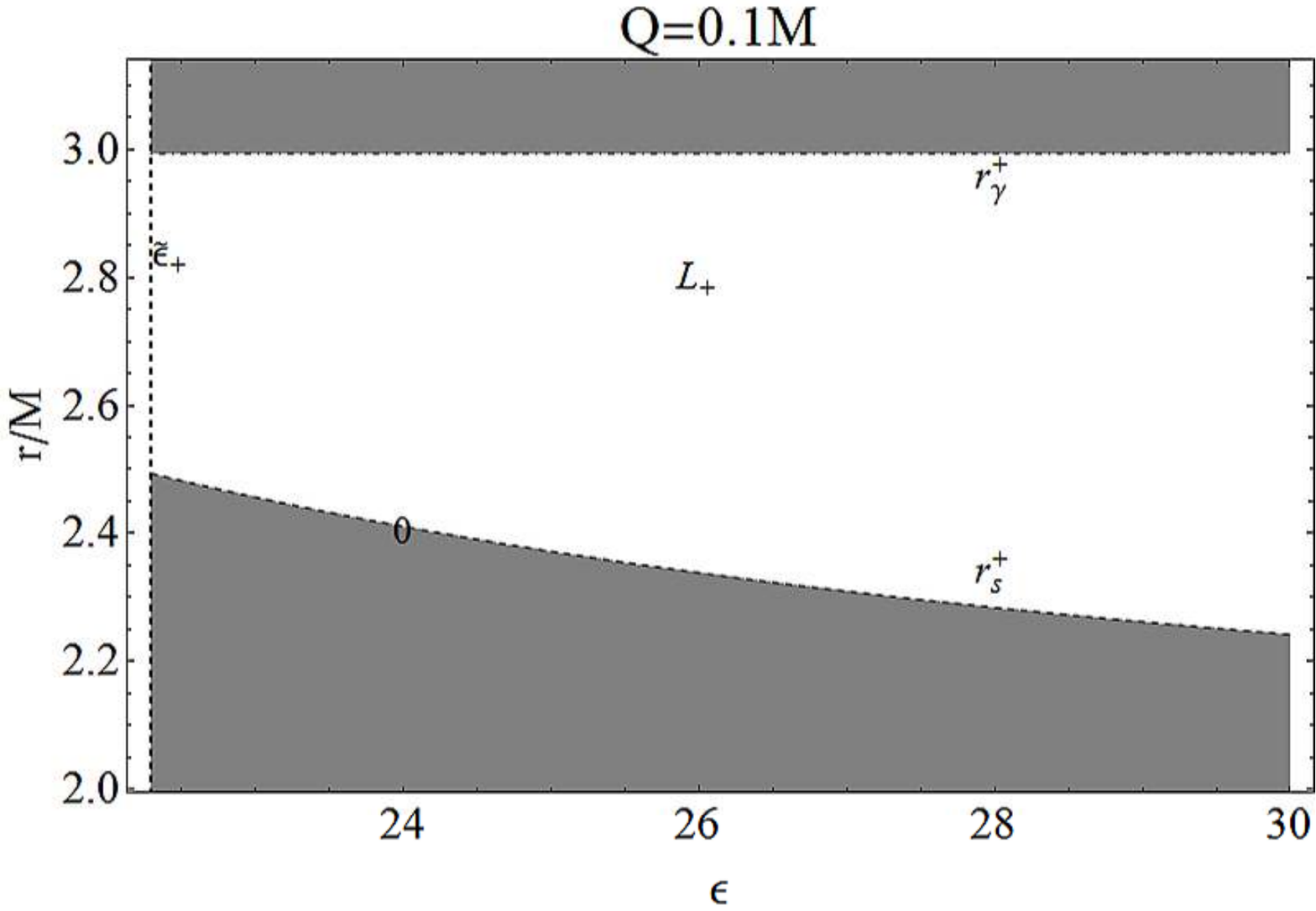}
\end{tabular}
\caption[font={footnotesize,it}]{\footnotesize{Black hole case with  $Q=0.1M$. Shaded regions are forbidden.  The radii $r_s^+$, $r_l^+$ and $r_{\gamma}^+$ are plotted as functions of the test particle charge-to-mass ratio $\epsilon$ in the region $[-10,30]$.}}
 \label{013}
\end{figure*}

First, we can recognize nine different situations listed in Table\il\ref{Tabdffau}. There are  orbits with angular momentum
$L=\pm L_{\pm}$, $L=\pm L_n$ or even $L=0$. As mentioned above, this last case occurs when the particle is located at rest with respect to an observer at infinity. Clearly, these particular ``orbits'' are the  consequence of the full balance between the attraction  and repulsion among   test particles and  sources. The infimum circular orbital radius  can be $r=r_{\gamma}^+$,
$r=r_{l}^+$ or also $r=r_s^+$.  Comparing  with the case $\epsilon<1$, we can clearly see that the main difference between the two cases  consists in the presence of a supremum circular orbital radius at $r=r_{\gamma}^+$  or at  $r=r_s^+$. This means that the repulsive electromagnetic effect, predominant at larger distances, does not  allow  circular orbits around the source.
The classifications of Table\il\ref{Tabdffaubis} and Table\il\ref{Tabdffautris} are  alternative  and equivalent. In the first one, if we fix the charge-to-mass ratio of the black hole, and we move towards increasing values of the particle charge-to-mass ratio, we propose  to divide all the possible scenarios into  four classes of objects characterized by
\bea\label{Iuno}
&&
{\mathbf{I}}: Q/M \in (0,1/2],\quad
{\mathbf{II}}: Q/M \in (1/2, \sqrt{13}/5]\ ,
\\&&\nonumber
{\mathbf{III}}: Q/M \in ( \sqrt{13}/5,\sqrt{2/3})\;
{\textbf{IV}}: Q/M \in ]\sqrt{2/3},1),
\eea
respectively. Thus, for a fixed charge of the  spacetime,  and following  Table\il\ref{Tabdffaubis} and  Table\il\ref{Tabdffau}, we can trace a picture of the dynamical properties of that spacetime.  An example is provided in figure\il\ref{01t} and figure\il\ref{013}, where the case of a spacetime of charge-to-mass ratio $Q/M=0.1$ is illustrated.
On the other hand, in Table\il\ref{Tabdffautris}, if we fix the charge-to-mass ratio of the test particle, and  move towards increasing values of the charge-to-mass ratio of the source, we find only two different cases: $\epsilon\leq2$ and $\epsilon>2$. In conclusion, using this classification, we can determine the orbital radius  followed by the selected particle in the fixed spacetime.

\section{Circular motion around a \textbf{RN} naked singularity}
\label{NSNSTRE}
\begin{figure}
\centering
\begin{tabular}{c}
\includegraphics[width=1\hsize,clip]{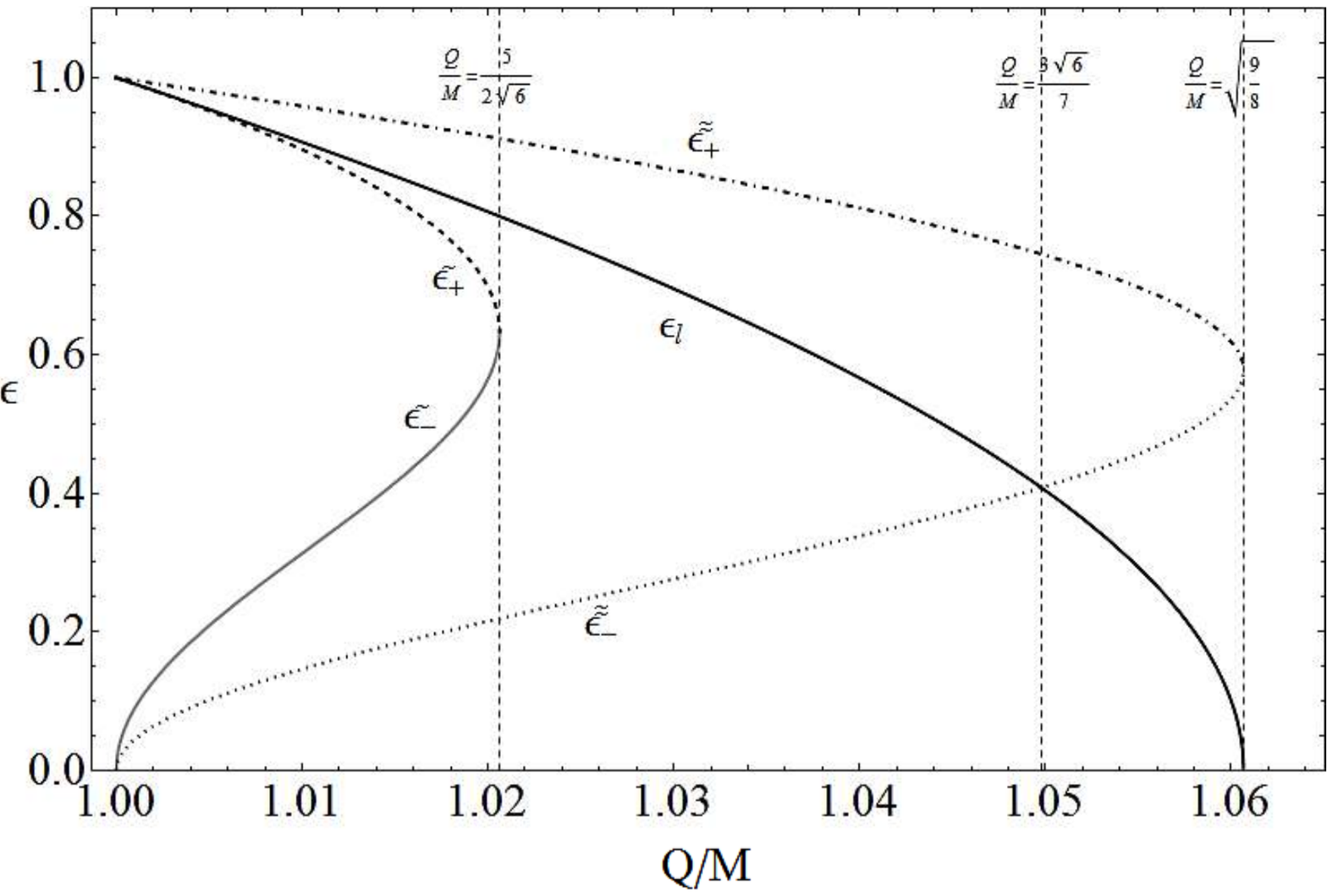}
\end{tabular}
\caption[font={footnotesize,it}]{\footnotesize{The charge parameters
$\epsilon_{l}$, ${\widetilde{\epsilon}}_{\pm}$, and $\widetilde{ \widetilde{\epsilon}}_{\pm}$
as functions of the charge--to--mass ratio of the \textbf{RN} naked singularity.
The special lines  $Q/M=5/\left(2\sqrt{6}\right)\approx 1.02$,
$Q/M=3\sqrt{6}/7\approx 1.05$, and $Q/M=\sqrt{9/8}\approx 1.06$ are also plotted.}}
 \label{Plotettmp}
\end{figure}

Equations \il(\ref{fg2}) govern the  circular motion around a \textbf{RN} naked singularity  {($Q\geq M$}) as well.
Because  (\ref{ECHEEUNAL}) and (\ref{Lagesp}) define the
angular momentum $L_{\pm}$  and the energy $E_{\pm}$
in terms of $r/M$, $Q/M$, and $\epsilon$, it is necessary to investigate several
intervals of values where circular motion is allowed.
To this end, it is useful to introduce  the following
notation
\bea
\label{20}
\epsilon _{l}&\equiv&\frac{\sqrt{9M^{2}-8Q^{2}}}{Q},
\\\label{Eq:second}
\widetilde{\epsilon}_{\pm}&\equiv&
\frac{1}{\sqrt{2}Q}\sqrt{5M^{2}\pm4Q^{2}+\sqrt{25M^{2}-24Q^{2}}}\ ,
\\
\label{r1novo}
\widetilde{\widetilde{\epsilon}}_{\pm}&\equiv& \frac{1}{\sqrt{2}Q} \sqrt{ 3M^2-2
Q^2\pm M\sqrt{9M^2-8 Q^2}} \ ,
\eea
where $\widetilde{\epsilon}_-=\widetilde{\epsilon}$ and $\widetilde{\widetilde{\epsilon}}_+=\epsilon_n$ as defined  in Eqs.(\ref{TTTilde}) and (\ref{TTTilden}), respectively{\footnote{{ The charge parameters  $\epsilon _{l}$,  $\widetilde{\epsilon}_{\pm}$ and
$ \widetilde{\widetilde{\epsilon}}_{\pm}$  in Eqs.\il(\ref{20},\ref{Eq:second},\ref{r1novo}) have been found by parameterizing the motion
of charged particles in naked singularity geometries for the dimensionless charge  $ Q/M $ of the central singularity. In this way, we define the  classes of naked singularities presented in Tables \ref{tabclassI},\ref{tabclassII},\ref{tabclassIII},\ref{tabclassIV},\ref{tabneg1}, and \ref{tabneg2}. Hence the particle motion is analyzed   according to the  appropriate restrictions  on the charge-to-mass ratio of the particle. The  study of the particle  angular momenta has led to the identifications of the charge limits (\ref{20}), (\ref{Eq:second}) and (\ref{r1novo}). An alternative analysis based on a different parametrization  can be found in  \cite{PuQueRu-charge}.}}}.
{These  limiting values for the  particle charge must be real and positive,  therefore, the ranges of definitions for the
charges $(\epsilon _{l},\widetilde{\epsilon}_{\pm},\widetilde{\widetilde{\epsilon}}_{\pm})$ are chosen in accordance with the positive roots of the Eqs.\il(\ref{20},\ref{Eq:second}) and Eq.\il(\ref{r1novo}) in terms of the charge-to-mass ratio $Q/M$ of the naked singularity.
For completeness, and following \cite{PuQueRu-charge}, we reproduce here in Fig. \il\ref{Plotettmp} the behavior of these parameters in terms of the ratio $Q/M>1$, where the definition domains for the charges  $(\epsilon _{l},\widetilde{\epsilon}_{\pm},\widetilde{\widetilde{\epsilon}}_{\pm})$  are also shown. }
The special points where ${\widetilde{\epsilon}}_{+}= {\widetilde{\epsilon}}_{-}$,
$\epsilon_l= \widetilde{ \widetilde{\epsilon}}_{-}$ and
$\widetilde{ \widetilde{\epsilon}}_{+} =\widetilde{ \widetilde{\epsilon}}_{-}$ correspond
to three special values of the ratio $Q/M$ that define four different intervals as follows:
\bea&&\label{I}
{\mathbf{I}}: Q/M \in (1,5/(2\sqrt{6})],\;
\\
&& \nonumber
{\mathbf{II}}: Q/M \in ( 5/(2\sqrt{6}),(3\sqrt{6})/7]\ ,\;
\\
&& \nonumber
{\mathbf{III}}:  Q/M \in ( (3\sqrt{6})/7,\sqrt{9/8}]\ ,\;
\\
&& \nonumber
{\mathbf{IV}}:  Q/M \in [\sqrt{9/8}, \infty).
\eea

Furthermore, it turns out that the properties of circular orbits drastically depends on the sign of the test charge. Moreover,
particles with charges within the interval $-1<\epsilon<1$ present a very rich structure of possible circular orbits. We therefore analyze separately positive and negative test charges in two different intervals.

\subsection{Positive test charges}
\label{sec:pos}

For $\epsilon>0$,  in general, circular orbits exist in the region  $r>r_{*}\equiv Q^{2}/M$.
This means that, in the repulsive case, even a small electric  charge generates a drastic change in the structure
of circular orbits in a \textbf{NS} spacetime, making this an extremely sensitive case.
We therefore introduce a classification of naked singularities which includes four classes ($\mathbf{I_<^+}$, $\mathbf{II_<^+}$, $\mathbf{III_<^+}$, $\mathbf{V_<^+}$) in the
interval of small charges, $0<\epsilon<1$, and two classes ($\mathbf{I>^+}$, $\mathbf{II>^+}$) for large charges, $\epsilon>1$.

In the case of small test charges, it is necessary to consider separately
all the possible values of the charge parameters
$\epsilon_l$, $\widetilde{\epsilon}_\pm$ and $\widetilde{\widetilde{\epsilon}}_\pm$
in all the regions determined by the four classes of naked singularities given in (\ref{I}).
The angular momentum of the test particles depends on the value of the charge-to-mass ration $\epsilon$
and the distance $r$ from the origin of coordinates. The results are schematically represented in the Tables
(\ref{tabclassI})--(\ref{tabclassIV}). A detailed analysis of the behavior of the energy and angular momentum
of positive test charges is presented in the figures of given in \ref{ap:figp}.

\begin{table*}
\caption[font={footnotesize,it}]{\footnotesize{\label{tabclassI}  \textbf{Class} $\mathbf{I_<^+}$: $M<Q\leq 5/(2\sqrt{6})M$ for a test particle charge-to-mass ratio $\epsilon\in]0,1[$.}}
%
\resizebox{1\textwidth}{!}{%
\begin{tabular}{@{}lll|ll|ll}
\hline
& \textbf{a)}: $0<\epsilon\leq\widetilde{\widetilde{\epsilon}}_{-}$ Fig.\il(\ref{Fig:Haidee})&
& \textbf{b)}: $\widetilde{\widetilde{\epsilon}}_{-}<\epsilon<\widetilde{\epsilon}_{-}$  Fig.\il(\ref{Fig:Edmo}) &
& \textbf{c)}: $\widetilde{\epsilon}_{-}\leq \epsilon\leq \widetilde{\epsilon}_{+}$  Fig.\il(\ref{Fig:Dandet})&
\\
\hline
&\textbf{Region}&\textbf{Momentum}
&\textbf{Region}&\textbf{Momentum}
&\textbf{Region}&\textbf{Momentum}
\\
\hline
&$r=r_{s}^+$&$ L=0$
&$(r_{\gamma}^-, r_s^+)$&$ L=\pm L_+$
&$(r_{\gamma}^-, r_s^+)$&$ L=\pm L_+$
\\
&$(r_{s}^+,r_{\gamma}^-) $&$ L=\pm L_{-}$
&$[r_{s}^+,r_{l}^-) $&$ L=\pm L_{\pm}$
&$r=r_s^+$&$ L=0$
\\
&$(r_{\gamma}^-,r_l^-] $&$ L=\pm L_{\pm}$
&$[r_l^+,r_{\gamma}^+)$&$ L=\pm L_{\pm}$
&$[r_l^+,r_{\gamma}^+)$&$L=\pm L_{\pm}$
\\
&$[r_{l}^+ ,r_{\gamma}^+)$&$ L=\pm L_{\pm}$
&$r\geq r_{\gamma}^+$&$ L=\pm L_-$
&$r\geq r_{\gamma}^+ $ &$L=\pm L_-$
\\
&$r\geq r_{\gamma}^+$&$ L=\pm L_{-}$
&&
&$$ &$$
\\
\hline
& \textbf{d)}: $\widetilde{\epsilon}_{+}<\epsilon\leq \epsilon_{l}$ Fig.\il(\ref{Fig:Mercedes})&
& \textbf{e)}: $\epsilon_{l}<\epsilon<\widetilde{\widetilde{\epsilon}}_{+}$ Fig.\il(\ref{Fig:Dandet})&
& \textbf{f)}: $\widetilde{\widetilde{\epsilon}}_{+}\leq \epsilon<M/Q$ Fig.\il(\ref{MC})&
\\
\hline
&\textbf{Region}&\textbf{Momentum}
&\textbf{Region}&\textbf{Momentum}
&\textbf{Region}&\textbf{Momentum}
\\
\hline
&$(r_{\gamma}^-, r_s^+)$&$ L=\pm L_+$
&$(r_{\gamma}^-, r_s^+)$&$L=\pm L_+$
&$(r_{\gamma}^-, r_{\gamma}^+)$&$ L=\pm L_+$
\\
&$[r_{s}^+,r_{l}^-) $&$ L=\pm L_{\pm}$
&$[r_{s}^+,r_{\gamma}^+) $&$  L=\pm L_{\pm}$
&$r>r_s^+ $&$ L=\pm L_{-}$
\\
&$[r_{l}^+,r_{\gamma}^+)$&$ L=\pm L_{\pm}$
&$r\geq r_{\gamma}^+$&$L=\pm L_- $
&$$&$ $
\\
&$r\geq r_{\gamma}^+ $&$ L=\pm L_-$
&$$ &$$
&$$&$ $
\\
\hline
&\textbf{ g)} $M /Q\leq \epsilon<1$  Fig.\il(\ref{Fig:Marsiglia})&
&     &
&  &
\\
\hline
&\textbf{Region}&\textbf{Momentum}
\\
\hline
&$(r_{\gamma}^-, r_{\gamma}^+)$&$ \pm L_+$
\\
\hline
\end{tabular}}
\end{table*}
\begin{table*}
\caption[font={footnotesize,it}]{\footnotesize{\label{tabclassII}  \textbf{Class} $\mathbf{II_<^+}$: $5/(2\sqrt{6})M<Q<(3\sqrt{6}/7)M$, for a test particle charge-to-mass ratio $\epsilon\in]0,1[$.}}
%
\resizebox{1\textwidth}{!}{%
\begin{tabular}{@{}lll|ll|ll}
\hline
& \textbf{a)}: $0<\epsilon<\widetilde{\widetilde{\epsilon}}_{-}$ Fig.\il(\ref{Fig:Morrel}) &
& \textbf{b)}: $\widetilde{\widetilde{\epsilon}}_{-}<\epsilon<\epsilon_{l}$ Fig.\il(\ref{Fig:Chalon})&
& \textbf{c)}: $\epsilon_{l}\le\epsilon \leq\widetilde{\widetilde{\epsilon}}_{+}$ Fig.\il(\ref{Fig:Barone}.)&
\\
\hline
&\textbf{Region}&\textbf{Momentum}
&\textbf{Region}&\textbf{Momentum}
&\textbf{Region}&\textbf{Momentum}
\\
\hline
&$r=r_{s}^+$&$ L=0$
&$(r_{\gamma}^+, r_s^+)$&$ L=\pm L_+$
&$(r_{\gamma}^+, r_s^+)$&$ L=\pm L_+$
\\
&$(r_{s}^+,r_{\gamma}^-] $&$ L=\pm L_{-}$
&$r=r_{s}^+$&$ L=0$
&$[r_{s}^+,r_{\gamma}^+)$&$ L=\pm L_{\pm}$
\\
&$(r_{\gamma}^-,r_{l}^-] $&$ L=\pm L_{\pm}$
&$(r_s^+, r_l^-)$&$ L=\pm L_{\pm}$
&$r\geq r_{\gamma}^+ $&$ L=\pm L_-$
\\
&$[r_{l}^+ ,r_{\gamma}^+)$&$ L=\pm L_{\pm}$
&$[r_{l}^+ ,r_{\gamma}^+)$&$ L=\pm L_{\pm}$
&$$ &$$
\\
&$r\geq r_{\gamma}^+$&$ L=\pm L_{-}$
&$r\geq r_{\gamma}^+$&$ L=\pm L_{-}$
&$$ &$$
\\
\hline
& \textbf{d)}: $\widetilde{\widetilde{\epsilon}}_{+}\leq \epsilon<M/Q$ Fig.\il(\ref{Contino})&
& \textbf{e)}: $M /Q\le\epsilon <1$ Fig.\il(\ref{Fig:Faraone})&
&      &
\\
\hline
&\textbf{Region}&\textbf{Momentum}
&\textbf{Region}&\textbf{Momentum}
&\textbf{Region}&\textbf{Momentum}
\\
\hline
&$(r_{\gamma}^-,r_{\gamma}^+)$&$ L=\pm L_+$
&$(r_{\gamma}^-,r_{\gamma}^+)$&$L=\pm L_+$
\\
&$r>r_{s}^+$&$ L=\pm L_{-}$
&&
\\
&$r=r_{s}^+ $&$ L=0$
&$$&$ $
\\
\hline
\end{tabular}}
\end{table*}
\begin{table*}
\caption[font={footnotesize,it}]{\footnotesize{\label{tabclassIII}
 \textbf{Class} $\mathbf{III_<^+}$: $(3\sqrt{6}/7)M\leq Q\leq\sqrt{9/8}M$, for a test particle charge-to-mass ratio $\epsilon\in]0,1[$.}}
%
\resizebox{1\textwidth}{!}{%
\begin{tabular}{@{}lll|ll|ll}
\hline
& \textbf{a)}: $0<\epsilon\le\epsilon_{l}$ Fig.\il(\ref{Fig:Vampa})&
& \textbf{b)}: $\epsilon_{l}<\epsilon\leq \widetilde{\widetilde{\epsilon}}_{-}$ Fig.\il(\ref{Caracoll})&
& \textbf{c)}: $\widetilde{\widetilde{\epsilon}}_{-}<\epsilon<\widetilde{\widetilde{\epsilon}}_{+}$ Fig.\il(\ref{Villart})&
\\
\hline
&\textbf{Region}&\textbf{Momentum}
&\textbf{Region}&\textbf{Momentum}
&\textbf{Region}&\textbf{Momentum}
\\
\hline
&$[r_{\gamma}^-,r_l^-]$&$ L=\pm L_-$
&$r=r_{s}^+$&$ L=0$
&$(r_{\gamma}^-,r_s^+)$&$ L=\pm L_+$
\\
&$r=r_s^+ $&$ L=0$
&$(r_{s}^+,r_{\gamma}^-) $&$ L=\pm L_{-}$
&$[r_{s}^+,r_{\gamma}^+) $&$ L=\pm L_{\pm}$
\\
&$(r_s^+,r_l^-)$&$ L=\pm L_{\pm}$
&$[r_{\gamma}^-, r_{\gamma}^+)$&$ L=\pm L_{\pm}$
&$r\geq r_{\gamma}^+ $&$ L=\pm L_-$
\\
&$[r_{l}^+ ,r_{\gamma}^+)$&$ L=\pm L_{\pm}$
&$r\geq r_{\gamma}^+ $&$ L=\pm L_-$
&$$ &$$
\\
&$r\geq r_{\gamma}^+$&$ L=\pm L_{-}$
&$$&$ $
&$$ &$$
\\
\hline
& \textbf{d)}: $\widetilde{\widetilde{\epsilon}}_{+}\leq \epsilon<M/Q$ Fig.\il(\ref{Fig:Catalani}) &
& \textbf{e)}: $M/Q\le\epsilon <1$ Fig.\il(\ref{Epinay})&
&  &
\\
\hline
&\textbf{Region}&\textbf{Momentum}
&\textbf{Region}&\textbf{Momentum}
&\textbf{Region}&\textbf{Momentum}
\\
\hline
&$(r_{\gamma}^-,r_{\gamma}^+)$&$ L=\pm L_+$
&$(r_{\gamma}^-,r_{\gamma}^+)$&$ L=\pm L_+$
\\
&$r=r_s^+$&$ L=0$
&$$&$ $
\\
&$r>r_{s}^+ $&$ L=\pm L_+;\ L=\pm L_-$
&$$&$ $
\\
\hline
&\textbf{Class}  $\mathbf{IV_{<}^+}$ $Q>\sqrt{9/8}M$ &&&
\\
\hline\hline
&$0<\epsilon<M/Q$&&&
\\
\hline
&\textbf{Region}&\textbf{Momentum}&&
\\\hline
&$r=r_s^+$& $L=0$&&
\\
&$r>r_s^+$& $L=\pm L_-$&&
\\
\hline
\end{tabular}}
\end{table*}

\begin{table*}
\caption[font={footnotesize,it}]{\footnotesize{\label{tabclassIV} Classes for large test charges ($\epsilon >1$). }}
%
\begin{tabular}{@{}llll|ll}
\hline
&\textbf{Class} \textbf{$\mathbf{I_>^+}$}: $M<Q<\sqrt{9/8} M$&
&&\textbf{Class} {\textbf{$\mathbf{II_>^+}$}}: $\sqrt{9/8} M<Q$&
\\
\hline
&\textbf{Region} &\textbf{Momentum}&
&\textbf{Region}&\textbf{Momentum}
\\
\hline
&$(r^-_\gamma,r^+_\gamma) $&$ L= L_+$&
&Forbidden&
\\
\hline
\end{tabular}
\end{table*}
We see that in the interval of small test charges several subclasses appear which are delimited by the value of the charge
parameters $\epsilon_l$, $\widetilde \epsilon _\pm$ and $\widetilde{\widetilde{\epsilon}} _ \pm$.
In general we can summarize the situation as follows.
There is always a minimum radius $r_{min}$ at which circular motion is allowed.
At the radius $r_\gamma^-$ the energy
 of the test
particle diverges, indicating that the hypersurface $r=r_\gamma^-$ is lightlike. In the simplest case, there is  a minimum radius $r_{min}$ so that circular orbits are allowed
in the  infinite interval $]r_{min},\infty)$.
Otherwise, this region is split by  a lightlike hypersurface situated at $r_\gamma^+>r_{min}$.

Another possible structure is that of a spatial configuration formed by two separated regions in which circular motion is allowed in a finite region filled with charged particles within the spatial
interval $(r_{min}=r_\gamma^-,r_{max}=r_\gamma^+)$. This region is usually
surrounded by an empty finite region in which no motion is allowed. Outside the empty region, we find a zone of allowed circular
motion in which either only neutral particles or neutral and charged particles can exist in circular motion.

The situation in the case
of large charges $(\epsilon >1)$ is simpler. There is only one region in class $\mathbf{I_>^+}$ in which circular motion can exist.
Naked singularities within the Class $\mathbf{II_>^+}$ do not allow any circular orbits.
In fact, since for $\epsilon Q>0$ the Coulomb interaction is repulsive, the configuration characterized by the values
 $Q\geq\sqrt{9/8}M$  and  $\epsilon>1$ corresponds to a repulsive electromagnetic effect that cannot be  balanced
by an attractive gravitational interaction.

The energy and angular momentum of circular orbits diverge as $r$ approaches the limiting orbits at $r_\gamma^\pm $;
 the boundary $r=r_\gamma^+$ in this case corresponds to a  lightlike hypersurface.
Finally, the energy of the states is always positive.
%
\subsection{Negative test charges}
\label{sec:neg}
In the case of test particles with negative charge, the dependence of circular orbits on the charge-to-mass ratio of the naked singularity
is simpler than in the case of positive charges. Indeed, we need to consider only two intervals defined as follows:
\be
\mathbf{I^-}: \ M<Q<\sqrt{9/8}M \quad \mathbf{II^-}: \ Q/M> \sqrt{9/8} .
\ee
\textbf{Large negative particle charge-to-mass ratio:}
For $\epsilon<-1$, the contribution of the electromagnetic interaction is always attractive. Hence,
the  repulsive force necessary to balance the attractive effects of the Coulomb interactions can be
generated only by a \textbf{RN} naked singularity. In particular,
for $\epsilon<-1$ and for $Q>\sqrt{9/8}M$ (\textbf{Class} $\mathbf{II^-}$)  circular orbits with $L=L_+$
always exist for $r>0$.
For $M<Q\leq\sqrt{9/8}M$ (\textbf{Class} $\mathbf{I^-}$), circular orbits exist with
$L=L_+$ in $0<r<r_\gamma^-$ and $r>r_\gamma^+ $.
Charged test particles with $\epsilon<-1$ can move along circular orbits also  in the region $(0,r_{*}]$.
The value of the energy on circular orbits increases  as $r$ approaches  $r=0$, and the angular momentum, as seen by an observer located at infinity, decreases as the radius of the orbit decreases.
In  the region $M<Q\leq\sqrt{9/8}M$, two limiting orbits appear at $r_\gamma^\pm $ (similar to  the  neutral particle case \cite{Pugliese:2010ps}).

\textbf{Small negative particle charge-to-mass ratio:}
In the case of small charges $(-1<\epsilon<0)$, it is necessary to split the analysis into three different intervals. The results
are summarized in Tables \ref{tabneg1} and \ref{tabneg2}. We classify naked singularities into two classes, according to the interval of $Q/M$ to which they belong. In general, two different configurations are allowed. For  $Q>\sqrt{9/8}M$ (\textbf{Class} $\mathbf{II^-}$)
a continuous region appears   from a minimum radius $r_{min}\in\{r_s^{\pm}, Q^2/(2M)\}$ to infinity in which circular orbits are allowed.
For  $M<Q\leq\sqrt{9/8}M$ (\textbf{Class} $\mathbf{I^-}$) there is a  non connected region $(r_{min},\infty)$  inside which there is
a forbidden region  $(r_\gamma^-,r_\gamma^+)$. The configuration is therefore composed of two
disconnected regions.

In  \ref{ap:fign}, we include several figures that depict the behavior of the angular momentum and energy of negative test charges.
\begin{table*}
\caption[font={footnotesize,it}]{\footnotesize{\label{tabneg1} \textbf{Class} $\mathbf{I^-}$  $M<Q\leq\sqrt{9/8}M$}}
%
\resizebox{1\textwidth}{!}{%
\begin{tabular}{@{}lll|ll|ll}
\hline
& \textbf{a)}: $\epsilon<-1 $  &
& \textbf{b)}: $-1< \epsilon<-M/Q$ Fig.\il(\ref{Reveng}) &
& \textbf{c)}: $\epsilon=-M/Q $  Fig.\il(\ref{Oro})&
\\
\hline
&\textbf{Region}&\textbf{Momentum}
&\textbf{Region}&\textbf{Momentum}
&\textbf{Region}&\textbf{Momentum}
\\
\hline
&$(0,r_{\gamma}^-)$&$ L=\pm L_+$
&$r=r_s^+$&$ L=0$
&$r=Q^2/2M$&$ L=0$
\\
&$r>r_{\gamma}^+ $&$ L=\pm L_{+}$
&$(r_s^+,r_{\gamma}^-)$&$ L=\pm L_+$
&$(Q^2/2M,r_{\gamma}^-) $&$ L=\pm L_{+}$
\\
&$$&$ $
&$r>r_{\gamma}^+ $&$ L=\pm L_+$
&$r>r_{\gamma}^+ $&$ L=\pm L_+$
\\
\hline
& \textbf{d)}: $-M/Q<\epsilon< 0$  Fig.\il(\ref{Grotta})&
&   &
&  &
\\
\hline
&\textbf{Region}&\textbf{Momentum}
\\
\hline
&$r=r_{s}^-$&$ L=0$
\\
&$(r_{s}^-,r_{\gamma}^-) $&$ L=\pm L_{+}$
\\
&$r>r_{\gamma}^+ $&$ L=\pm L_+$
\\
\hline
\end{tabular}}
\end{table*}
\begin{table*}
\caption[font={footnotesize,it}]{\footnotesize{\label{tabneg2} \textbf{Class} $\mathbf{II^-}$ $Q>\sqrt{9/8}M$}}
%
\resizebox{1.1\textwidth}{!}{%
\begin{tabular}{@{}lll|ll|ll}
\hline
& \textbf{a)}: $\epsilon<-1 $ &
& \textbf{b)}: $-1< \epsilon<-M/Q$ Fig.\il(\ref{Franz})&
& \textbf{c)}: $\epsilon=-M/Q $  Fig.\il(\ref{Beuch})&
\\
\hline
&\textbf{Region}&\textbf{Momentum}
&\textbf{Region}&\textbf{Momentum}
&\textbf{Region}&\textbf{Momentum}
\\
\hline
&$r>0$&$ L=\pm L_+$
&$r=r_{s}^+$&$ L=0$
&$r=Q^2/2M$&$ L=0$
\\
&$$&$ $
&$r>r_s^+$&$ L=\pm L_{+}$
&$r>Q^2/2M$&$ L=\pm L_{+}$
\\
\hline
& \textbf{d)}: $-M/Q<\epsilon< 0$ Fig.\il(\ref{Beuchnuovo}) &
\\
\hline
&\textbf{Region}&\textbf{Momentum}
\\
\hline
&$r=r_{s}^-$&$ L=0$
\\
&$r>r_{s}^- $&$ L=\pm L_{+}$
\\
\hline
\end{tabular}}
\end{table*}
%
%
\section{Discussion and future perspectives}
\label{Sec:m.l-ka}
In this work,  we explored  the motion of charged test particles along circular orbits in the spacetime described by the
Reissner--Nordstr\"om  metric.
A detailed discussion of the dynamics for the  black hole and  naked singularity cases has been performed. Circular orbits have been classified in detail for a complete set of cases.

We   adopt the effective potential  approach to
study test particle orbits in the Reissner--Nordstr\"om spacetime, focusing on the  equatorial orbits.
We explore  the morphology of the orbital
regions.  This analysis leads to a  clear differentiation between  naked singularities and black holes. A remarkable implication of this description is that the circular orbit configuration  in the black hole  case
is not allowed in the naked singularity regime. Instead, the study of the circular orbits trace out a possible way to distinguish the two physical situations.

In a series of  previous works \cite{Pugliese:2010ps,Pugliese:2010he,PuQueRu-charge}, we analyzed the dynamics of the \textbf{RN} spacetime, and  studied the motion of neutral and charged test particles, by using the   effective
potential approach.
We showed that in the case of charged test particles the term
$\epsilon Q/r$  drastically changes the behavior of the effective potential. The study shows the existence of stability regions
whose geometric structure clearly distinguishes naked singularities from black holes (see also \cite{Virbhadra:2002ju, Virbhadra:2007kw} and \cite{do1,Dabrowski:2002iw}).
In \cite{PuQueRu-charge}, in particular, we studied the spatial regions of the \textbf{RN} spacetime where circular motion is allowed around either black holes or naked singularities.  We showed that the geometric structure of stable accretion disks allows us to clearly distinguish between black holes and naked singularities.
In this work, we presented  the complete classification of  circular motion around a \textbf{RN} black hole, and  around a \textbf{RN} naked singularity
with $0<\epsilon<1$ and $-1<\epsilon<0$.

Clearly, this analysis could be used
to construct  an accretion disk with disconnected rings made of test particles.
A  precise
characterization of matter configurations
surrounding a charged compact astrophysical
object  can account for significant
astrophysical processes observed in the electromagnetic band, like the jet emissions. Therefore understanding the dynamics around compact objects  is important for the understanding of the accretion disk phenomena and  the classification of their  general properties.

In the naked singularity case,  gravity can assume a repulsive nature that produces a complicated picture of dynamics around the source. We distinguish four regions of charge-to-mass ratio values of the source that characterized these configurations. In each region, different cases can occur:
an infinite, continuum, region or otherwise a disconnected region.
In the black hole case the circular orbits configuration strongly varies if $\epsilon\leq1$ or $\epsilon>1$. An  infimum radius always appears, and in the case $\epsilon>1$ a supremum circular orbits radius can appear.

The existence of special schemes for circular orbits  in \textbf{RN} geometries as outlined in Tables\il \ref{Tabdffautrismeno},\ref{Tabdffau},\ref{Tabdffaubis},\ref{Tabdffautris},\ref{tabclassI},\ref{tabclassII},\ref{tabclassIII},\ref{tabclassIV},\ref{tabneg1}, and \ref{tabneg2}
show a complicated  scenario able to distinguish classes of attractors and particles in motion identified through  their charge-to-mass ratios.
More importantly, this investigation has enlighten particular limiting values for the \textbf{BH}  and  \textbf{NS} metric parameters    and, remarkably, notable values for the charge-to mass ratio of the orbiting particle.
The importance of these studies lies also  in the fact that for a long time it has been discussed  about the existence and interpretation of the limits of the charge-to-mass ratio and of the dimensionless spin emerging from the General Relativistic formulation of
non-quantum self-gravitating objects with  electric charge or intrinsic  spin. These descriptions have been performed  by using   the relativistic  exact solutions of  Kerr, Reissner-Nordstr\"om and Kerr-Newman. In the case of black holes, considered in  Tables\il \ref{Tabdffautrismeno},\ref{Tabdffau},\ref{Tabdffaubis}, and \ref{Tabdffautris},  the situation is very clear. For negatively charged elementary particles,  $\epsilon <0$  (with attractive force $Q \epsilon<0$) the  region of circular motion  is bounded from below by the photon circular orbit.
The situation is definitely complicated for particles with small positive charges, $\epsilon\in ]0,1[$,
where an articulated structure, determined by the new limiting radius $r_l$, appears.
The role of the radius  $r_*=Q^2/M$ for {zero angular momentum particles} (\textbf{ZAMPs}) in \textbf{BH} geometries and  $r_{\star}\equiv Q^2/2M$ for   \textbf{ZAMPs} in \textbf{NS} spacetimes are thoroughly  considered, finding a  generalization with  radius $r_s^+(Q)$. It is interesting to note that $r_*$  corresponds to  the value of the classical
radius of an electric charge\footnote{We  note that $
 r_{s}= {2GM}/{c^{2}}$ is the Schwarzschild radius
$r_{Q}^{2}= {Q^{2}G}/{4\pi \epsilon _{0}c^{4}}$  is the scale radius corresponding to the central mass $M$ with an  electric charge $Q$, where $G$  the gravitational constant  and
 $(4\pi\epsilon_0)^{-1}$ is the Coulomb constant ($\epsilon_0$ is here the  electric permittivity).}. However this proves that   high negative charges (attractive interaction)  have some similarities with small charges $\epsilon\in]0,1[$   with  repulsive force
$(\epsilon Q/M <1)$.
The repulsive case however, $Q\epsilon>0$, shows a  richer  classification of possible cases  as $\epsilon $ increases (see
Tables\il\ref{Tabdffau} and \ref{Tabdffaubis}).
From the rich structure
  of the cases shown in Table \il\ref{Tabdffaubis},  we point out the presence of particular limiting values for
black holes, namely  $\left.Q/M\right|_{\lim}\in (1/2, \sqrt{13}/5, \sqrt{2/3})$, and for charged  particles, namely
 $\left. \epsilon\right|_{\lim}\in(1,2, M/Q,\epsilon_n(Q),\tilde{\epsilon}(Q))$.
  However,  Table\il\ref{Tabdffautris}  clearly shows  that $\epsilon =1$ and $\epsilon=2$ are particularly important limiting particle charges.
  Naked singularities,  on the other hand, must  be classified as weak $(Q/M< \sqrt{9/8})$ or  strong  $(Q/M> \sqrt{9/8})$ compact objects. We recognize four classes for the charge parameters with limiting values
 $\left.Q/M\right|_{\lim}\in (1, 5/ (2 \sqrt{6}),3 \sqrt{6}/7, \sqrt{9/8})$,
  and different values for the particle charge, for instance, $\epsilon_{\lim}=M /Q$\footnote{
   A similar limit, present also in  the analysis of the \textbf{BH} regimes,  has been  found with respect to the angular momentum and rotational frequency for an orbiting particle in different regions of the Kerr geometry. In this  case, the correspondence is between
the charge  and the spin of the central source, namely,  a limiting value of $\approx M/a$ \cite{Ergon,Observers1,Pugliese:2011xn,Pugliese:2013zma}.}. These results are important in the characterization  of interactions between self-gravitating
charged objects and charged  particles, particularly for parameters close to the limiting values pointed out here.
Further limits have been studied, for example, in
\cite{Zakharov:2014lqa}. {
In \cite{Blaschke:2016uyo}, some limiting values have been  found in the context of  Keplerian accretion in braneworld Kerr-Newman spacetimes. In particular, in the limit of \textbf{RN}-solutions, the parameter values  $Q/M=\sqrt{5/4}$ and $Q/M=\sqrt{9/8}$ have already been identified.}

More generally, these studies can also be applied in the investigation of the coupling between  gravitational  and electromagnetic waves, which is a well-known feature of this geometry,  and of  the role of the electric charge in the gravitational collapse. Indeed, the
Reissner-Nordstr\"om metric is a static electrovacuum, spherically symmetric exact solution of Einstein-Maxwell equations
with a radial electric field. This metric can also be interpreted  as  describing  the exterior gravitational and electromagnetic fields
of a static, expanding or collapsing,  or oscillating  spherically symmetric,
electrically charged body.  Stability and instability of Reissner-Nordstr\"om black holes and, particularly,  the extreme case
are investigated,  for  example,  in \cite{Aretakis:2011hc,Lucietti:2012xr,Aretakis:2011ha,Aretakis:2010gd} and in
\cite{117,118,Dain:2012qw}. Quasinormal modes of nearly extreme Reissner-Nordstr\"om black holes are studied in
\cite{Andersson:1996xw}.
The investigation  of this solution still leaves open the problem of finding a precise astrophysical situation in which
the metric is of considerable relevance. Indeed, it is usually
 assumed that, if formed, a  highly charged object, such as a  black hole with $(Q/M\in]0,1[)$, would  in short time  be  neutralized by   some matter-field environments (see, for instance, \cite{Hubeny:1998ga}).
Despite this, the interest towards the  Reissner-Nordstr\"om solution  is still huge. For instance,
several applications have been found also as an extension  in super-symmetric  analysis and  super-string theories
\cite{117,118,119}, where it is
 considered  even as a general relativistic (non-quantum) model  of  charged elementary
particles ($\epsilon\gtrless\pm 1$).
The \textbf{RN} metric has been also considered for a number of stellar objects  as a limiting  electrovacuum solution
determined by matching an interior solution with an exterior
naked singularity. The interior solution can be a boson star  \cite{Pugliese:2013gsa},
a solution with  cosmological constant   \cite{conLam} or
the class of astrophysical Tolman-Bayin solutions \cite{Pi-las}.
{An interesting comparison of the results found here could be carried out by considering similarities with braneworld scenarios. Indeed,   assuming a spherically symmetric metric induced on the 3-brane, the constrained
effective gravitational field equations on the brane can be  solved, giving \textbf{NS} and
\textbf{BH}  solutions  with a braneworld parameter playing the role of a tidal charge (see \cite{Schee:2008kz} and \cite{Stuchlik:2008fy}).
  Further applications may be found by considering the  regular spacetimes related to non-linear electrodynamics solutions  discussed in
   \cite{Stuchlik:2014qja,Schee:2015nua}}.
Furthermore, the role of the
electric charge during the  collapse  of compact stars  remains still to be fully understood \cite{B,z,Pugliese:2014yla}.
  In general, the study of collapsing stars permits to investigate   the  relationship between the gravitational and electromagnetic fields in a geometrized approach to unification where, for example,   the  electromagnetic wave
 gives rise to gravitational waves,  constituting  therefore one  of the  most intriguing  aspects of the   possible conversion of electromagnetic energy into
gravitational energy and  \emph{viceversa}. This is a very remarkably feature of  the coupling between the electromagnetic and gravitational perturbations \cite{Bicak:2000ea,Chandra,B-PRD,Cardoso:2016olt}.

We expect that   future analysis in these directions  can  place more precisely in this context  the parameter limits  for the central source and the charged particles; these additional analysis  could give a clear explanation of the existence and role of these limits.

\begin{acknowledgements}
One of us (DP) acknowledges support from a Junior GACR Grant of the Czech Science Foundation No:16-03564Y. DP also
gratefully acknowledges financial support from  the  Blanceflor Boncompagni-Ludovisi n\'ee Bildt Foundation
in the first part of this work.
This work has been supported by the UNAM-DGAPA-PAPIIT, Grant  No. IN111617.
\end{acknowledgements}

\appendix

\section{Behavior of the angular momentum and energy of positive test charges}
\label{ap:figp}
\begin{figure}
\centering
\begin{tabular}{cc}
\includegraphics[width=0.55\hsize,clip]{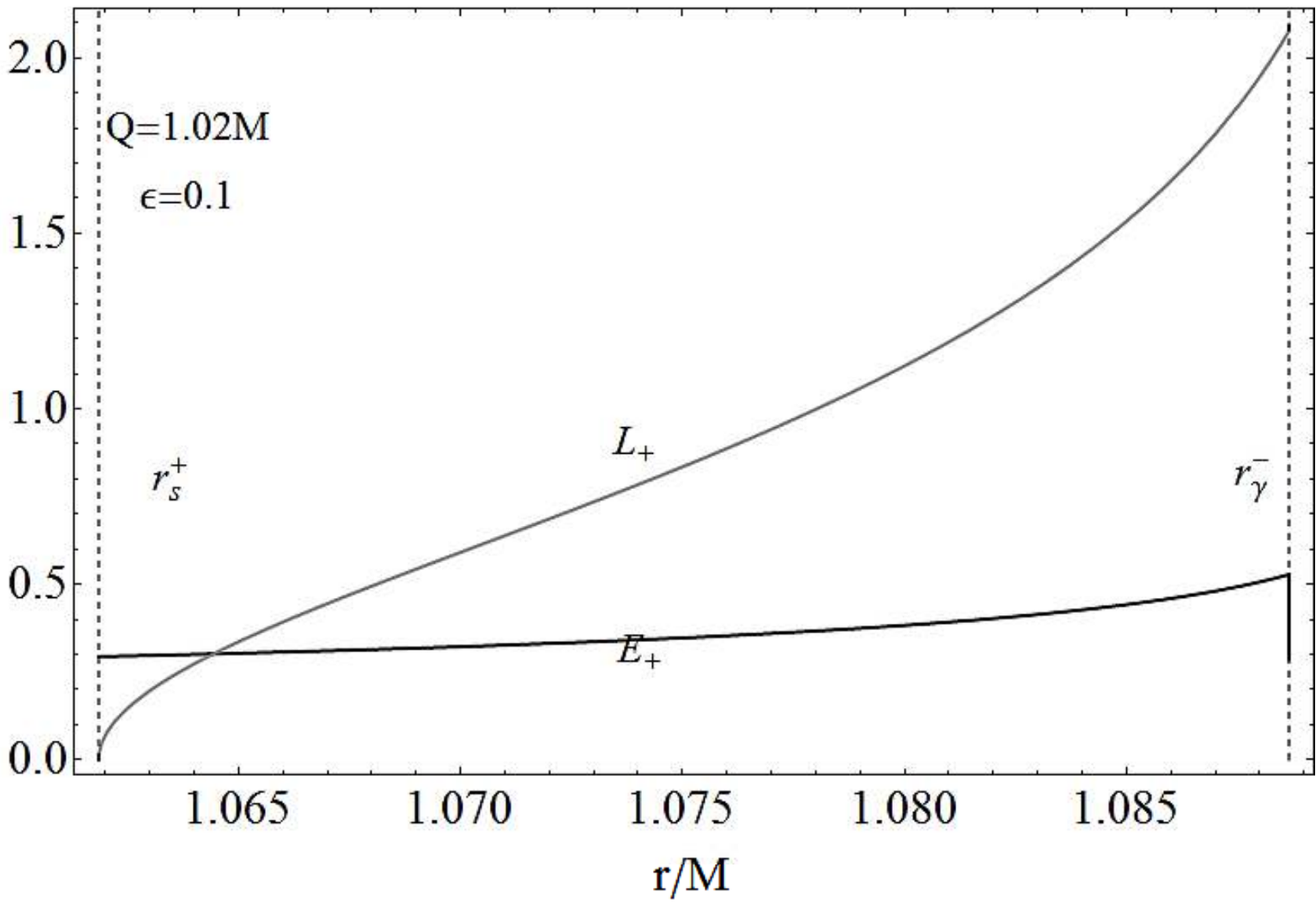}
\includegraphics[width=0.55\hsize,clip]{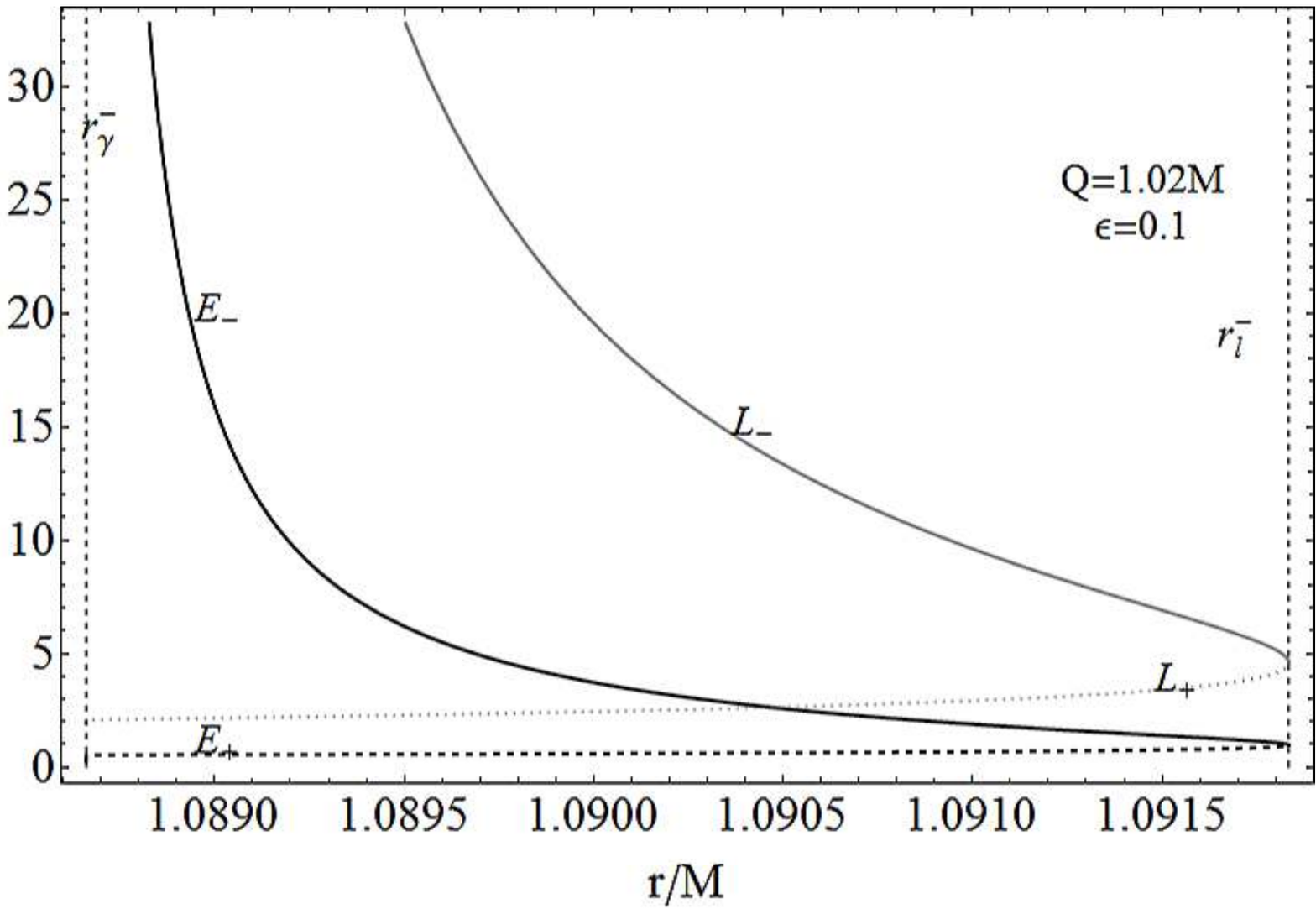}\\
\includegraphics[width=0.55\hsize,clip]{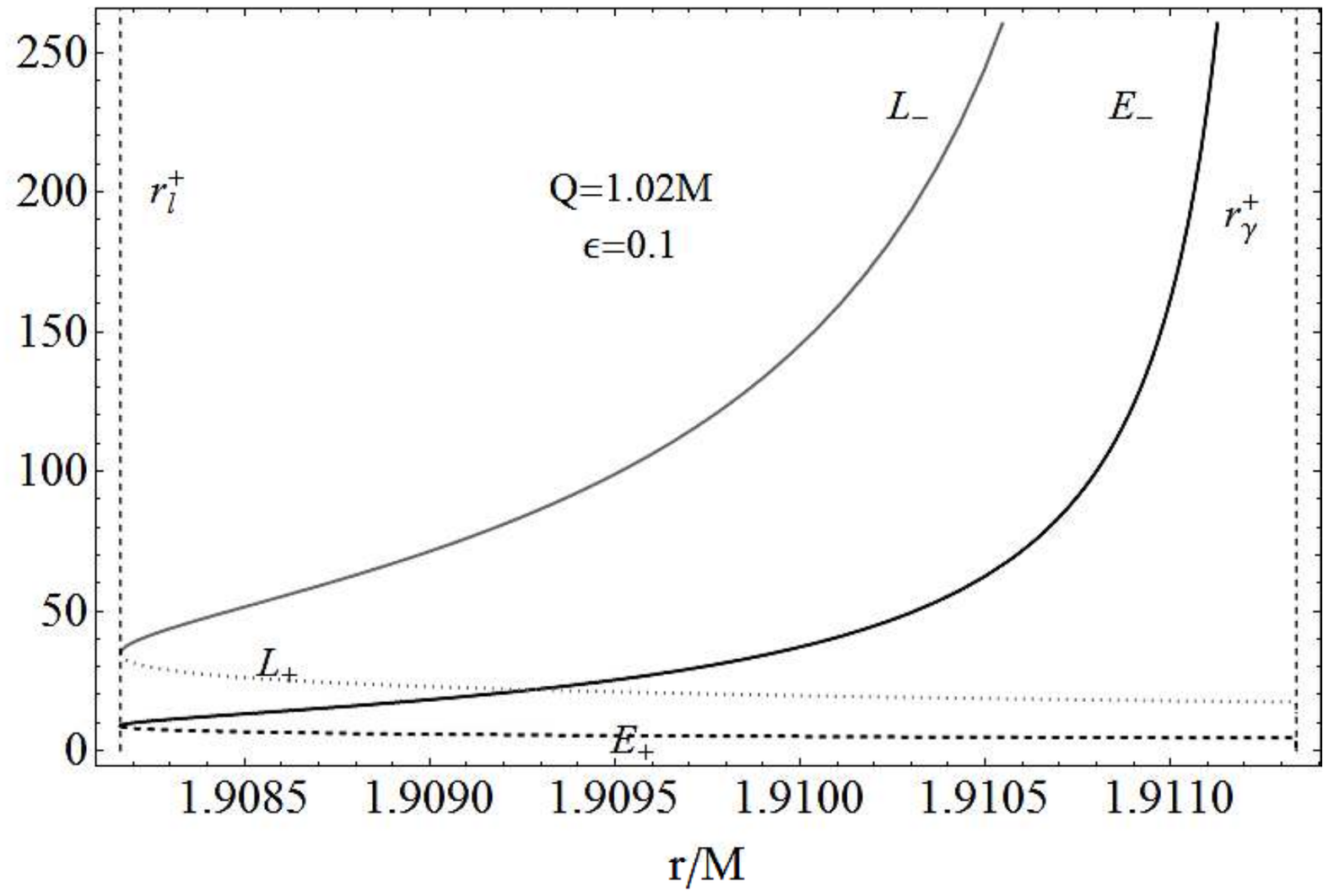}
\includegraphics[width=0.55\hsize,clip]{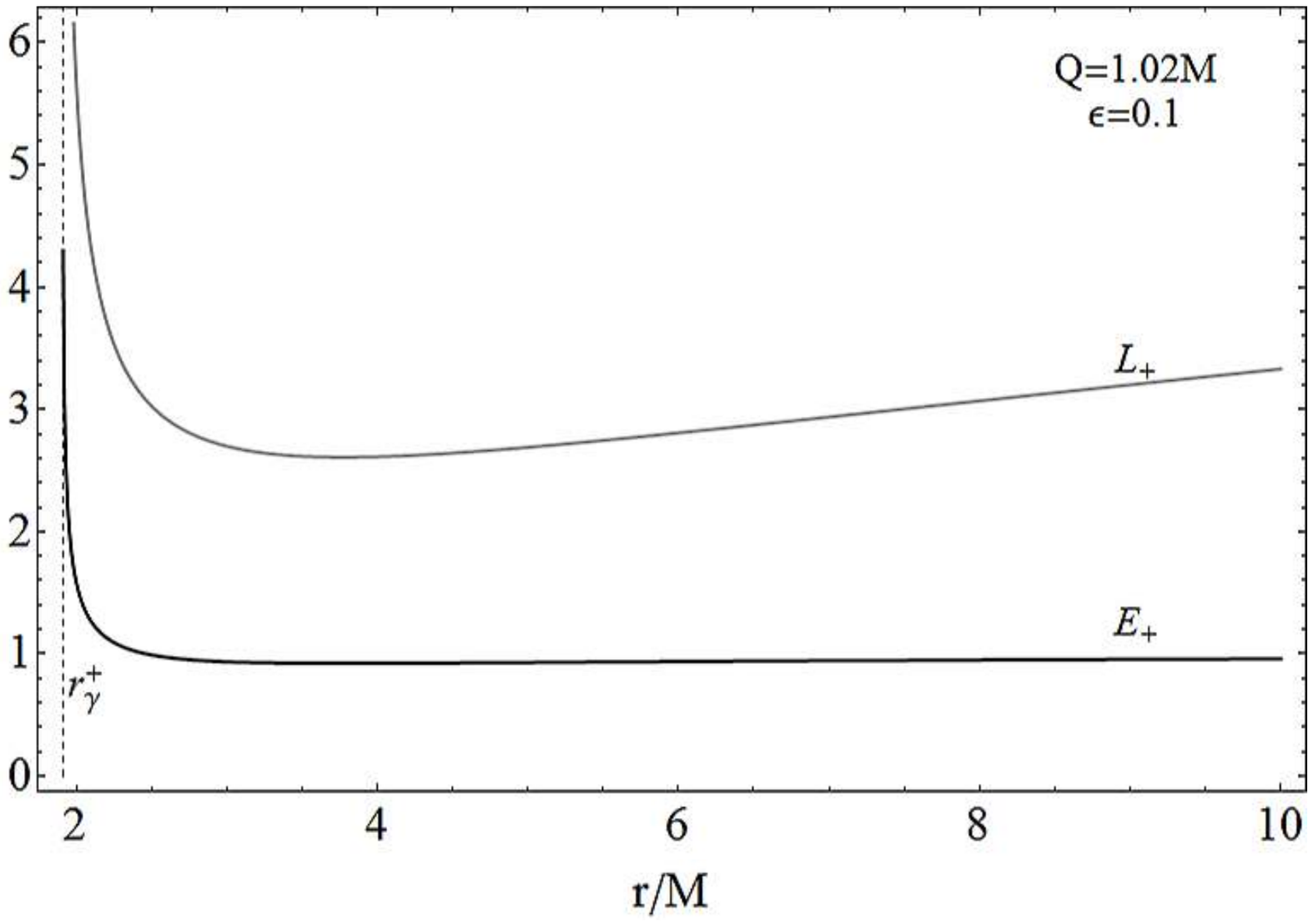}
\end{tabular}
\caption[font={footnotesize,it}]{\footnotesize{\textbf{Class}:  $M<Q\leq 5/(2\sqrt{6})M$ and $0<\epsilon\leq\widetilde{\widetilde{\epsilon}}_{-}$.
Parameter choice: $Q= 1.02M$ and $\epsilon=0.1$. Then $\widetilde{\widetilde{\epsilon}}_{-}=0.215376$,
$r_{s}^+ =1.06185M$, $r_\gamma^-=1.08866M$, $r_\gamma^+ =1.91134M$, $r_{l}^{-} =1.09183M$, and
$r_l^+  =1.90817M$.
Circular orbits exist with angular momentum
$L=L^{-}$ (gray curve) and energy $E=E^{-}$ (black curve) in $r_{s}^+ <r<r_\gamma^-$;
$L=0$ at $r=r_{s}^+ $ (upper left plot);
$L=L_{\pm}$ in  $r_\gamma^-<r\leq r_{l}^{-}$ (upper right plot) and  in $r_l^+ \leq r<r_\gamma^+ $ (bottom left plot);
$L=L^{-}$  in $r\geq r_\gamma^+ $ (bottom right plot).
The angular momentum $L_+$ is represented by a gray dotted curve and the energy $E_+$ by a black dashed curve. }}
\label{Fig:Haidee}
\end{figure}
\begin{figure}
\centering
\begin{tabular}{cc}
\includegraphics[width=0.55\hsize,clip]{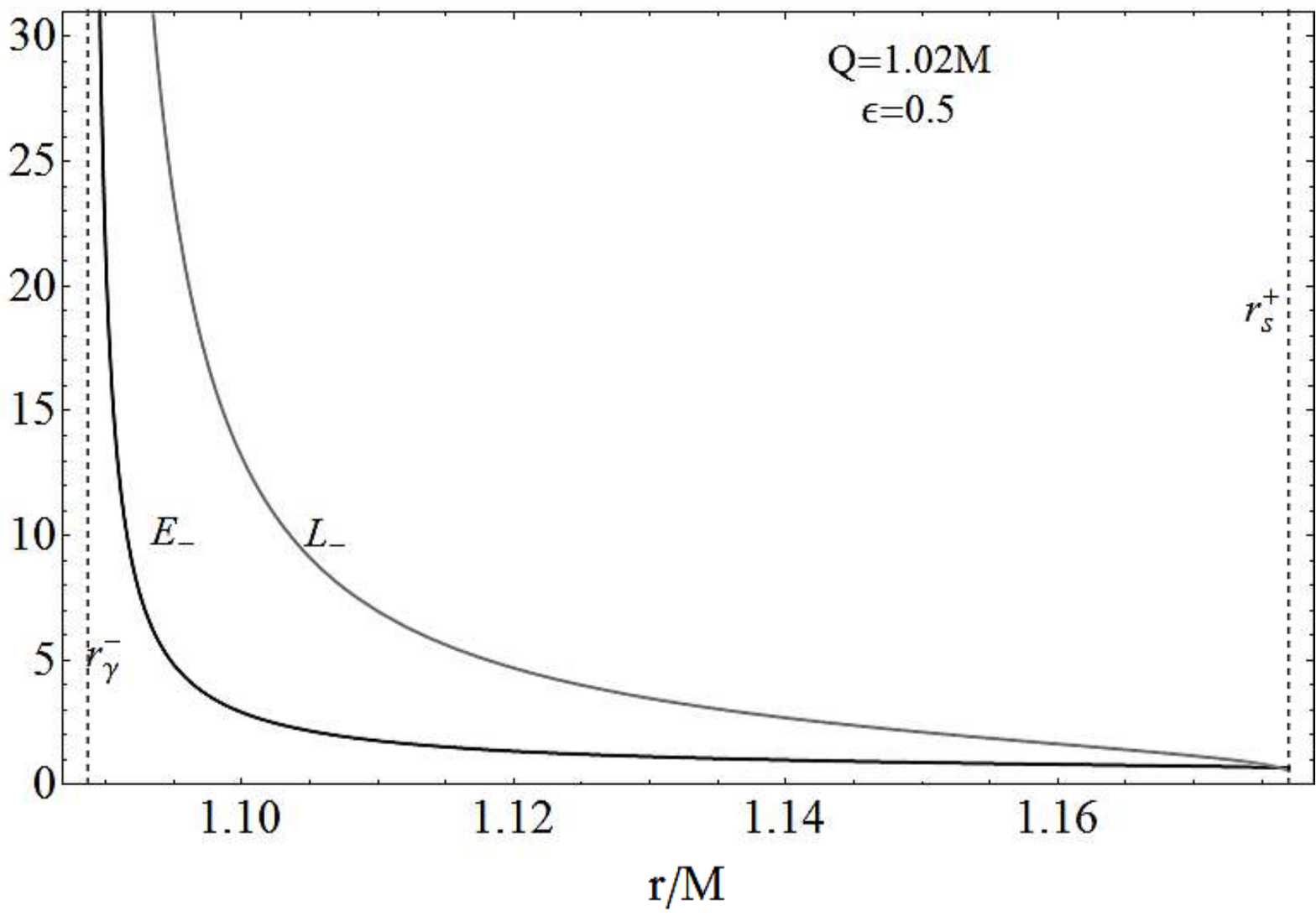}
\includegraphics[width=0.55\hsize,clip]{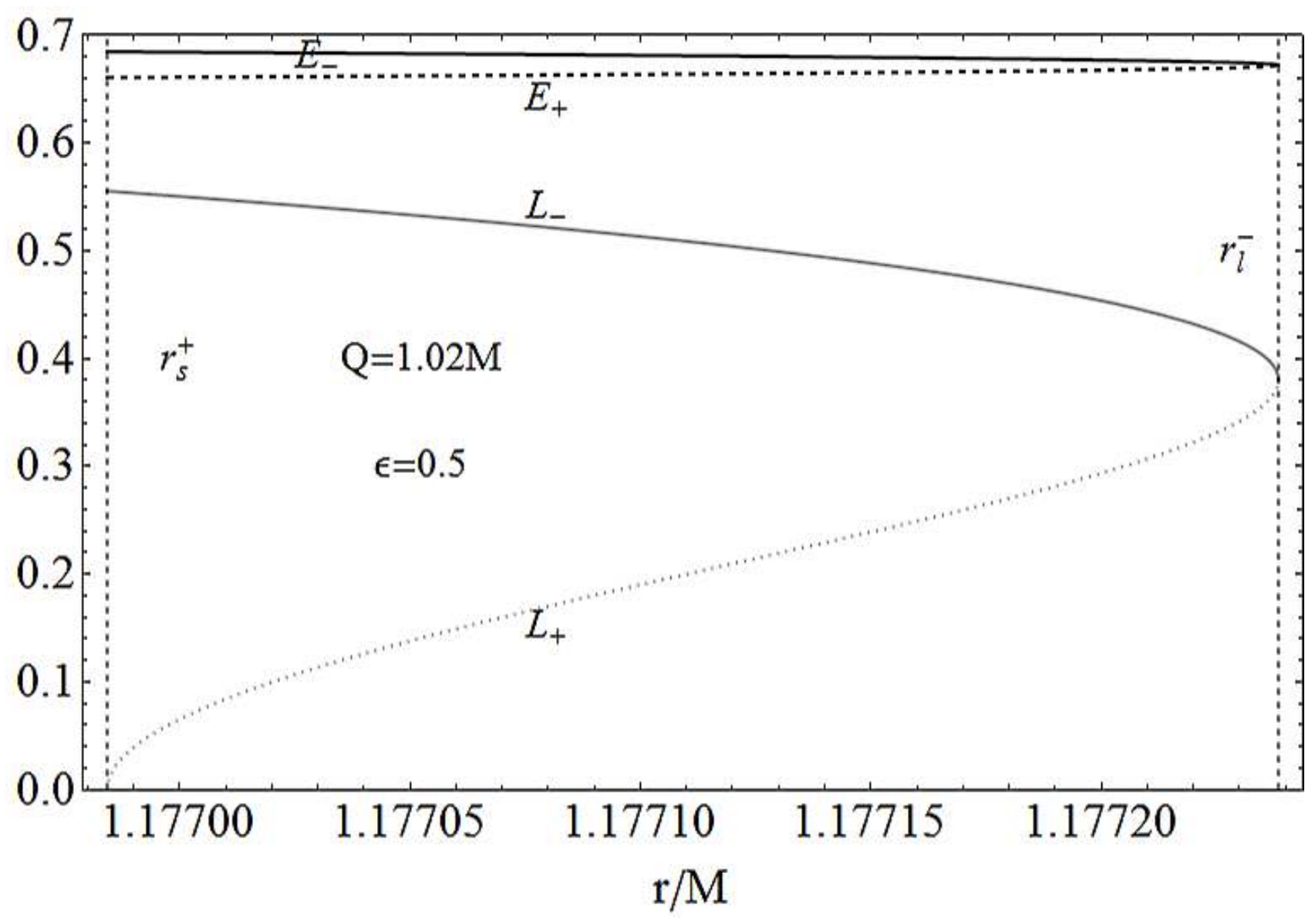}\\
\includegraphics[width=0.55\hsize,clip]{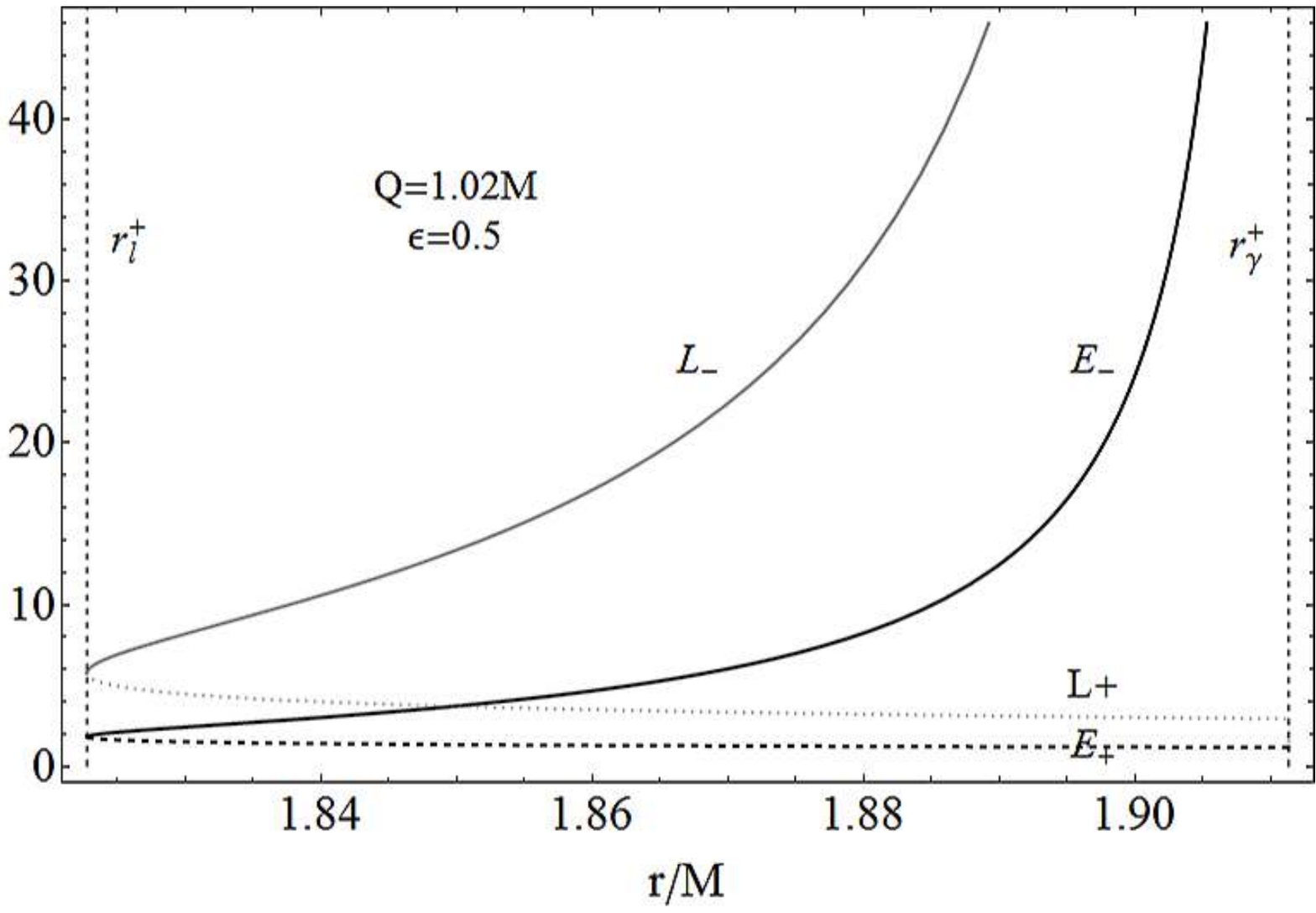}
\includegraphics[width=0.55\hsize,clip]{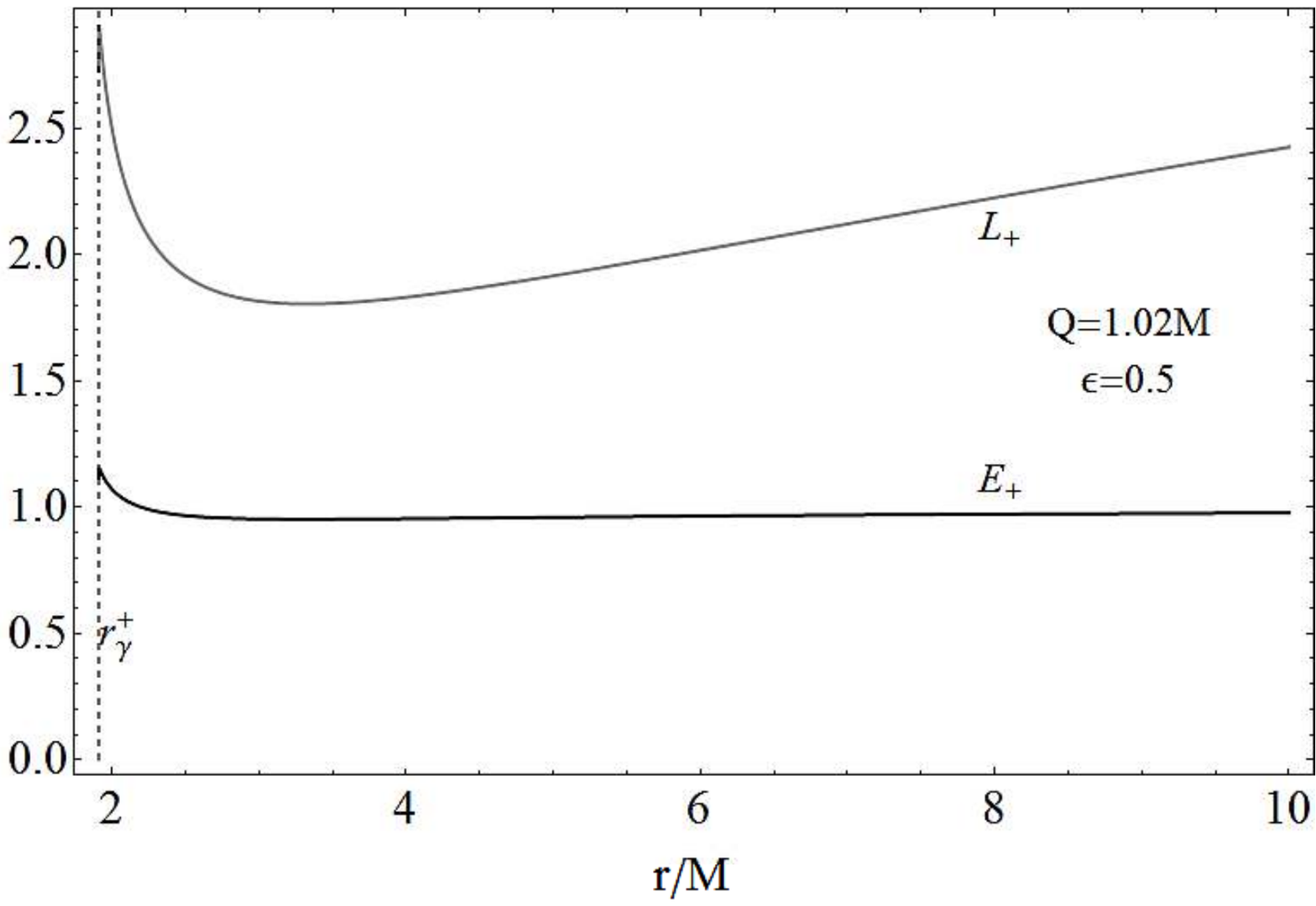}
\end{tabular}
\caption[font={footnotesize,it}]{\footnotesize{\textbf{Class}: $M<Q\leq 5/(2\sqrt{6})M$ and  $\widetilde{\widetilde{\epsilon}}_{-}<\epsilon<\widetilde{\epsilon}_{-}$.
Parameter choice: $Q= 1.02M$ and $\epsilon =0.5$.
Then $\widetilde{\widetilde{\epsilon}}_{-}=0.215376$,  $\widetilde{\epsilon}_{-}=0.564915$, $r_{s}^+ =1.17698M$, $r_\gamma^-=1.08866M$, $r_\gamma^+ =1.91134M$, $r_l^- =1.17724M$, and $r_l^+  =1.82276M$.
Circular orbits exist with angular momentum
$L=L_+$ (gray curve) and energy  $E=E_+$ (black curve) in $r_\gamma^-<r<r_{s}^{+}$ (upper left plot);
$L=L_{\pm}$ in  $r_{s}^{+}\leq r<r_l^-$ (upper right plot) and $r_l^+ \leq r<r_\gamma^+ $ (bottom left plot);
$L=L^{-}$ in $r\geq r_\gamma^+ $    (bottom right plot).
The angular momentum $L^{-}$ (gray dotted curve)  and  the energy $E^{-}$ (black dashed curve) are also plotted.
}}
\label{Fig:Edmo}
\end{figure}
\begin{figure*}
\centering
\begin{tabular}{lcr}
\includegraphics[width=0.3\hsize,clip]{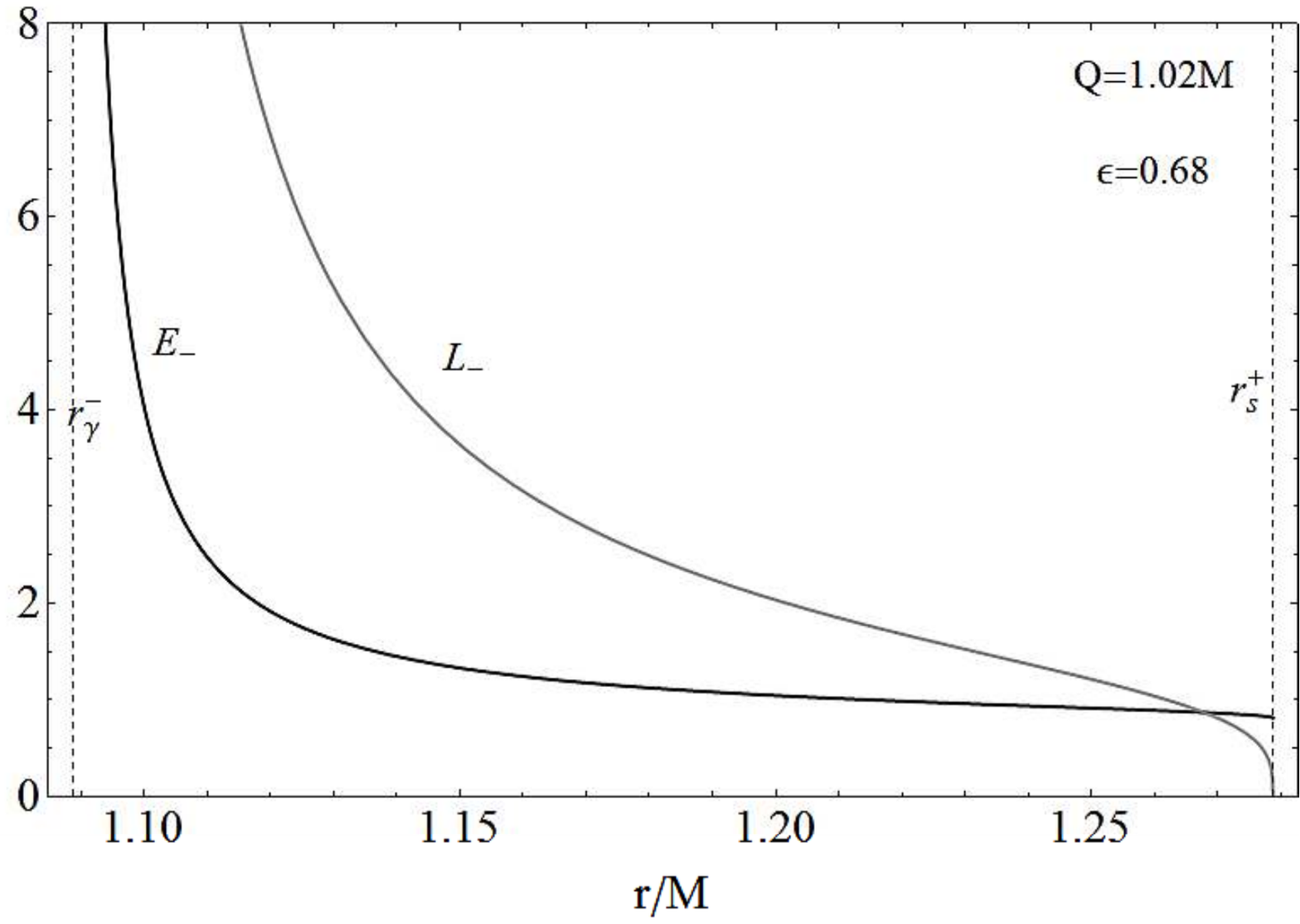}
\includegraphics[width=0.3\hsize,clip]{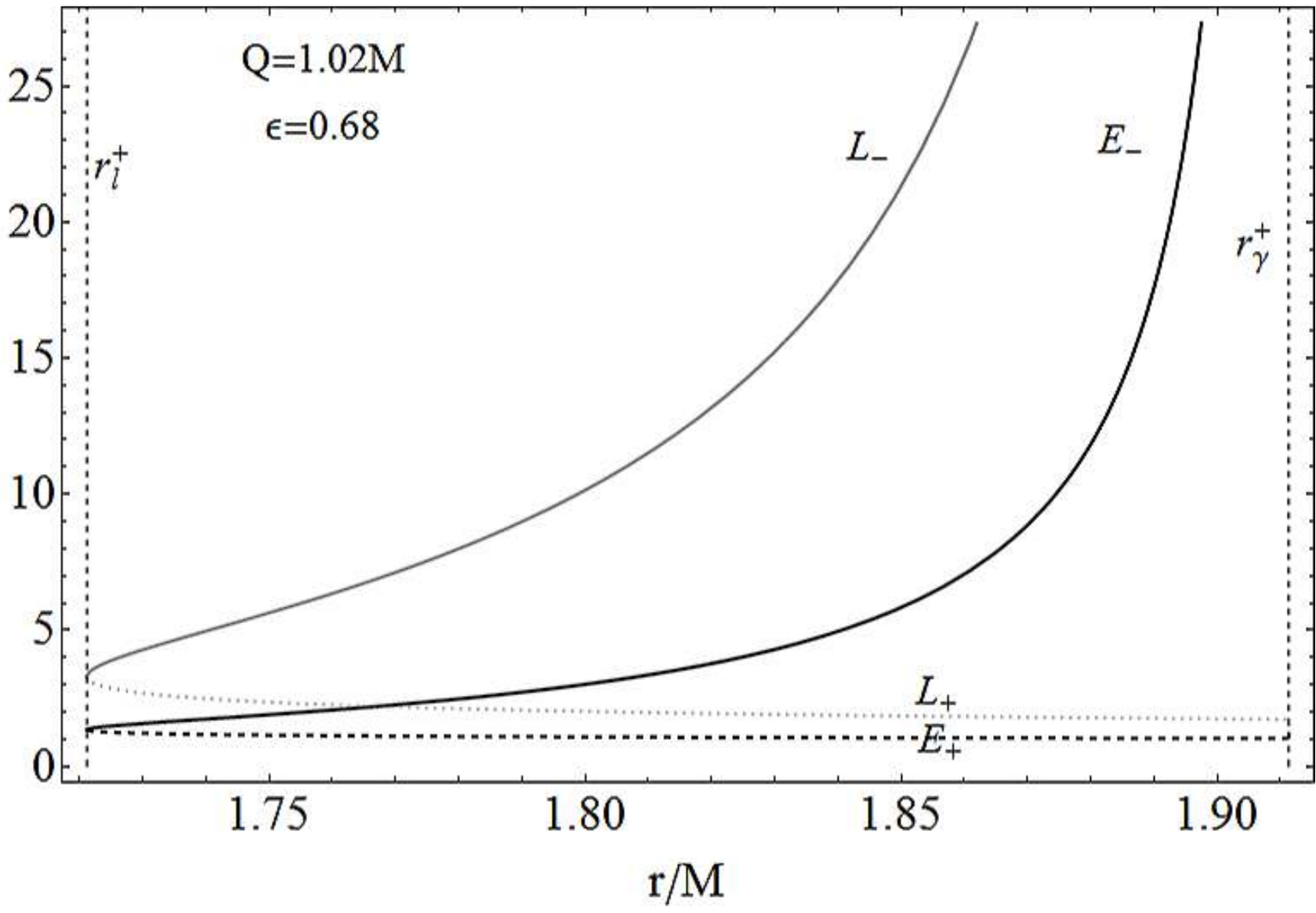}
\includegraphics[width=0.3\hsize,clip]{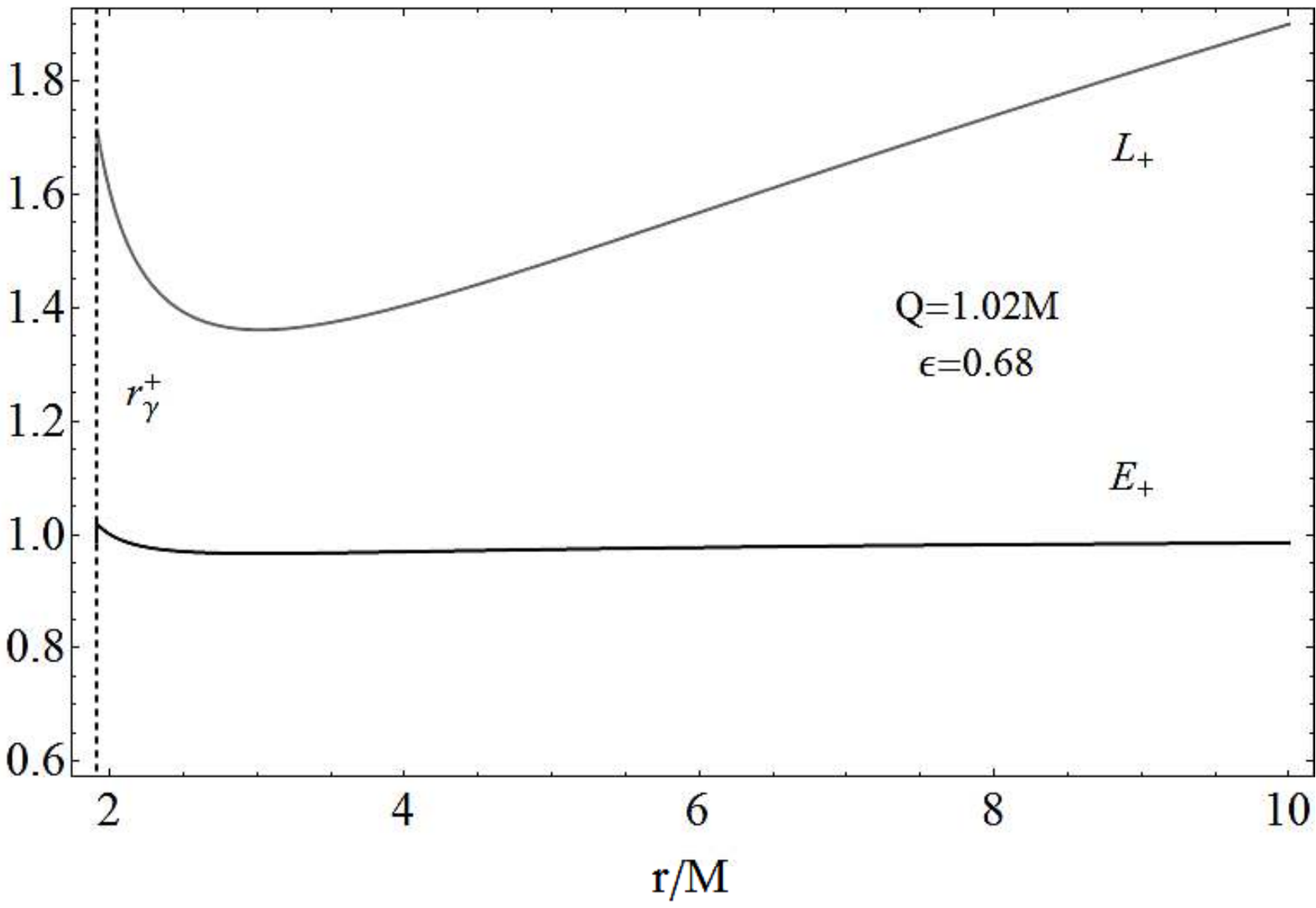}
\end{tabular}
\caption[font={footnotesize,it}]{\footnotesize{\textbf{Class}: $M<Q\leq 5/(2\sqrt{6})M$ and  $\widetilde{\epsilon}_{-}\leq \epsilon\leq \widetilde{\epsilon}_{+}$.
Parameter choice: $Q= 1.02M$ and  $\epsilon =0.68$.
Then $\widetilde{\epsilon}_{-}=0.564915$, $\widetilde{\epsilon}_{+}=0.697649$ $r_{s}^+ =1.27878M$, $r_\gamma^-=1.08866M$, $r_\gamma^+ =1.91134M$,  $r_l^- =1.2788M$
and $r_l^+  =1.7212M$.
Circular orbits exist with angular momentum
$L=L_+$ (gray curve) and energy  $E=E_+$ (black curve) in $r_\gamma^-<r<r_{s}^{+}$ (left plot);
$L=0$ at $r=r_{s}^{+}$;
$L=L_{\pm}$ in $r_l^+ \leq r<r_\gamma^+ $ (center plot);
$L=L^{-}$  in $r\geq r_\gamma^+ $ (right  plot).
}}
\label{Fig:Dandet}
\end{figure*}
\begin{figure}
\centering
\begin{tabular}{cc}
\includegraphics[width=0.451\hsize,clip]{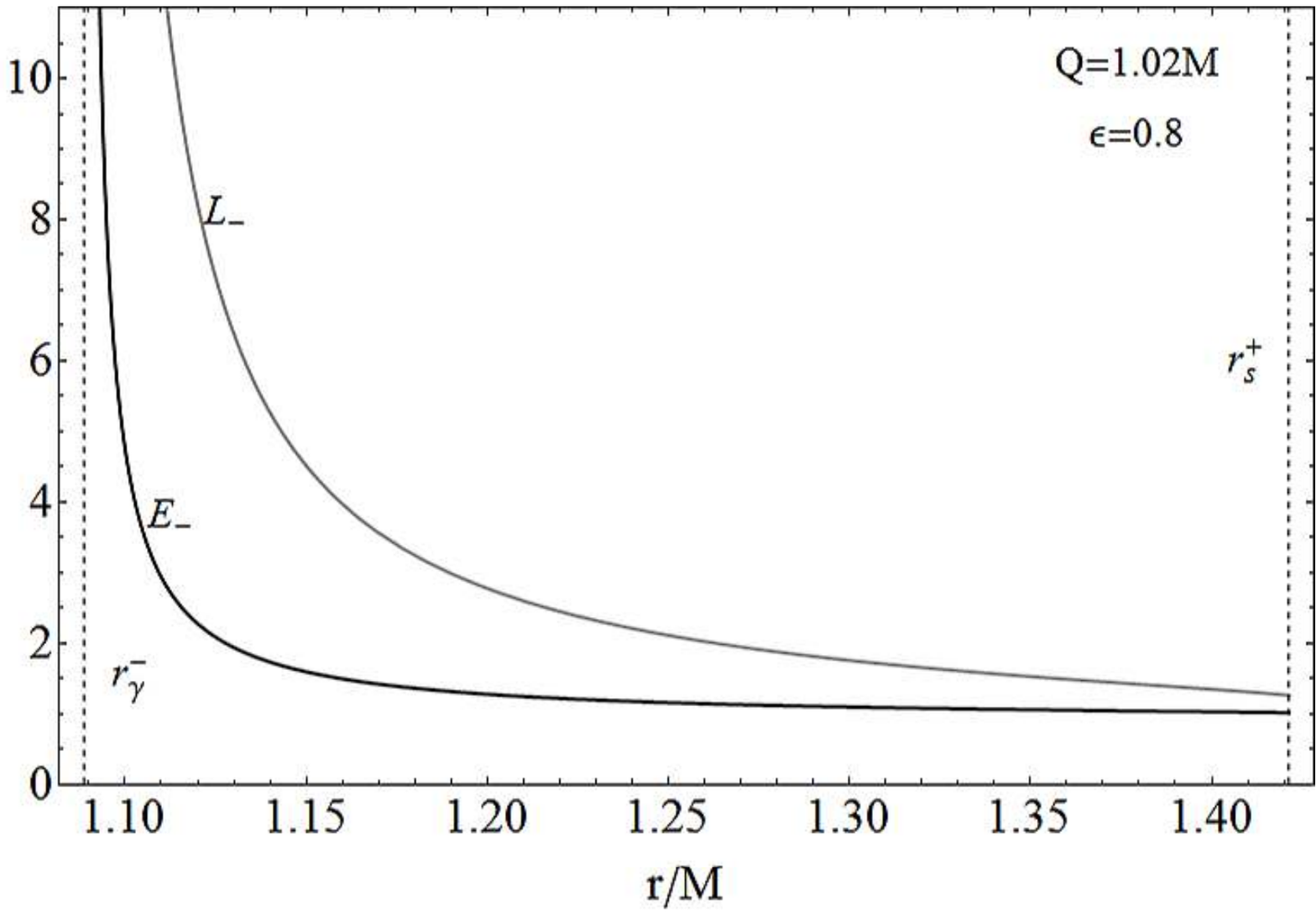}
\includegraphics[width=0.451\hsize,clip]{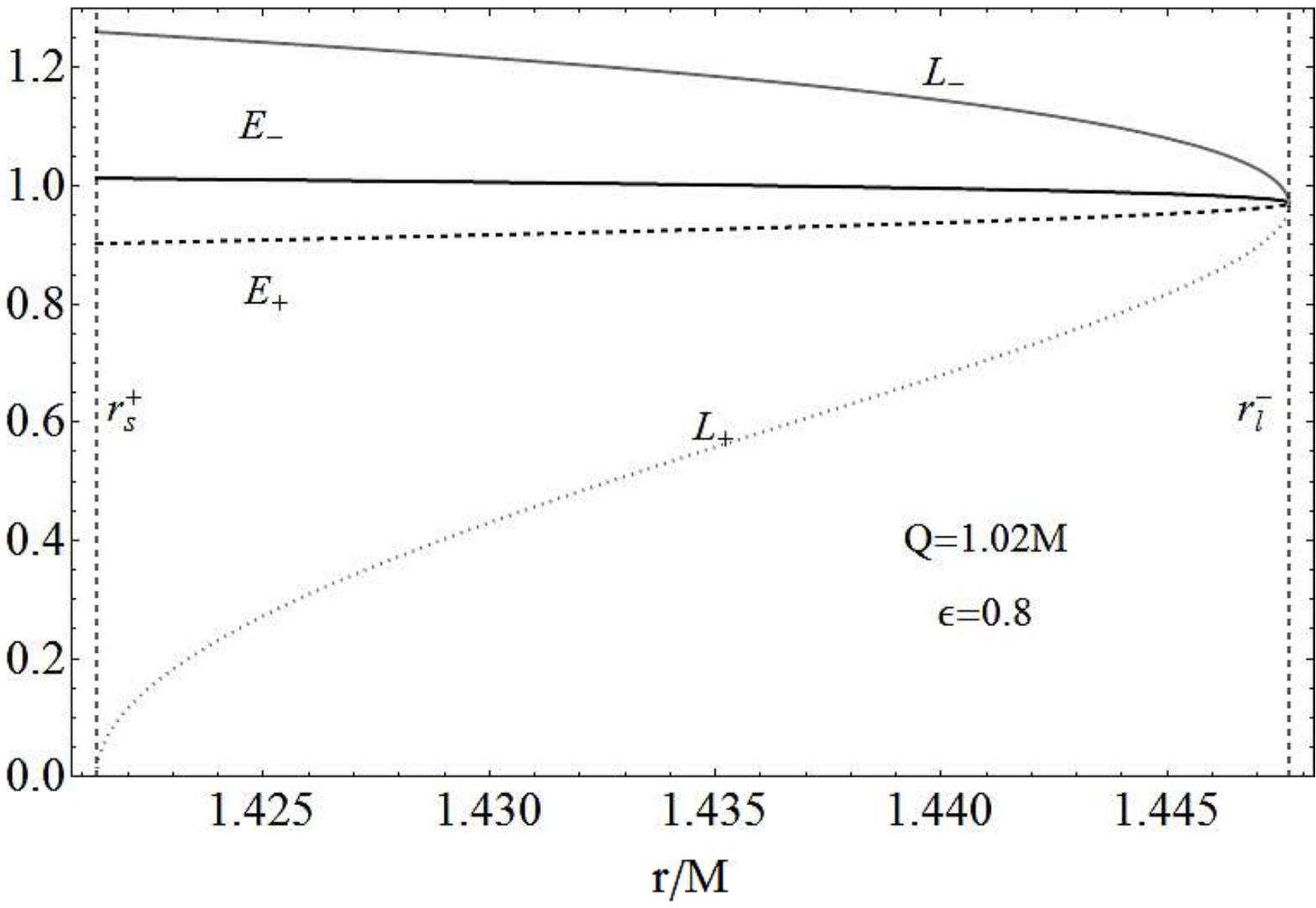}\\
\includegraphics[width=0.451\hsize,clip]{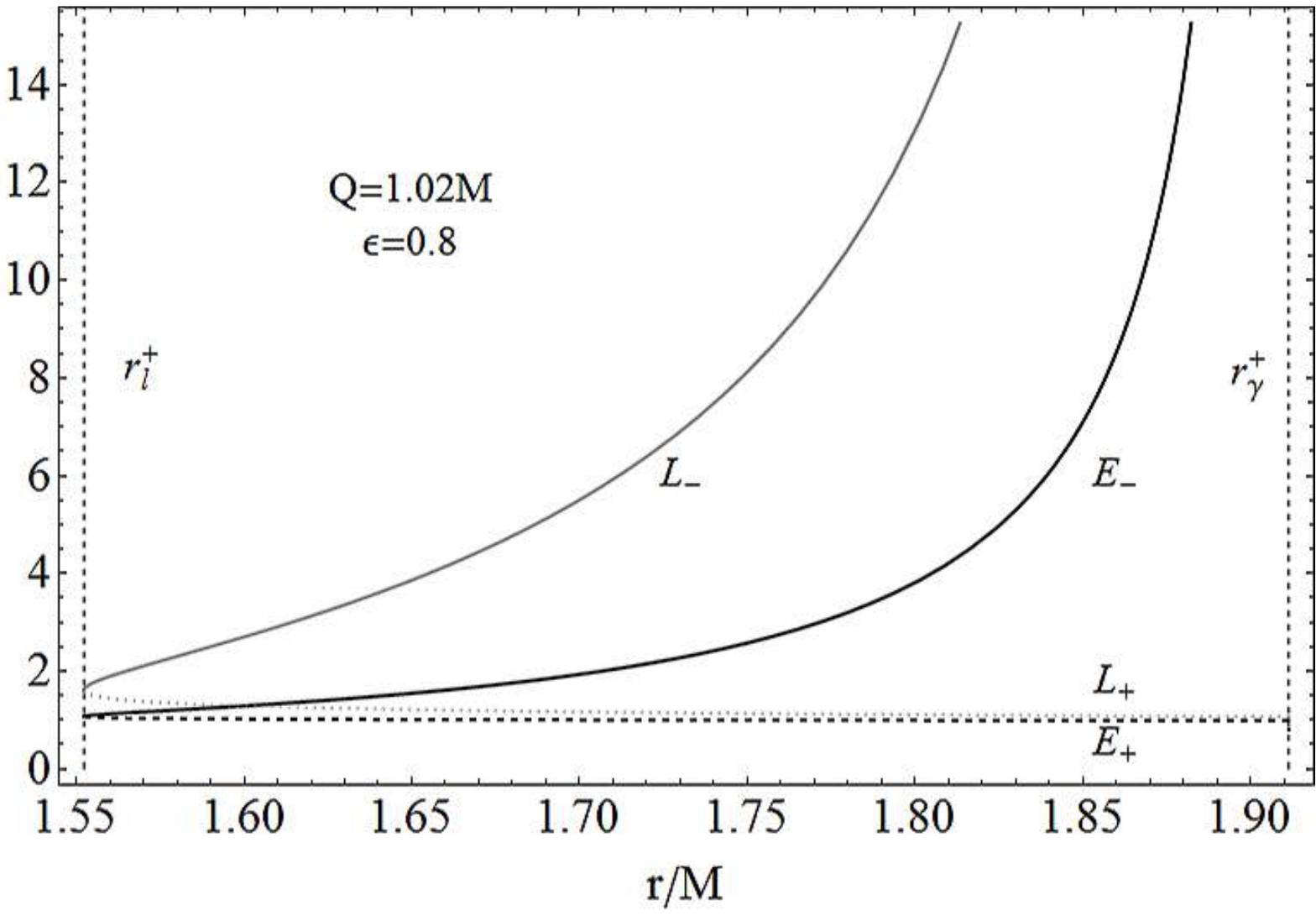}
\includegraphics[width=0.451\hsize,clip]{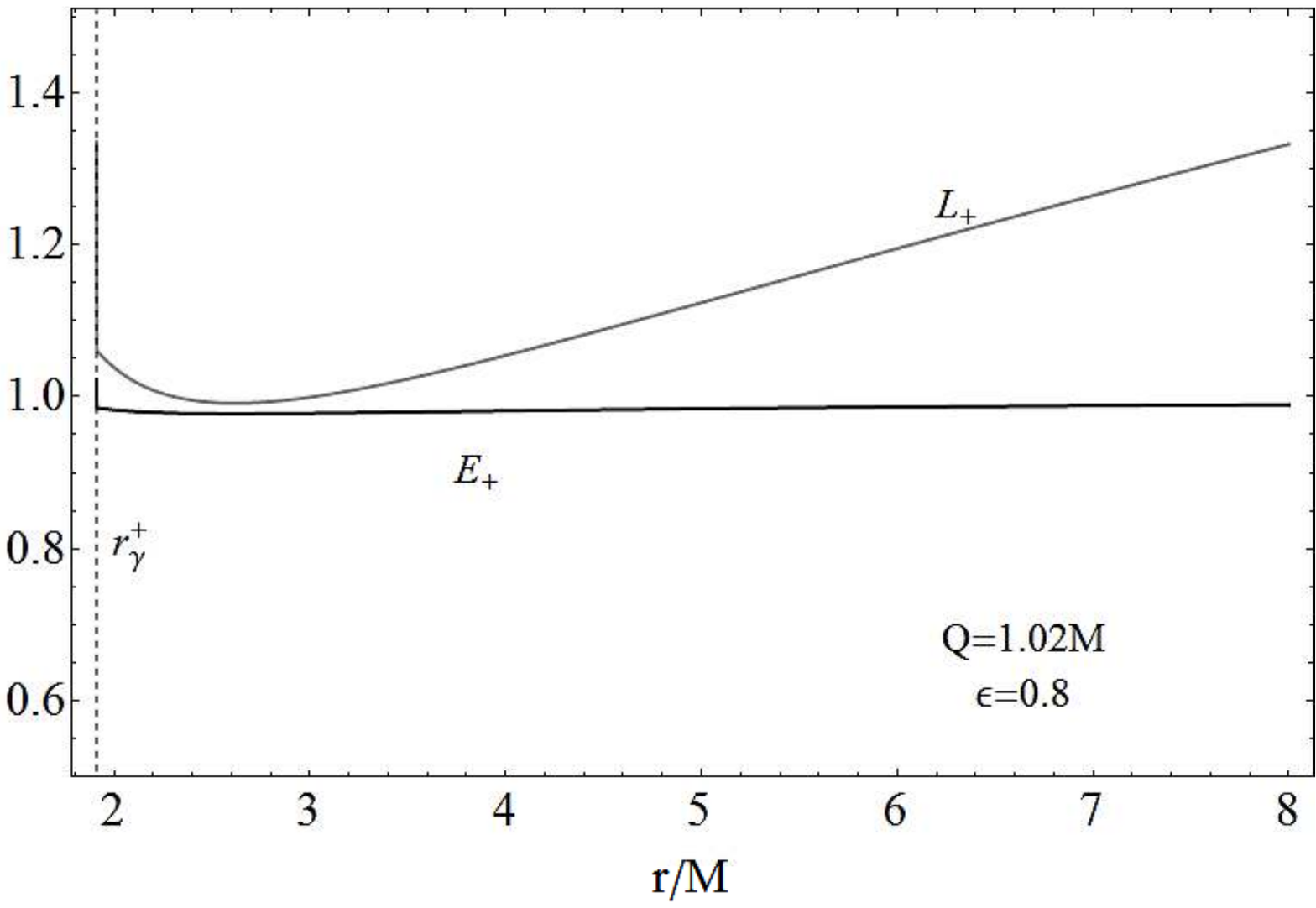}
\end{tabular}
\caption[font={footnotesize,it}]{\footnotesize{\textbf{Class}: $M<Q\leq 5/(2\sqrt{6})M$ and  $\widetilde{\epsilon}_{+}<\epsilon\leq \epsilon_{l}$.
Parameter choice: $Q= 1.02M$ and $\epsilon =0.8$. Then $\epsilon_{l}=0.806548$, $\widetilde{\epsilon}_{+}=0.697649$, $r_{s}^+ =1.42131M$, $r_\gamma^-=1.08866M$,
$r_\gamma^+ =1.91134M$, $r_l^- =1.44769M$, and $r_l^+  =1.55231M$.
Circular orbits exist with angular momentum
$L=L_+$ (gray curves) and energy  $E=E_+$ (black curves) in $r_\gamma^-<r<r_{s}^{+}$ (upper left plot);
$L=L_{\pm}$ in  $r_{s}^{+}\leq r<r_l^-$ (upper right plot) and  $r_l^+ \leq r<r_\gamma^+ $ (bottom left plot);
$L=L^{-}$  in $r\geq r_\gamma^+ $ (bottom right plot).
}}
\label{Fig:Mercedes}
\end{figure}
\begin{figure*}
\centering
\begin{tabular}{lcr}
\includegraphics[width=0.31\hsize,clip]{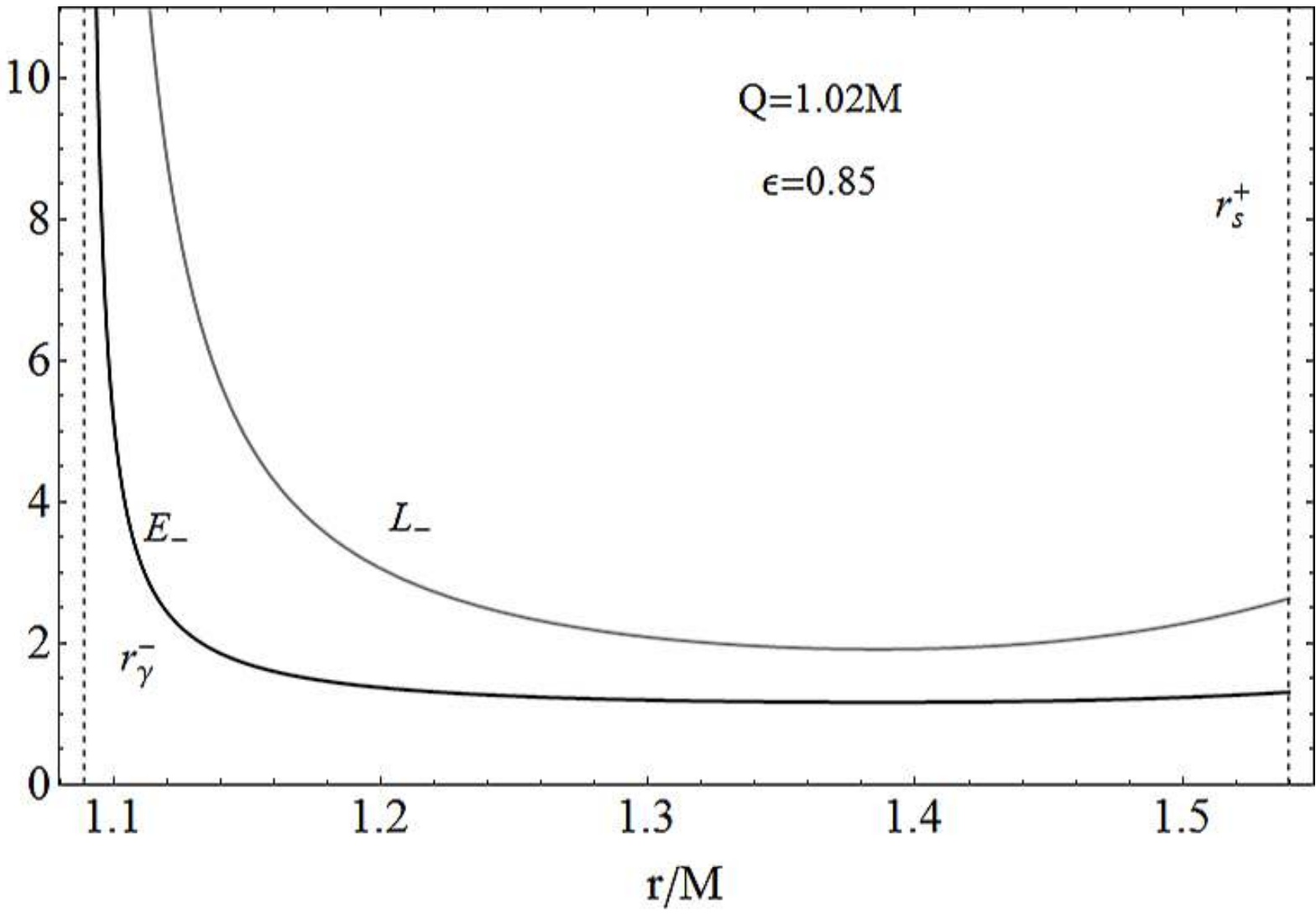}
\includegraphics[width=0.31\hsize,clip]{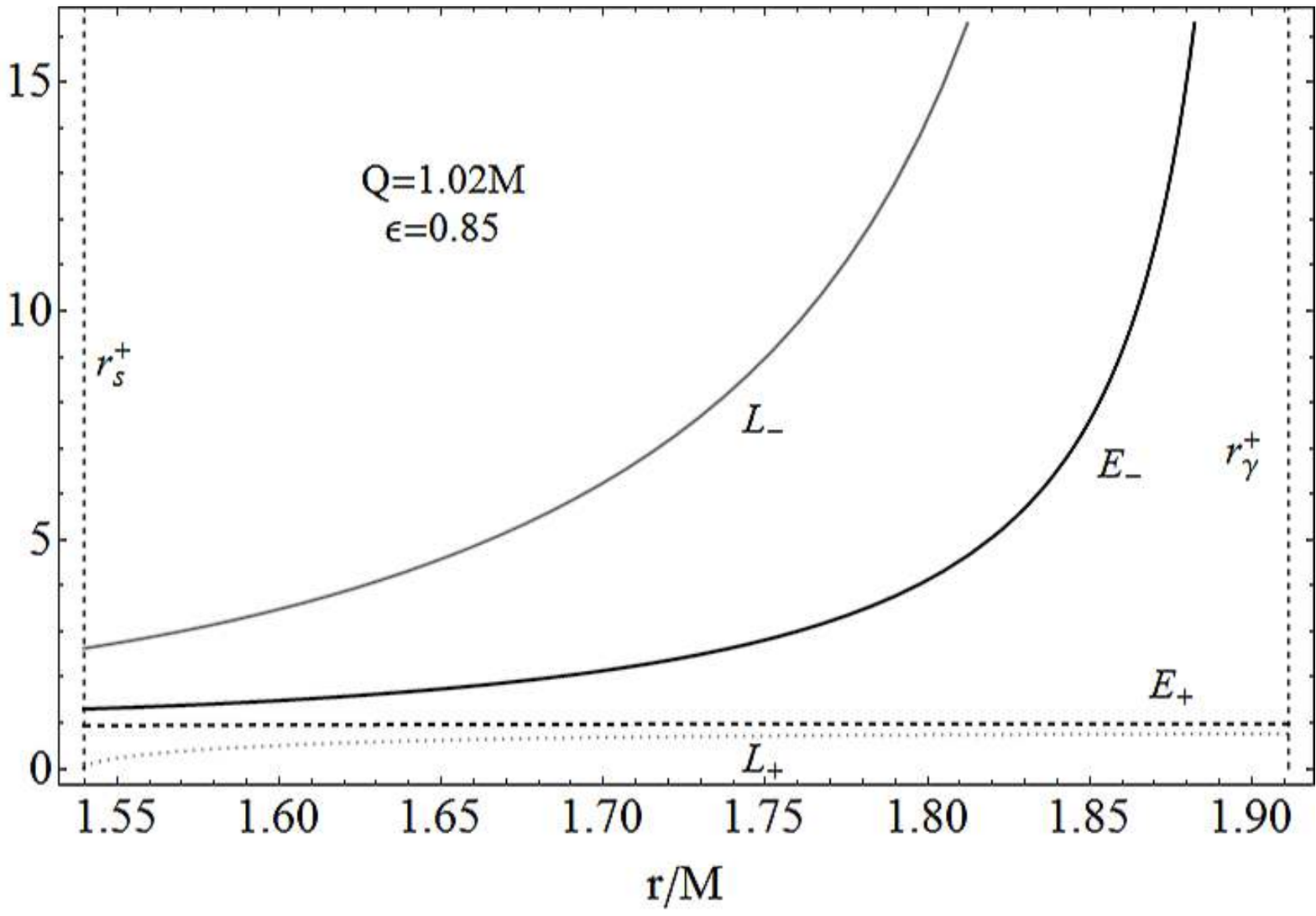}
\includegraphics[width=0.31\hsize,clip]{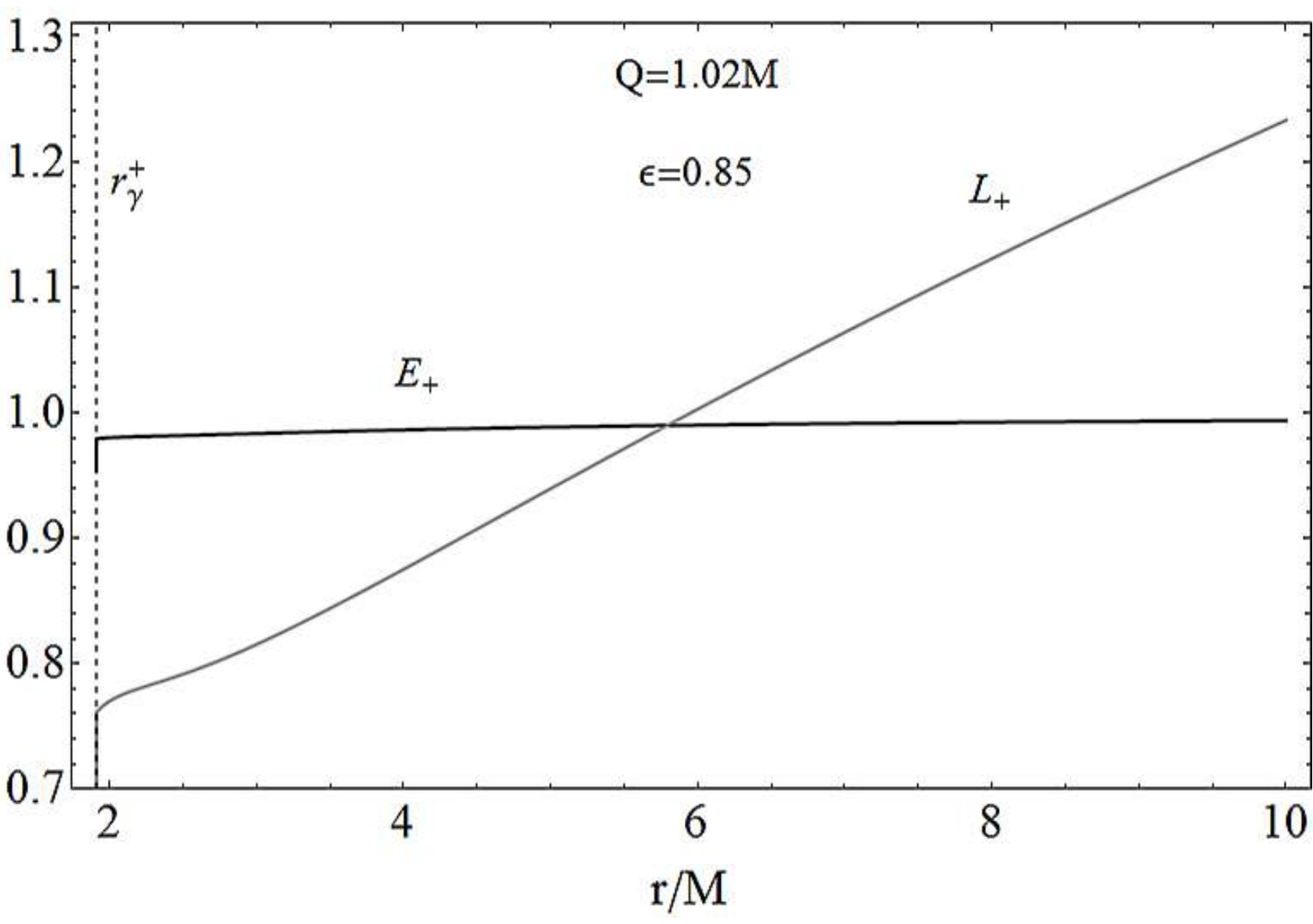}
\end{tabular}
\caption[font={footnotesize,it}]{\footnotesize{\textbf{Class}: $M<Q\leq 5/(2\sqrt{6})M$ and  $\epsilon_{l}<\epsilon<\widetilde{\widetilde{\epsilon}}_{+}$.
Parameter choice: $Q= 1.02M$ and $\epsilon =0.85$.
Then $\epsilon_{l}=0.806548$, $\widetilde{\widetilde{\epsilon}}_{+}=0.914942$, $r_{s}^+ =1.53979M$, $r_\gamma^-=1.08866M$, and $r_\gamma^+ =1.91134M$.
Circular orbits exist with angular momentum
$L=L_+$ (gray curves) and energy  $E=E_+$ (black curves) in $r_\gamma^-<r<r_{s}^{+}$ (left plot);
$L=L_{\pm}$ in  $r_{s}^{+}\leq r<r_\gamma^+ $ (center plot); and
$L=L^{-}$  in $r\geq r_\gamma^+ $ (right plot).
}}
\label{Fig:Dagla}
\end{figure*}
\begin{figure}
\centering
\begin{tabular}{cc}
\includegraphics[width=0.51\hsize,clip]{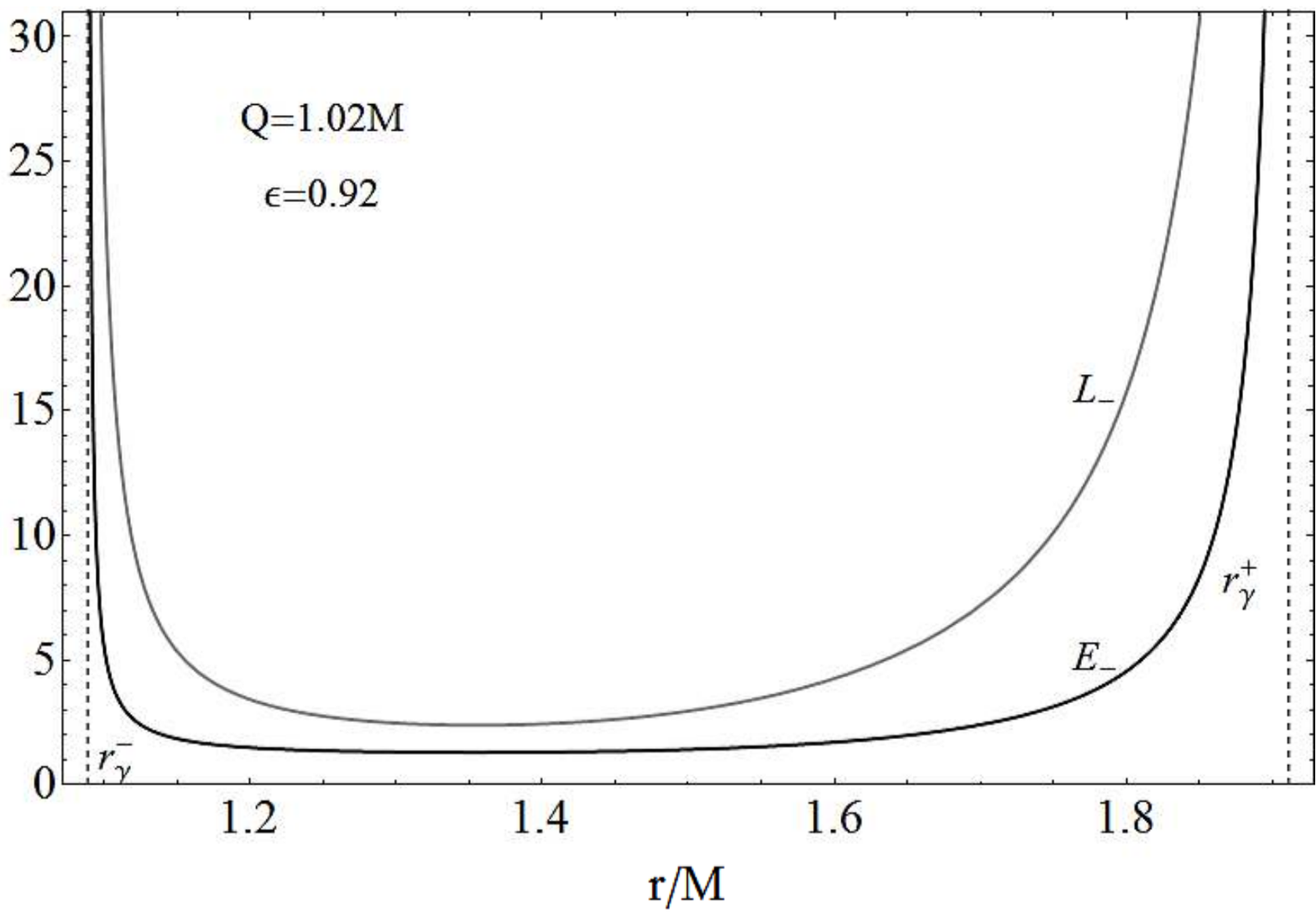}
\includegraphics[width=0.51\hsize,clip]{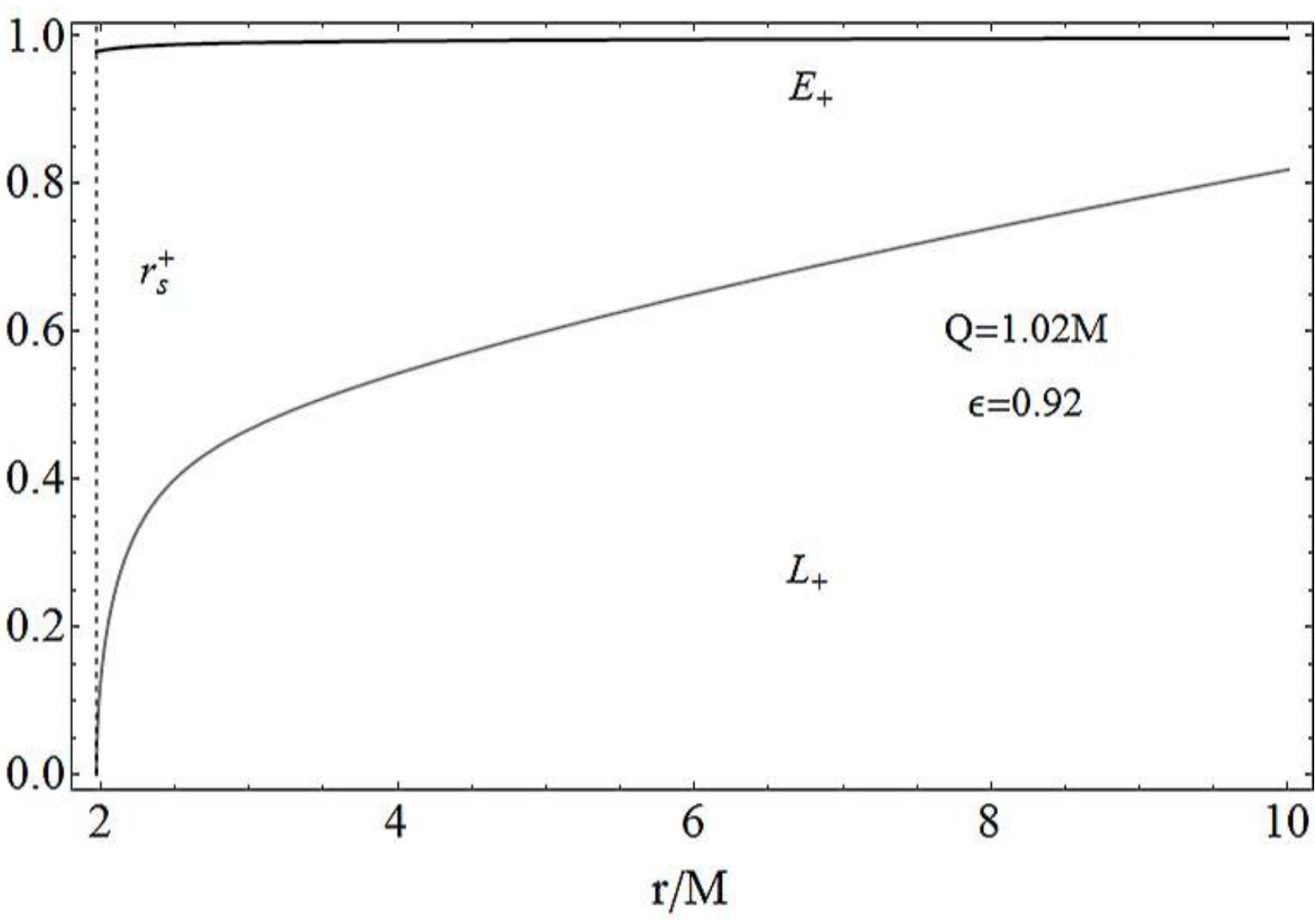}
\end{tabular}
\caption[font={footnotesize,it}]{\footnotesize{\textbf{Class}: $M<Q\leq 5/(2\sqrt{6})M$ and  $\widetilde{\widetilde{\epsilon}}_{+}\leq \epsilon<M/Q$.
Parameter choice: $Q= 1.02M$ and $\epsilon =0.92$.
Then $\widetilde{\widetilde{\epsilon}}_{+}=0.914942$, $M/Q=0.980392,$ $r_{s}^+ =1.96981M$, $r_\gamma^-=1.08866M$, and $r_\gamma^+ =1.91134M$.
Circular orbits exist with angular momentum
$L=L_+$ (gray curves) and energy  $E=E_+$ (black curves) in $r_\gamma^-<r<r_\gamma^+ $ (left plot) and
$L=L^{-}$  in $r>r_{s}^{+}$ (right plot).
}}
\label{MC}
\end{figure}
\begin{figure}
\centering
\begin{tabular}{c}
\includegraphics[width=0.551\hsize,clip]{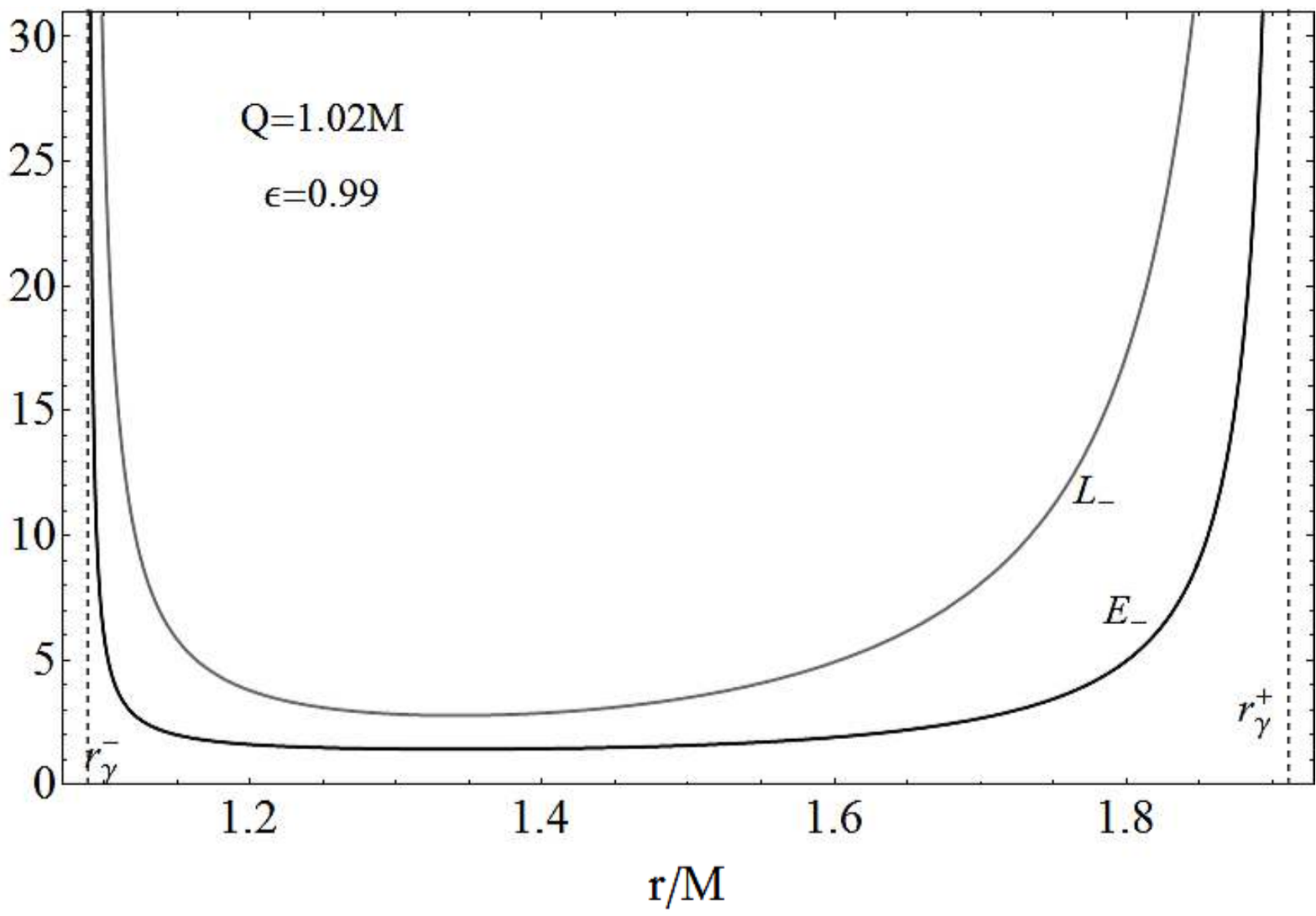}
\end{tabular}
\caption[font={footnotesize,it}]{\footnotesize{\textbf{Class}: $M<Q\leq 5/(2\sqrt{6})M$ and  $M /Q\leq \epsilon<1$.
Parameter choice: $Q= 1.02M$ and $\epsilon =0.99$.
Then $\widetilde{\widetilde{\epsilon}}_{+}=0.914942$, $r_{s}^+ =0.431596M$, $r_\gamma^-=1.08866M$, and $r_\gamma^+ =1.91134M$.
Circular orbits exist with angular momentum
$L=L_+$ (gray curve) and energy  $E=E_+$ (black curve) in $r_\gamma^-<r<r_\gamma^+ $.
}}
\label{Fig:Marsiglia}
\end{figure}
%

\begin{figure}
\centering
\begin{tabular}{rr}
\includegraphics[width=0.41\hsize,clip]{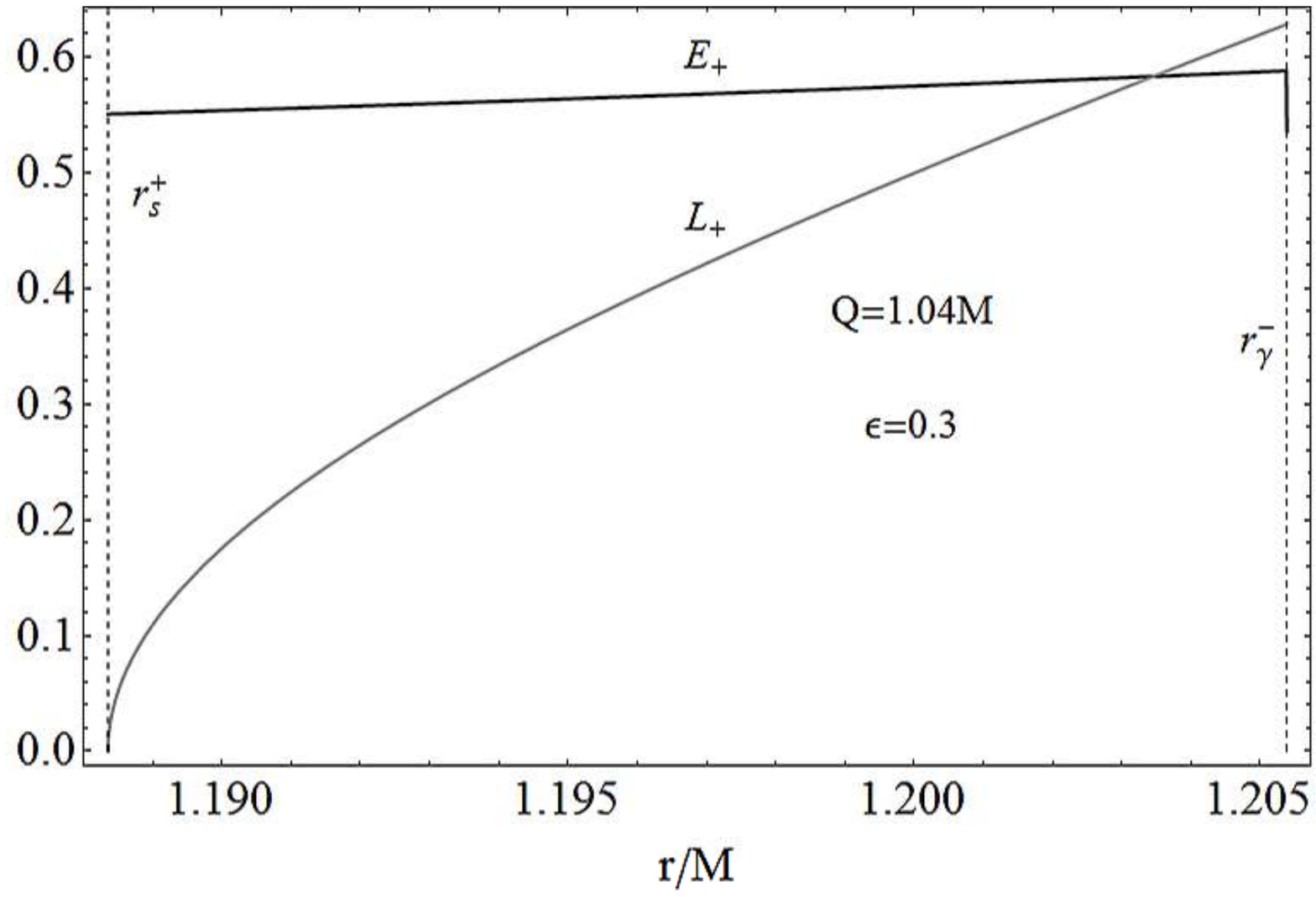}
\includegraphics[width=0.41\hsize,clip]{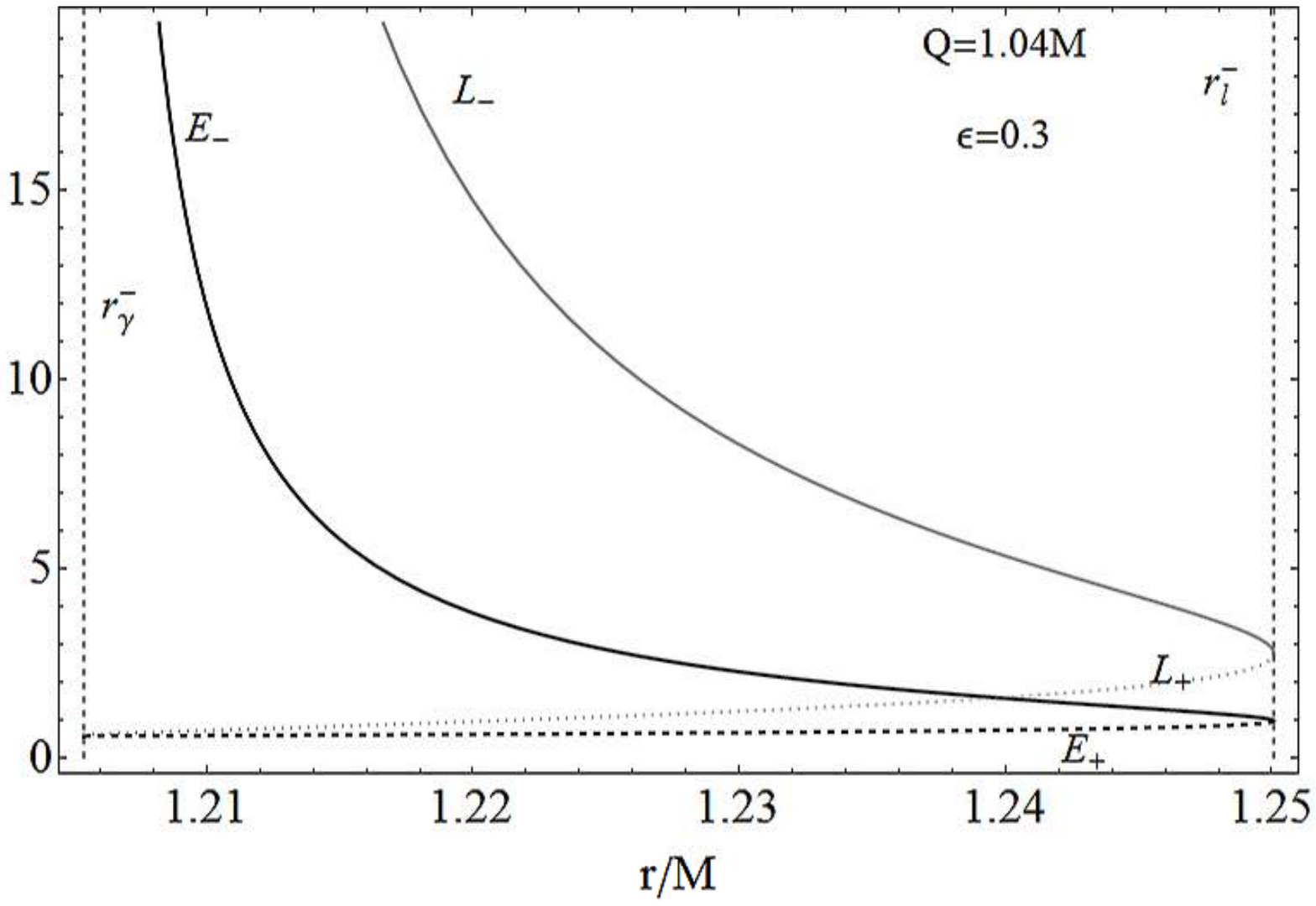}\\
\includegraphics[width=0.41\hsize,clip]{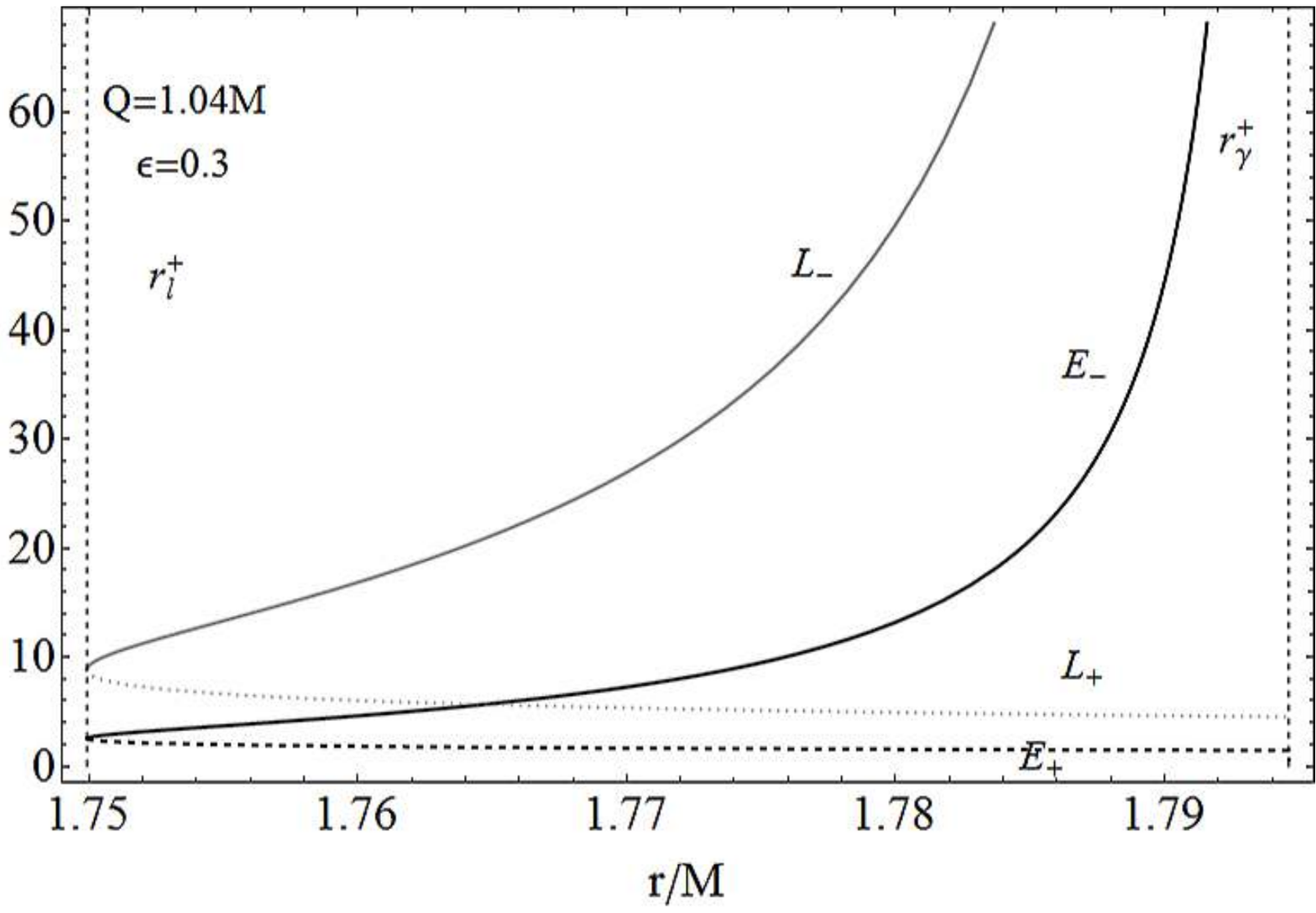}
\includegraphics[width=0.41\hsize,clip]{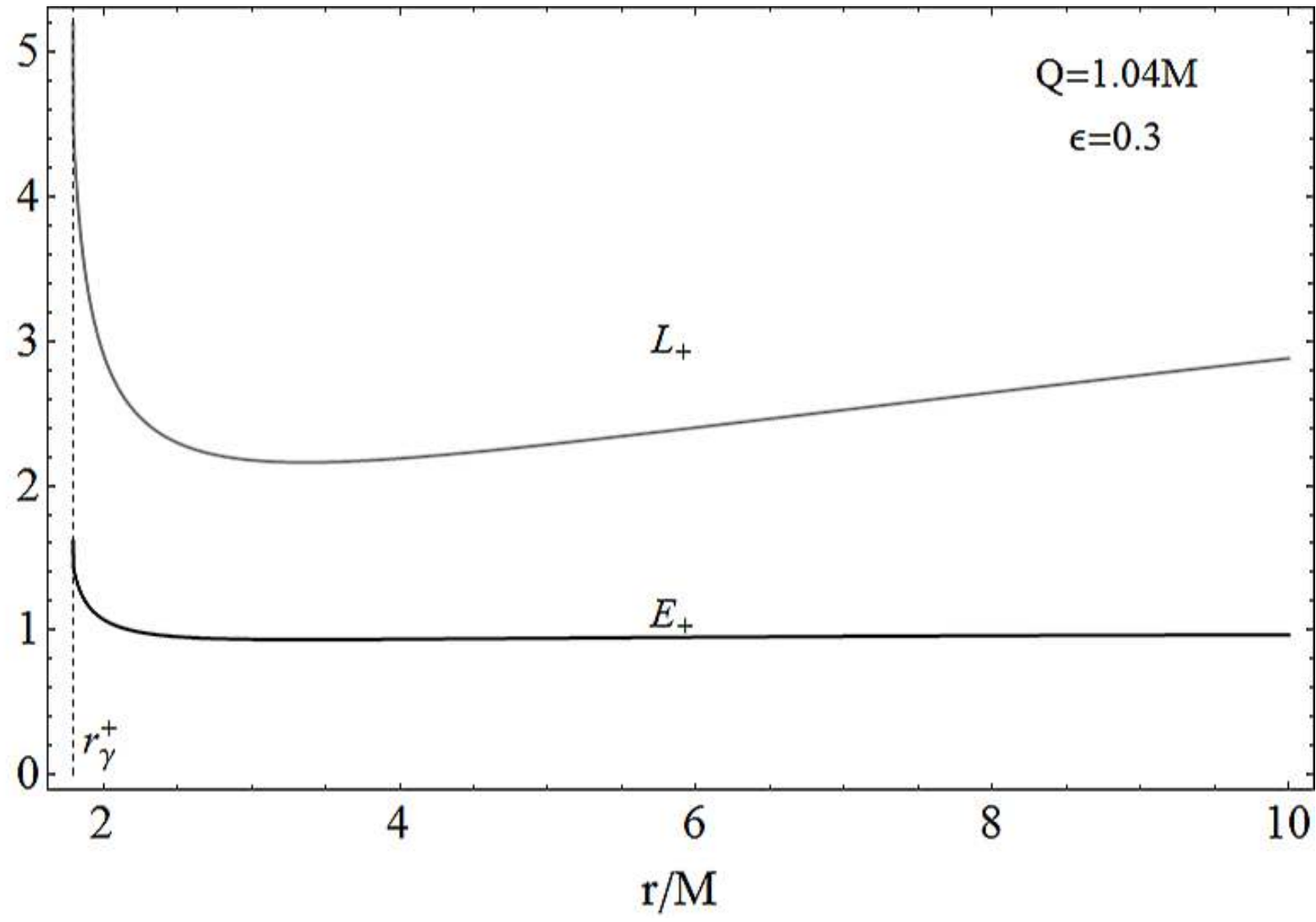}\\
\end{tabular}
\caption[font={footnotesize,it}]{\footnotesize{\textbf{Class}: $5/(2\sqrt{6})M<Q<(3\sqrt{6}/7)M$ and  $0<\epsilon<\widetilde{\widetilde{\epsilon}}_{-}$.
Parameter choice: $Q= 1.04M$ and $\epsilon =0.3$.
Then $\widetilde{\widetilde{\epsilon}}_{-}=0.338294$, $r_{s}^+ =1.18836M$, $r_\gamma^-=1.20538M$, $r_\gamma^+ =1.79462M$,  $r_l^- =1.25007M$, and $r_l^+  =1.74993M$.
Circular orbits exist with angular momentum
$L=L^{-}$ (gray curves) and energy  $E=E^{-}$ (black curves) in $r_{s}^+ <r\leq r_\gamma^-$ (upper left plot);
$L=0$ at $r=r_{s}^+ $;
$L=L_{\pm}$ in $r_\gamma^-<r\leq r_l^-$ (upper right plot) and  $r_l^+ \leq r<r_\gamma^+ $ (bottom left plot);
$L=L^{-}$ in $r\geq r_\gamma^+ $ (bottom right plot).
}}
\label{Fig:Morrel}
\end{figure}
\begin{figure}
\centering
\begin{tabular}{cc}
\includegraphics[width=0.41\hsize,clip]{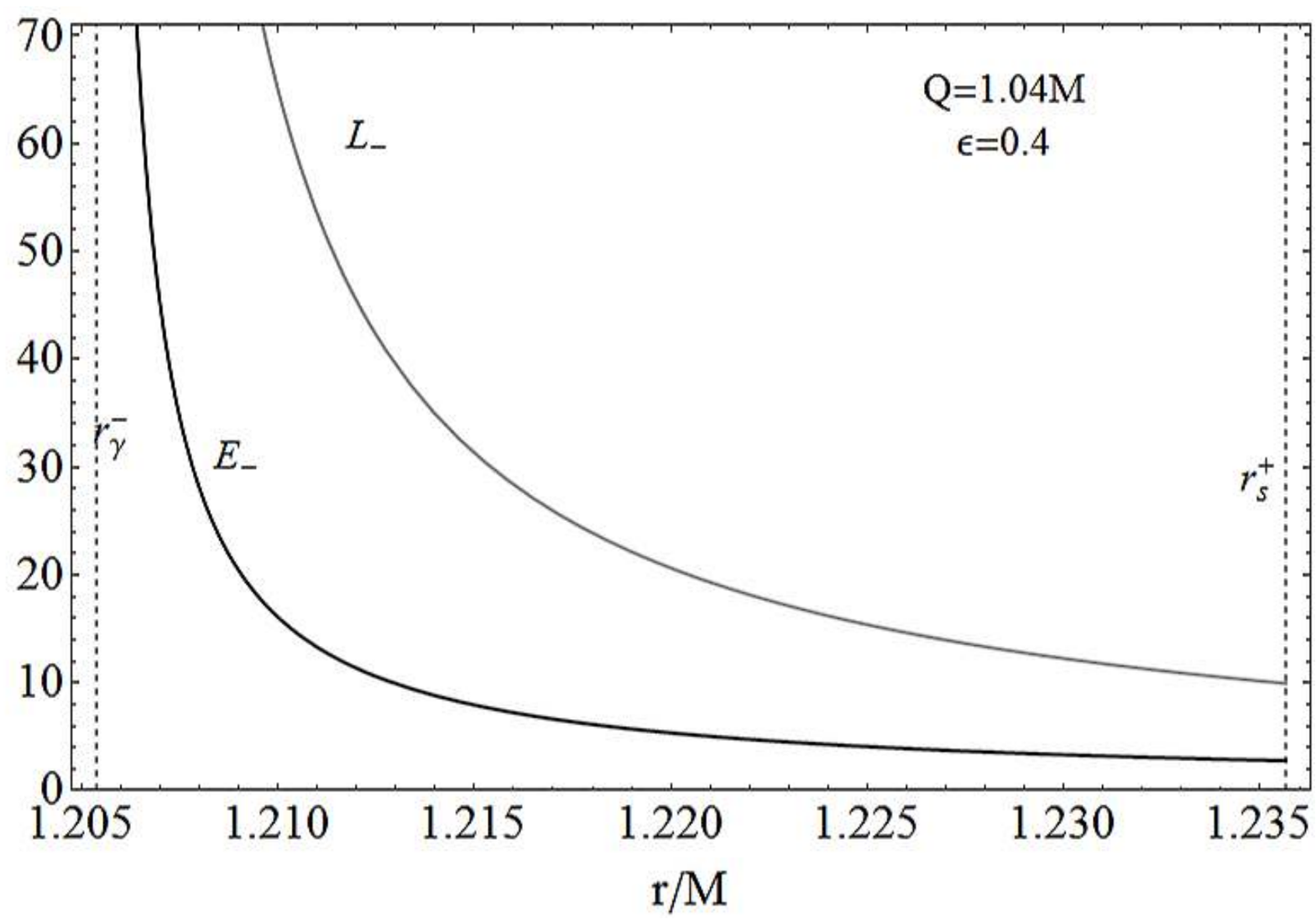}
\includegraphics[width=0.41\hsize,clip]{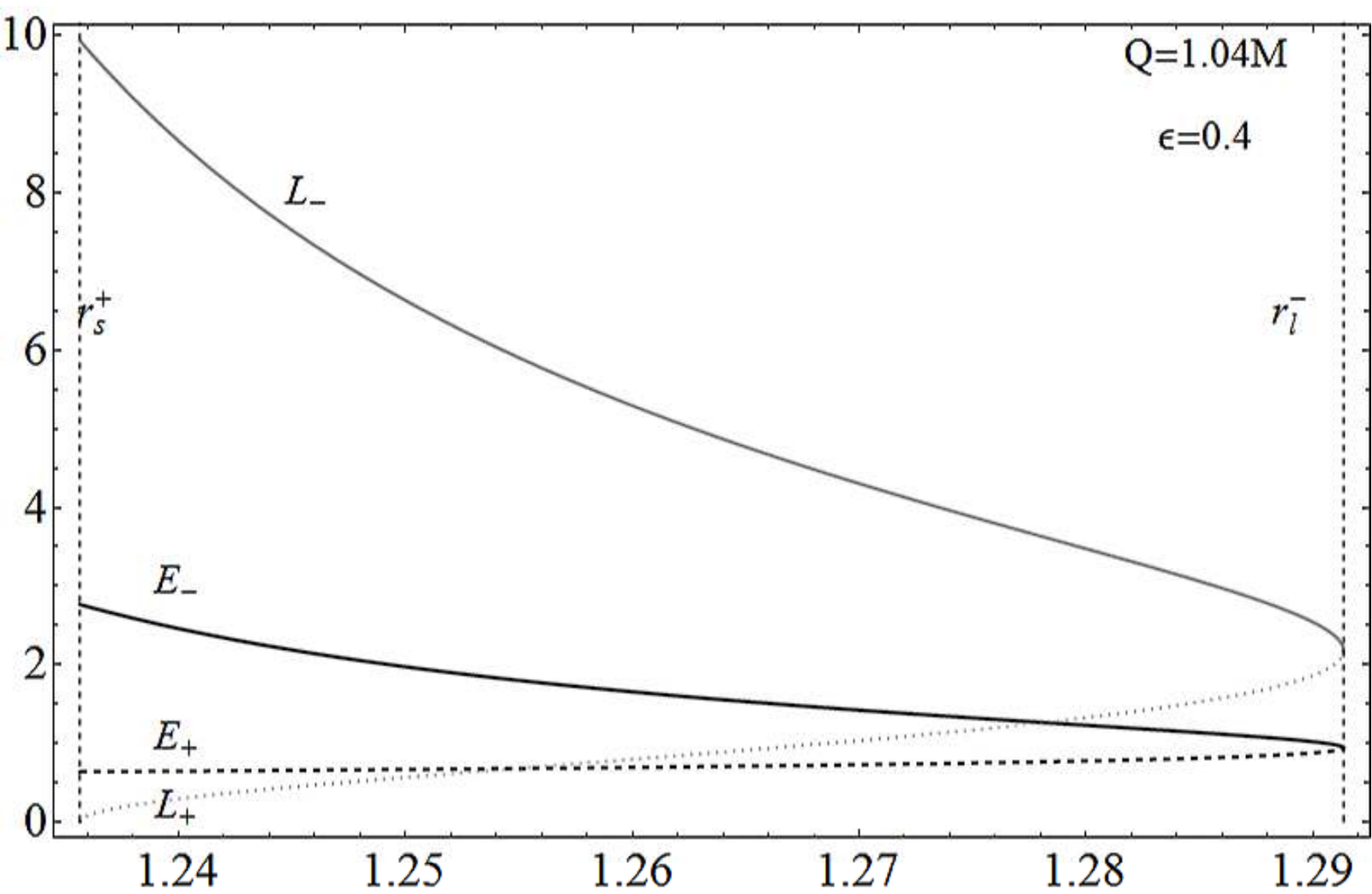}\\
\includegraphics[width=0.41\hsize,clip]{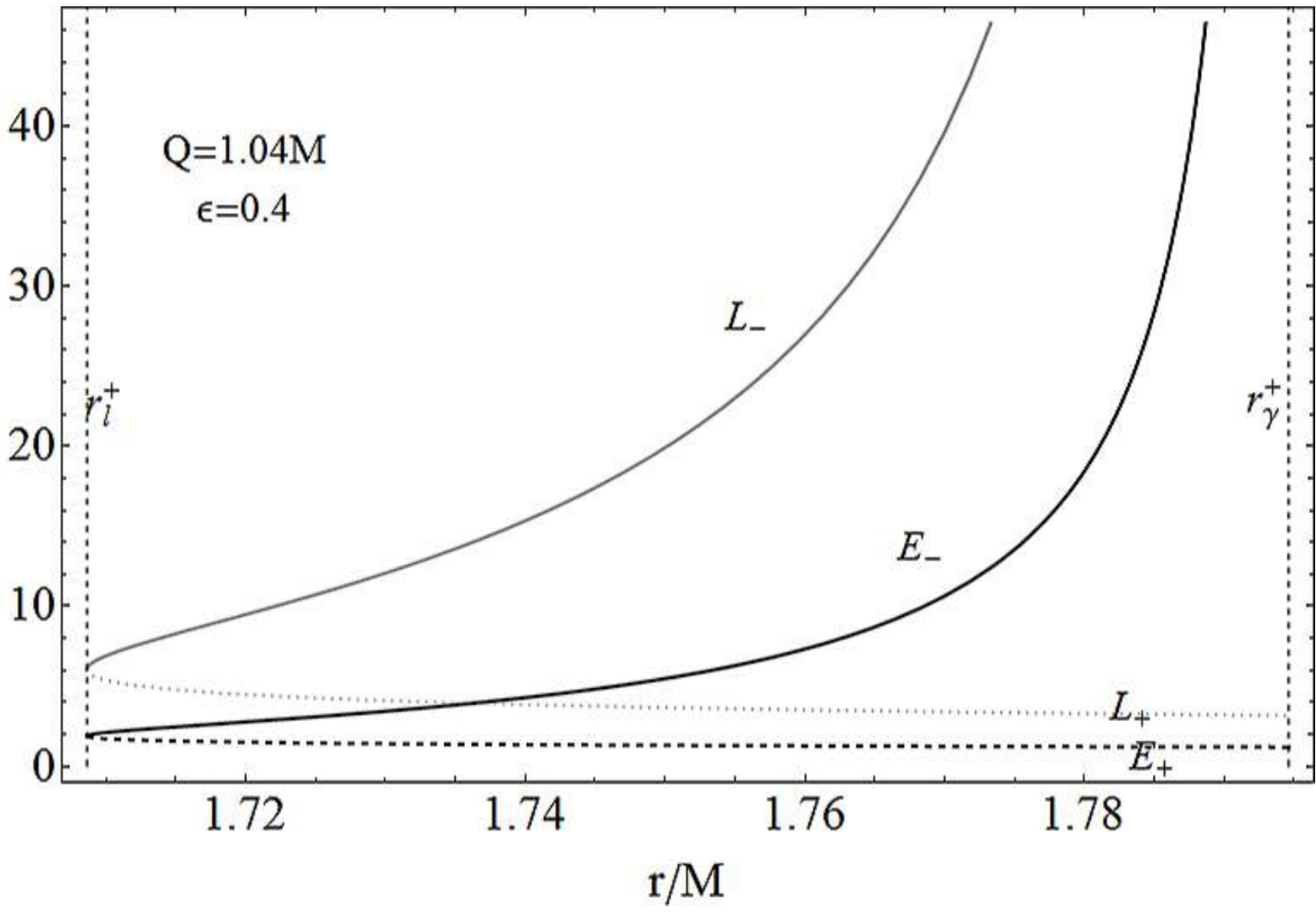}
\includegraphics[width=0.41\hsize,clip]{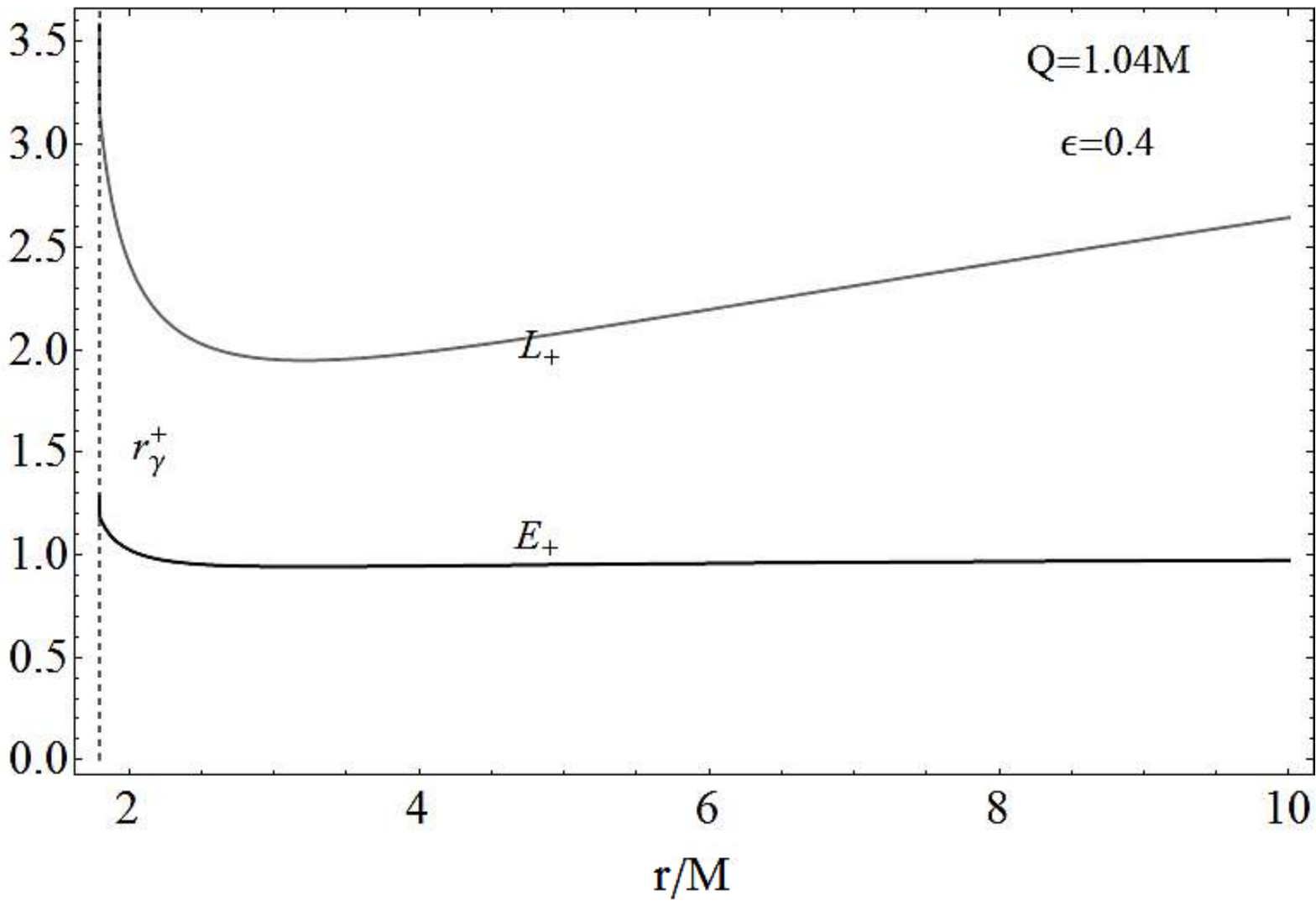}
\end{tabular}
\caption[font={footnotesize,it}]{\footnotesize{\textbf{Class}: $5/(2\sqrt{6})M<Q<(3\sqrt{6}/7)M$ and  $\widetilde{\widetilde{\epsilon}}_{-}<\epsilon<\epsilon_{l}$.
Parameter choice: $Q=1.04M$ and $\epsilon =0.4$.
Then $\widetilde{\widetilde{\epsilon}}_{-}=0.338294$, $\epsilon_{l}=0.566574$, $r_{s}^+ =1.23565M$, $r_\gamma^-=1.20538M$, $r_\gamma^+ =1.79462M$,  $r_l^- =1.29135M$, and $r_l^+  =1.70865M$.
Circular orbits exist with angular momentum
$L=L_+$ (gray curve) and energy  $E=E_+$ (black curve) in $r_\gamma^-<r<r_{s}^+ $ (upper left plot);
$L=0$ at $r=r_{s}^+ $;
$L=L_{\pm}$ in $r_{s}^+ <r<r_l^-$ (upper right plot) and  $r_l^+ \leq r<r_\gamma^+ $ (bottom left plot);
$L=L^{-}$ in $r\geq r_\gamma^+ $ (bottom right plot).
}}
\label{Fig:Chalon}
\end{figure}
\begin{figure*}
\centering
\begin{tabular}{lcr}
\includegraphics[width=0.31\hsize,clip]{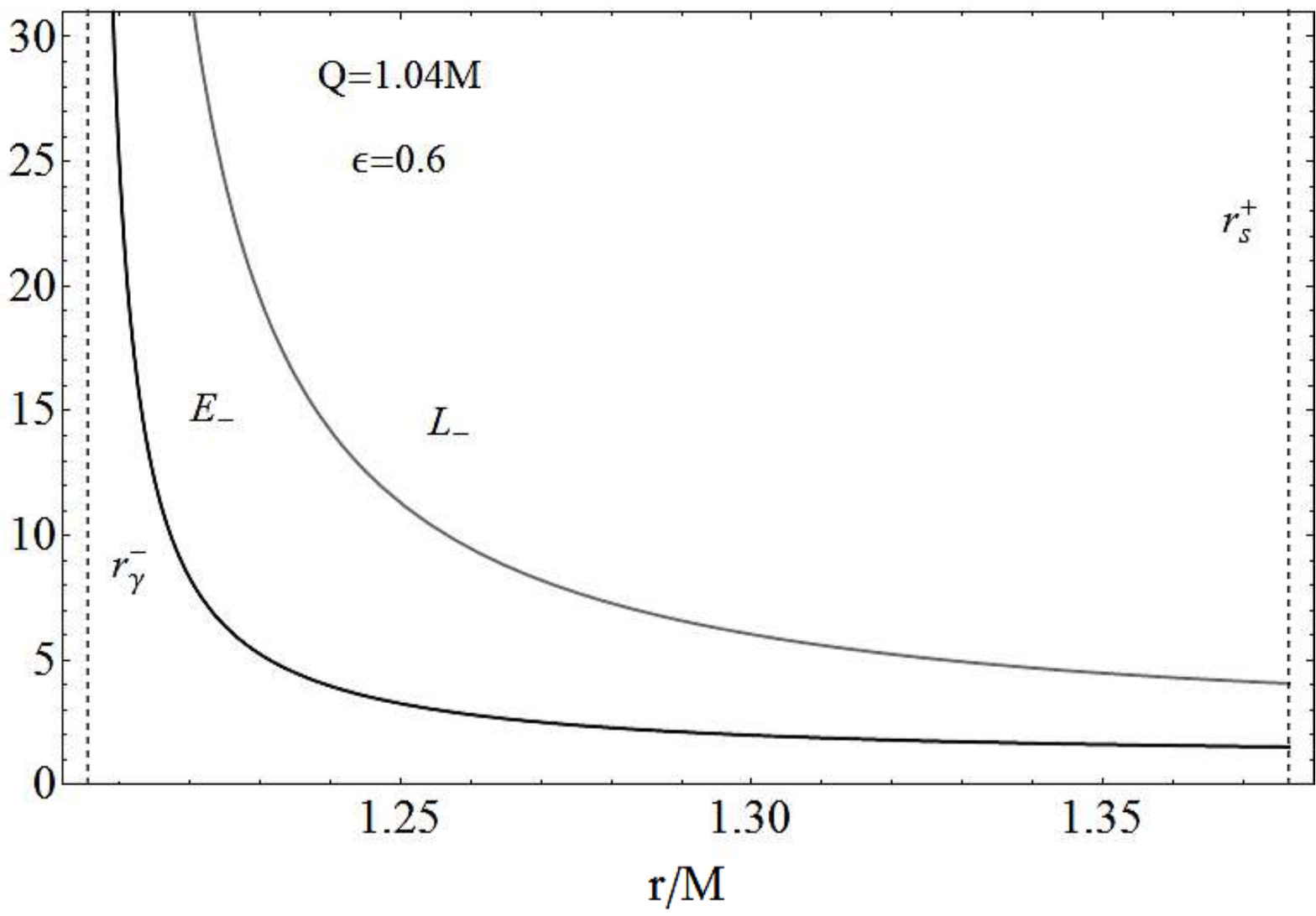}
\includegraphics[width=0.31\hsize,clip]{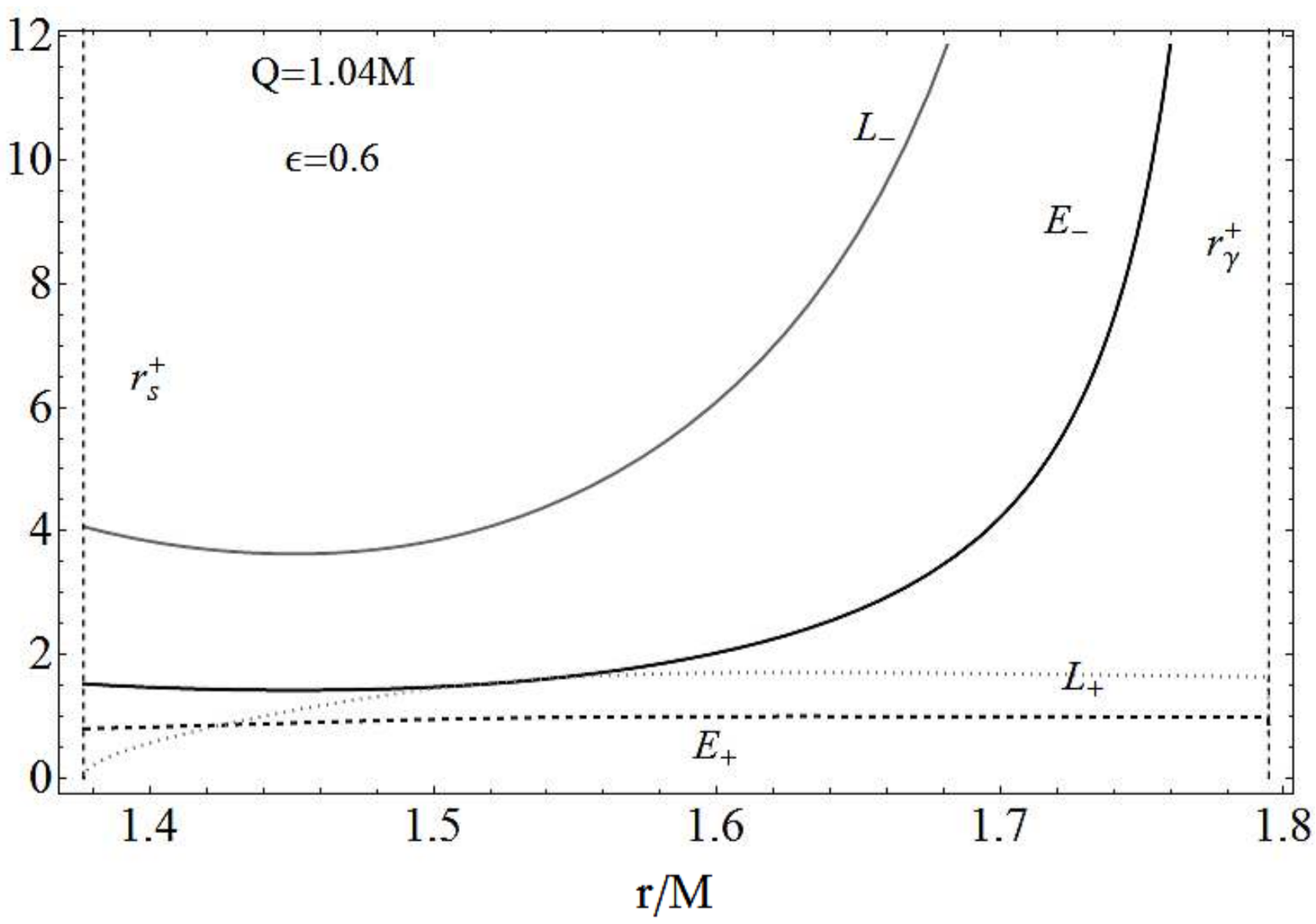}
\includegraphics[width=0.31\hsize,clip]{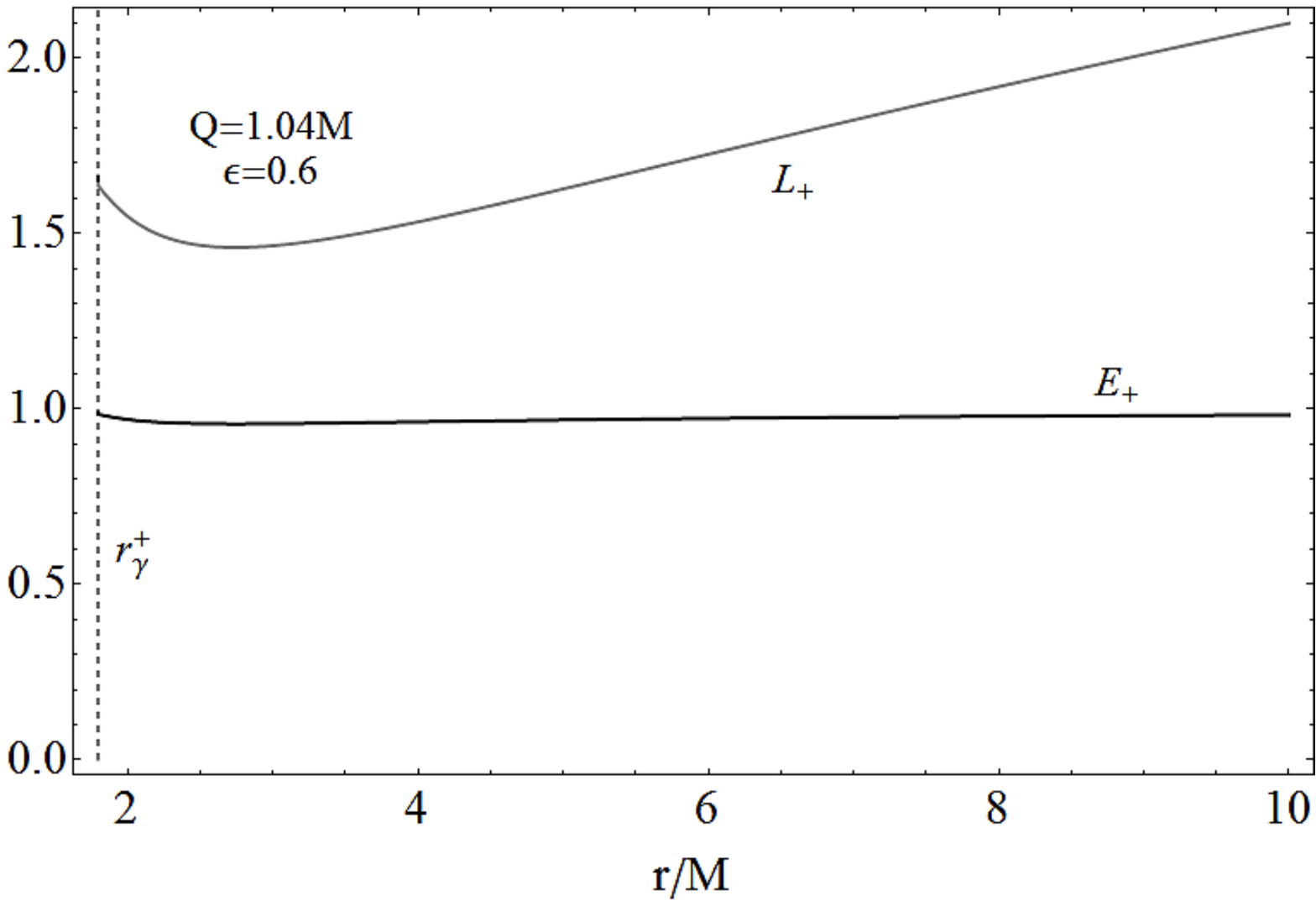}
\end{tabular}
\caption[font={footnotesize,it}]{\footnotesize{\textbf{Class}: $5/(2\sqrt{6})M<Q<(3\sqrt{6}/7)M$ and  $\epsilon_{l}\le\epsilon \leq\widetilde{\widetilde{\epsilon}}_{+}$.
Parameter choice: $Q= 1.04M$ and $\epsilon =0.6$.
Then $\widetilde{\widetilde{\epsilon}}_{+}=0.811927$ and  $\epsilon_{l}=0.566574$, $r_{s}^+ =1.37651M$, $r_\gamma^-=1.20538M$, and $r_\gamma^+ =1.79462M$.
Circular orbits exist with angular momentum
$L=L_+$ (gray curve) and energy  $E=E_+$ (black curve) in $r_\gamma^-<r<r_{s}^+ $ (left plot);
$l=L_{\pm}$ in $r_{s}^+ \leq r<r_\gamma^+ $ (center plot); and
$L=L^{-}$ in $r\geq r_\gamma^+ $ (right  plot).
}}
\label{Fig:Barone}
\end{figure*}
\begin{figure}
\centering
\begin{tabular}{cc}
\includegraphics[width=0.51\hsize,clip]{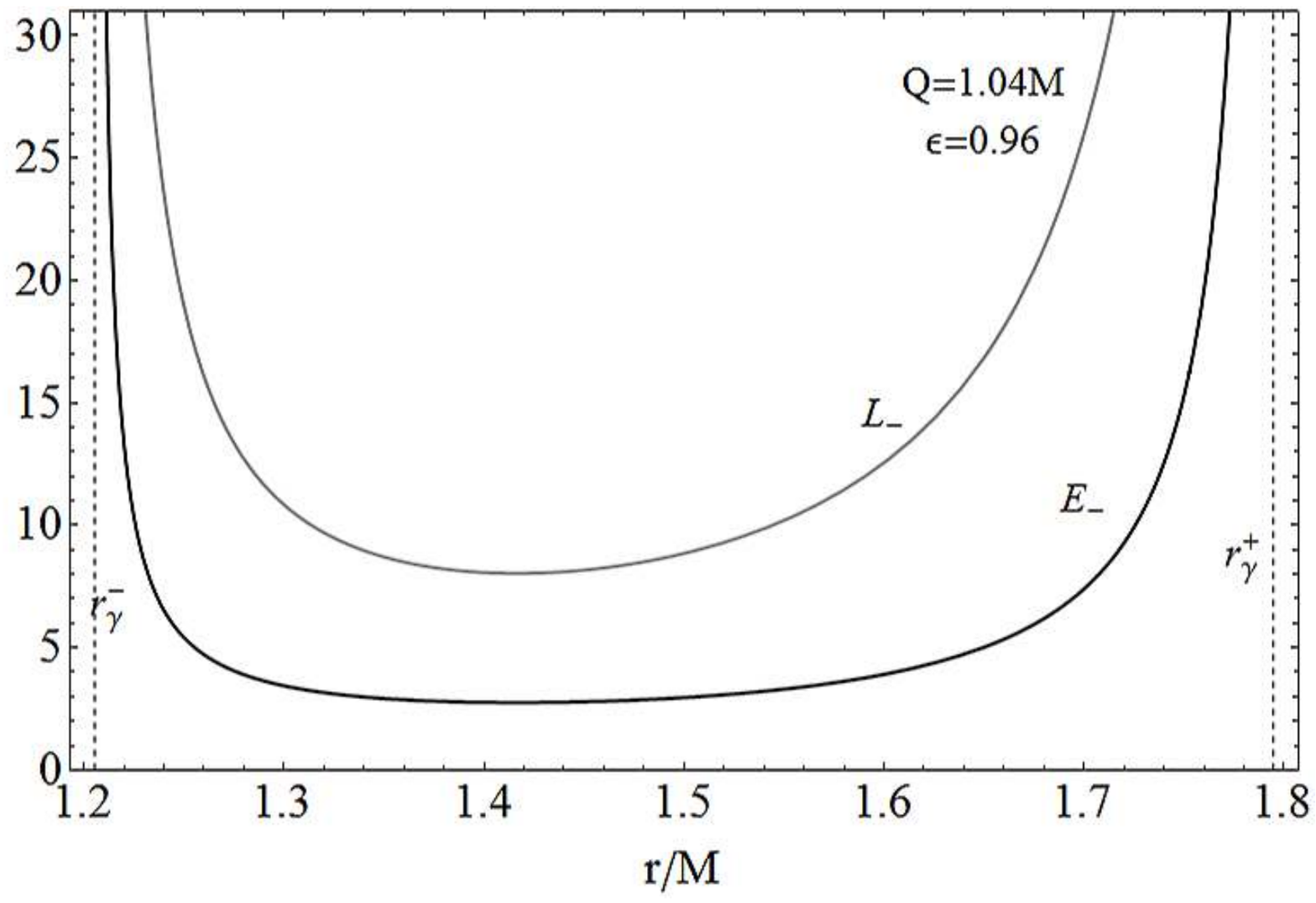}
\includegraphics[width=0.51\hsize,clip]{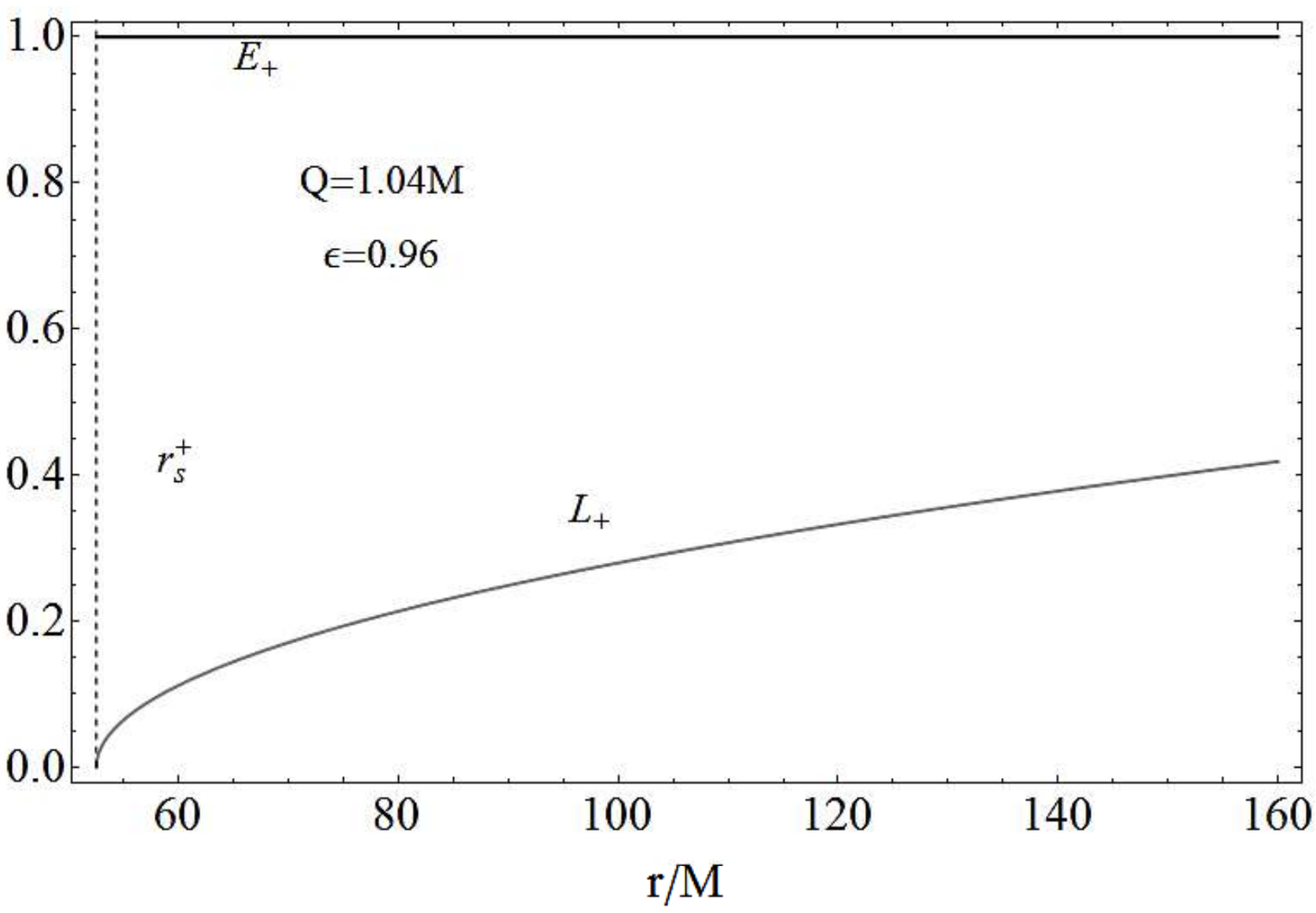}
\end{tabular}
\caption[font={footnotesize,it}]{\footnotesize{\textbf{Class}: $5/(2\sqrt{6})M<Q<(3\sqrt{6}/7)M$ and  $\widetilde{\widetilde{\epsilon}}_{+}\leq \epsilon<M/Q$.
Parameter choice: $Q= 1.04M$ and $\epsilon =0.96$.
Then $\widetilde{\widetilde{\epsilon}}_{+}=0.811927$, $M/Q=0.961538$ $r_{s}^+ =52.4944M$, $r_\gamma^-=1.20538M$, and  $r_\gamma^+ =1.79462M$.
Circular orbits exist with angular momentum
$L=L_+$ (gray curve) and energy  $E=E_+$ (black curve) in $r_\gamma^-<r<r_\gamma^+ $ ( left plot)   and
$L=L^{-}$  in $r>r_{s}^+ $;
$L=0$ at $r=r_{s}^{+}$ (right  plot).
}}
\label{Contino}
\end{figure}
\begin{figure}
\centering
\begin{tabular}{c}
\includegraphics[width=0.71\hsize,clip]{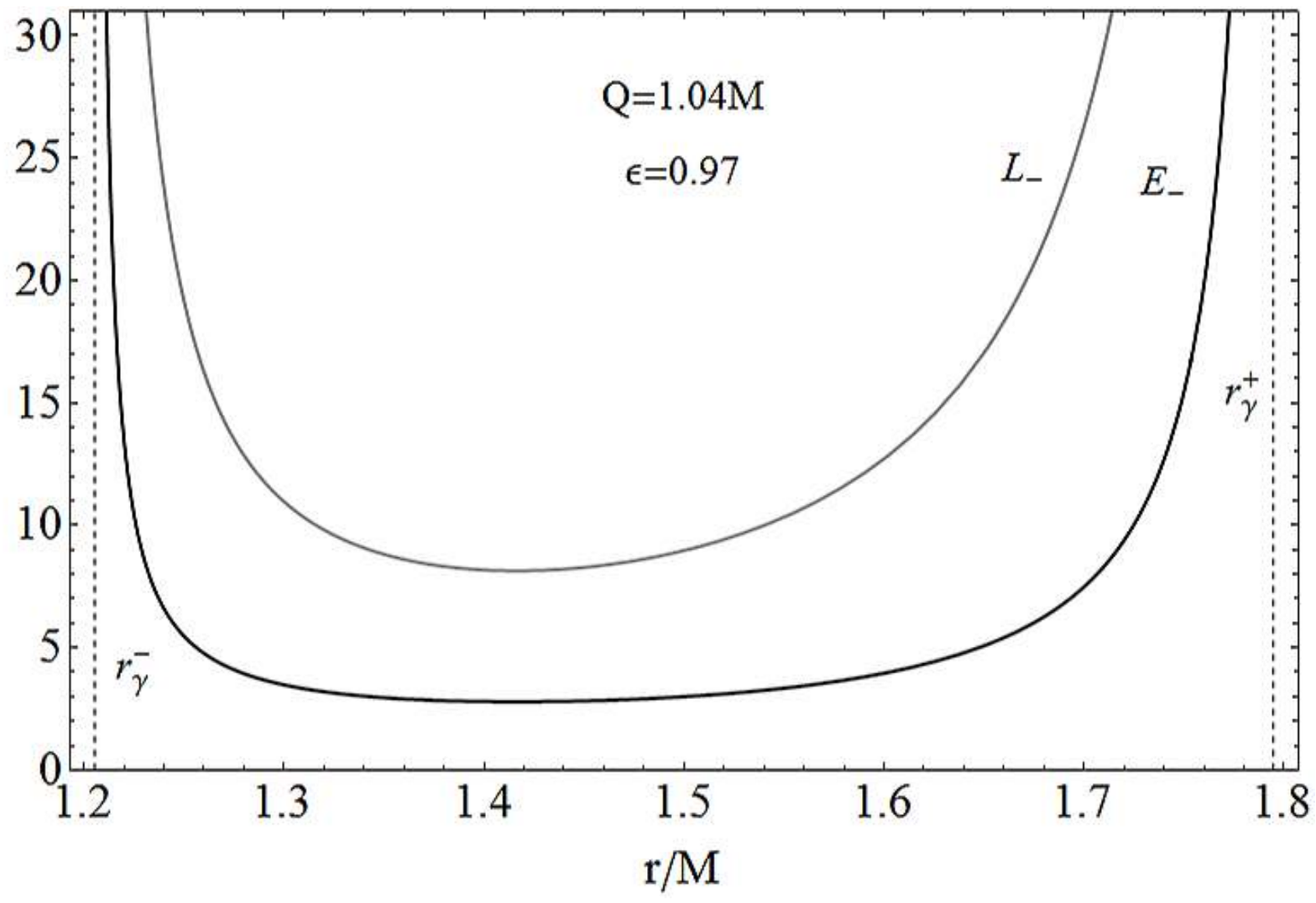}
\end{tabular}
\caption[font={footnotesize,it}]{\footnotesize{\textbf{Class}: $5/(2\sqrt{6})M<Q<(3\sqrt{6}/7)M$ and  $M /Q\le\epsilon <1$.
Parameter choice: $Q= 1.04M$ and $\epsilon =0.97$,
Then $\widetilde{\widetilde{\epsilon}}_{+}=0.866828$, $M/Q=0.961538$,  $r_\gamma^-=1.20538M$, and $r_\gamma^+ = 1.79462M$.
Circular orbits exist with
$L=L_+$  (gray curve) and   $E=E_+$ (black curve) in $r_\gamma^-<r<r_\gamma^+ $.
}}
\label{Fig:Faraone}
\end{figure}
%

\begin{figure}
\centering
\begin{tabular}{cc}
\includegraphics[width=0.51\hsize,clip]{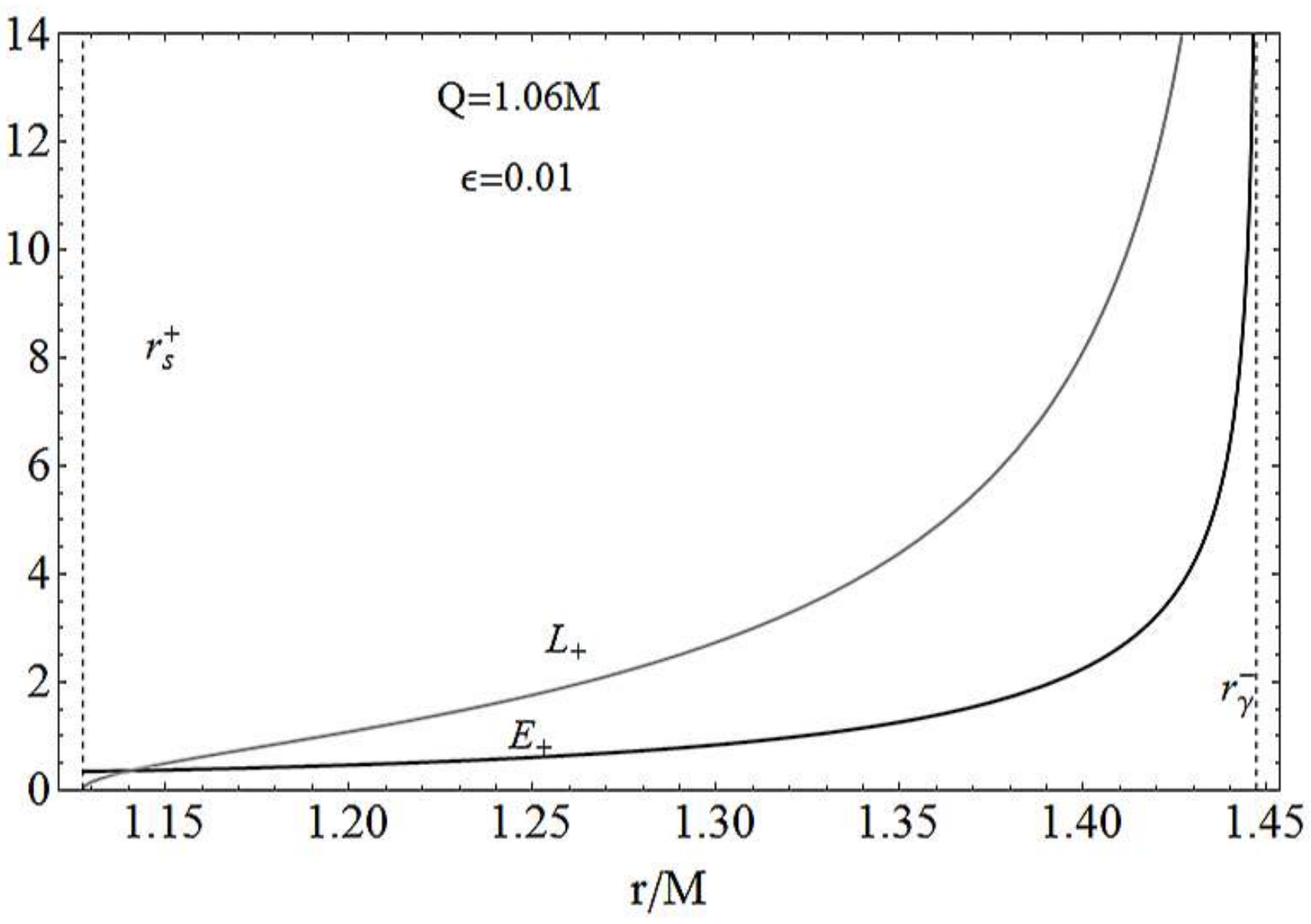}
\includegraphics[width=0.51\hsize,clip]{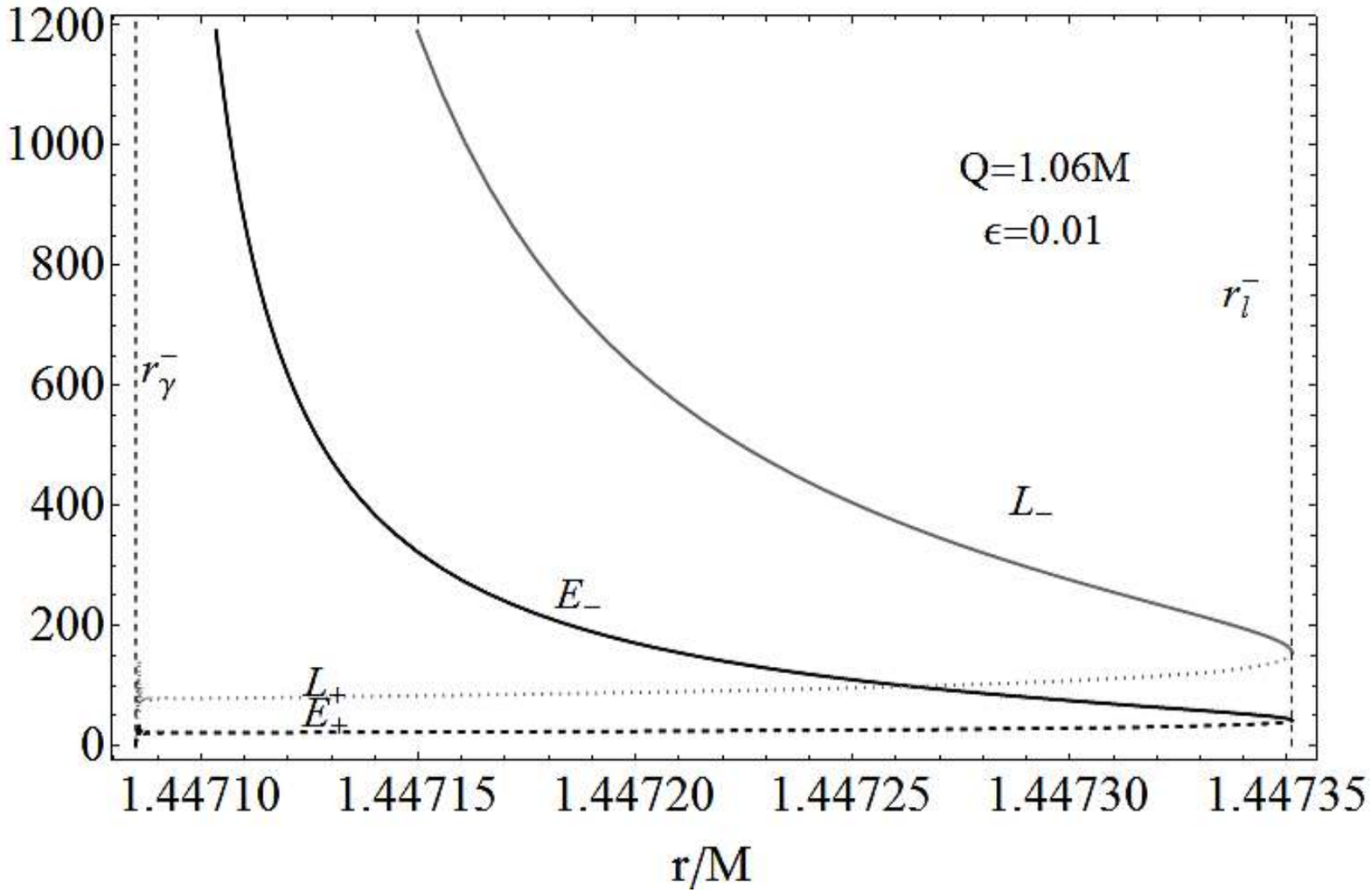}\\
\includegraphics[width=0.51\hsize,clip]{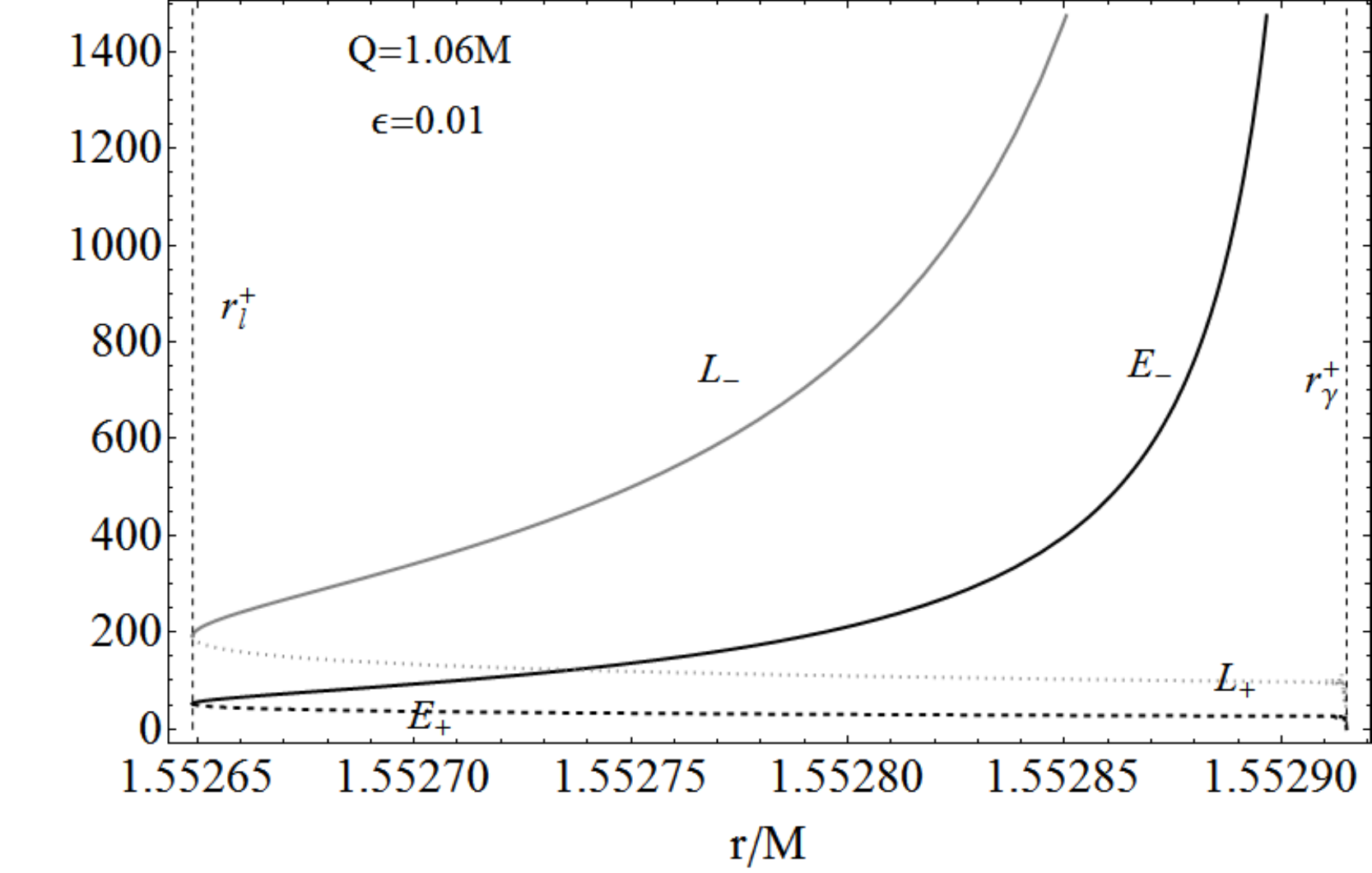}
\includegraphics[width=0.51\hsize,clip]{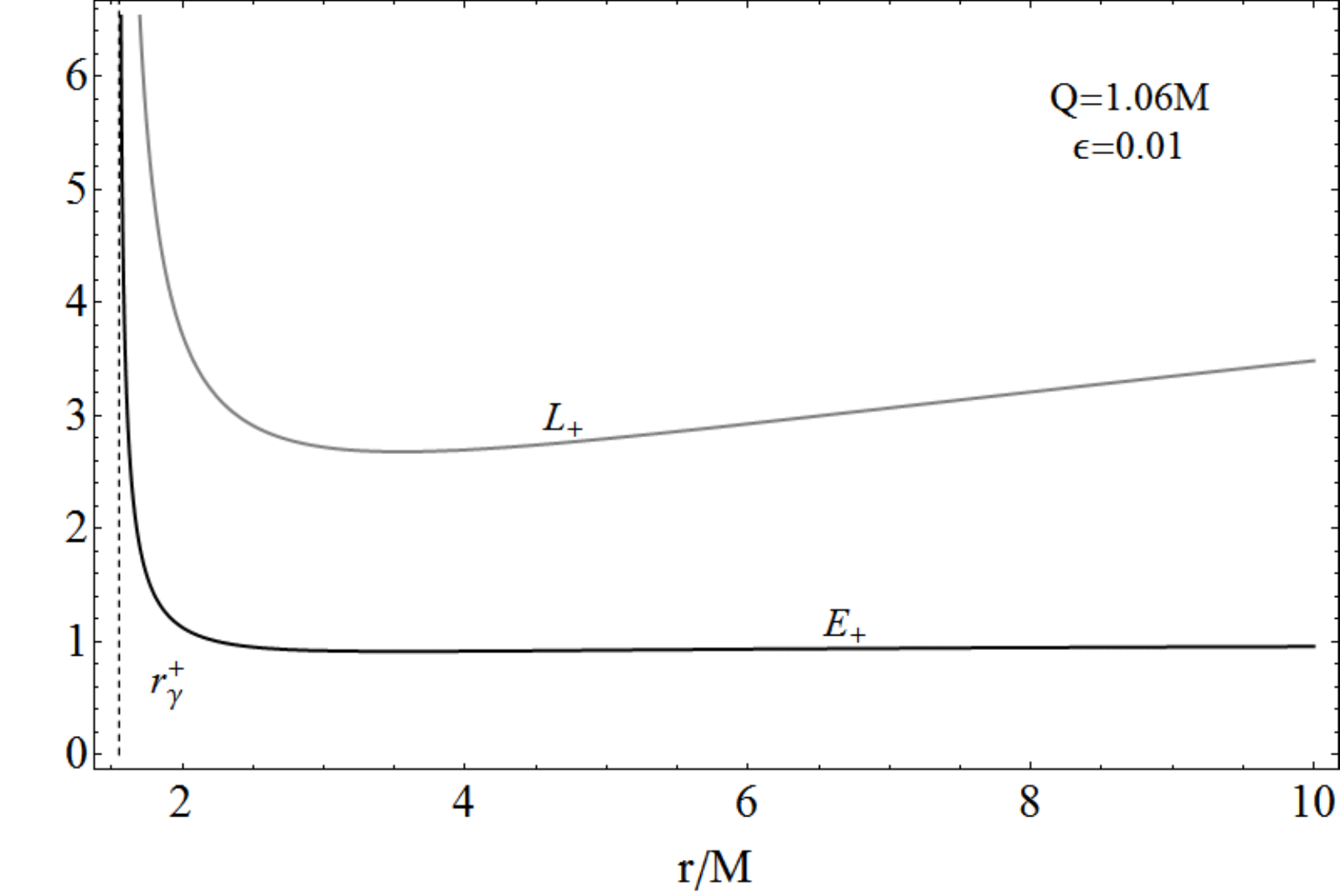}
\end{tabular}
\caption[font={footnotesize,it}]{\footnotesize{\textbf{Class}: $(3\sqrt{6}/7)M\leq Q\leq\sqrt{9/8}M$ and  $0<\epsilon\le\epsilon_{l}$
Parameter choice: $Q= 1.06M$ and $\epsilon =0.01$.
Then $\epsilon_{l}=0.0998397$,  $r_{s}^+ =1.12756M$, $r_\gamma^-=1.44708M$, $r_\gamma^+ =1.55292M$ $r_l^-= 1.44735M$, and $r_l^+ =1.55265M$.
Circular orbits exist with angular momentum
$L=L^{-}$ (gray curve) and energy  $E=E^{-}$ (black curve) in $r_\gamma^-\leq r\leq r_l^-$;
$L=0$ at $r=r_{s}^{+}$ (upper left plot);
$L=L_{\pm}$ in $r_{s}^{+}<r<r_l^-$ (upper right plot) and  $r_l^+ \leq r<r_\gamma^+ $ (bottom left plot);
$L=L^{-}$ in $r\geq r_\gamma^+ $ (bottom right plot).
}}
\label{Fig:Vampa}
\end{figure}
\begin{figure*}
\centering
\begin{tabular}{lcc}
\includegraphics[width=0.31\hsize,clip]{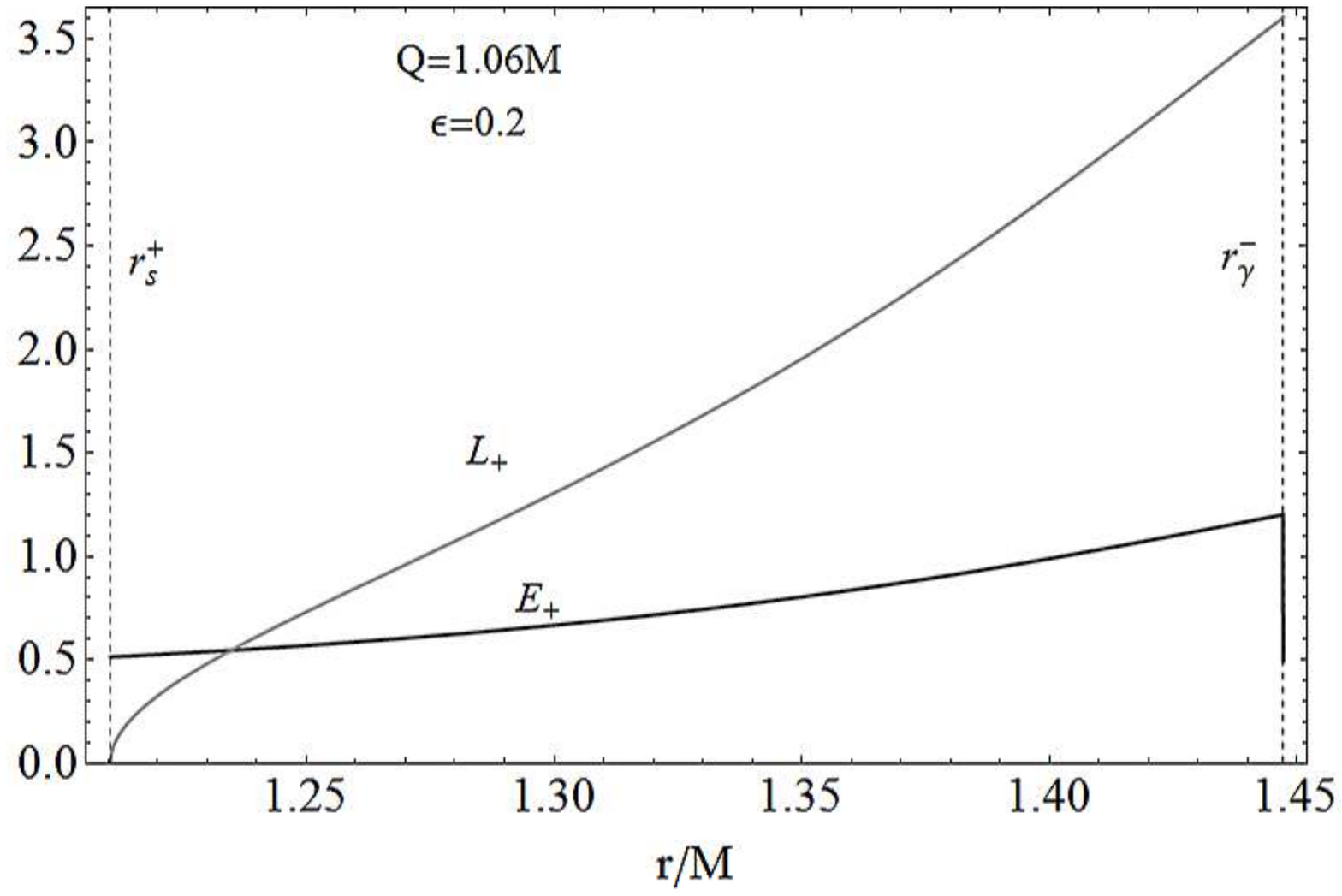}
\includegraphics[width=0.31\hsize,clip]{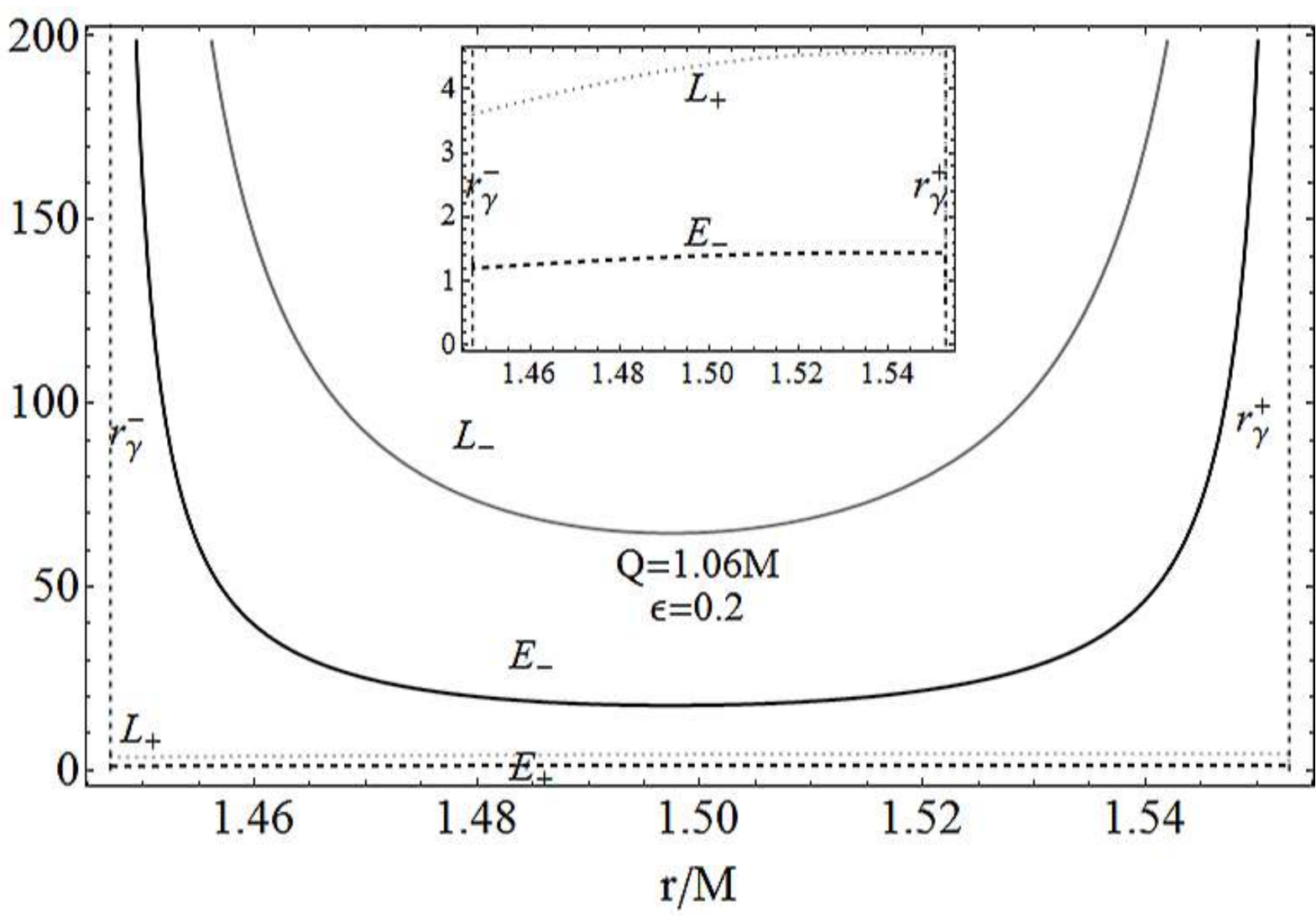}
\includegraphics[width=0.31\hsize,clip]{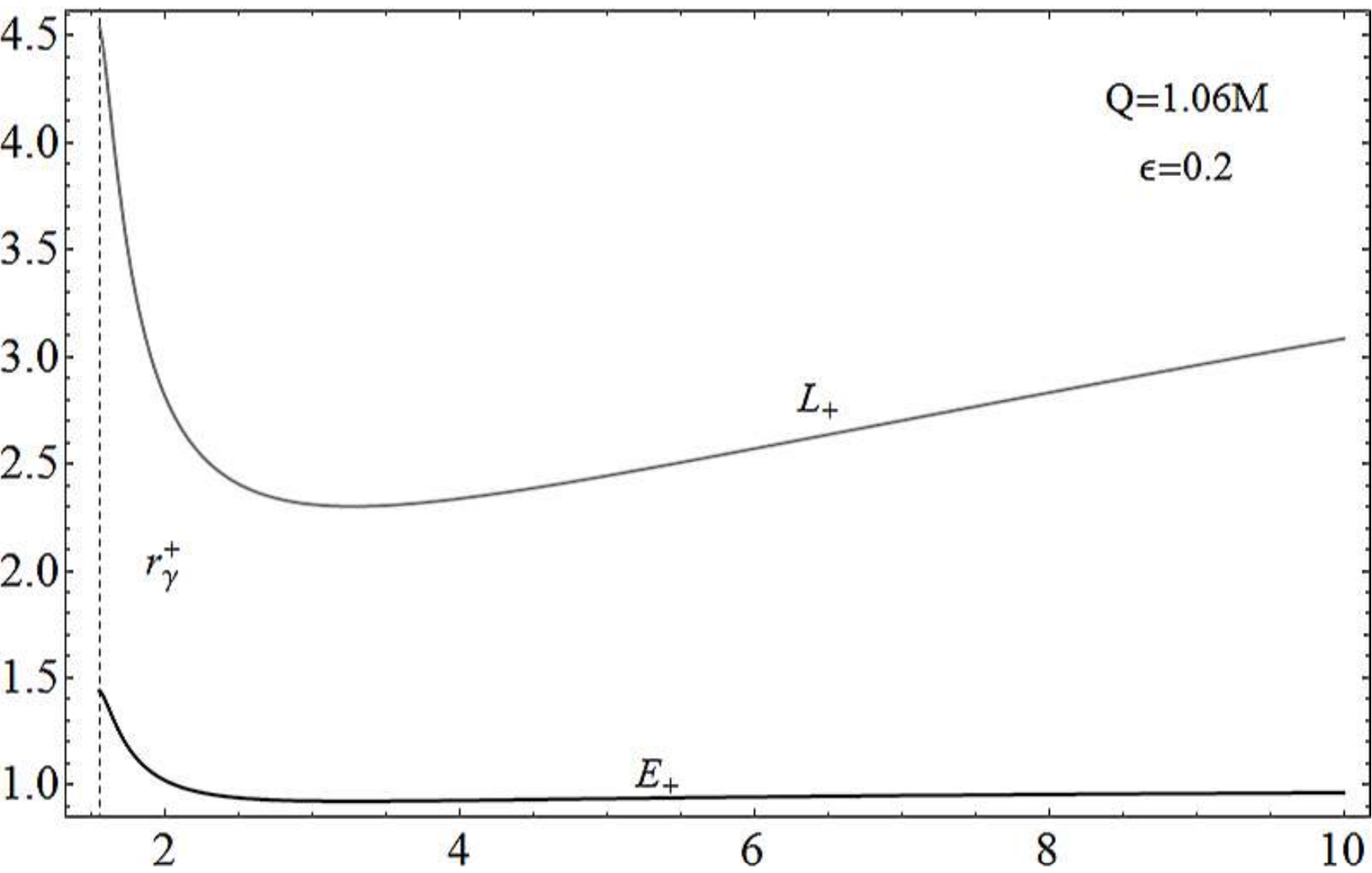}
\end{tabular}
\caption[font={footnotesize,it}]{\footnotesize{\textbf{Class}: $(3\sqrt{6}/7)M\leq Q\leq\sqrt{9/8}M$ and  $\epsilon_{l}<\epsilon\leq \widetilde{\widetilde{\epsilon}}_{-}$. Parameter choice: $Q= 1.06M$ and $\epsilon =0.2$.
Then $\epsilon_{l}=0.0998397$, $\widetilde{\widetilde{\epsilon}}_{-}=0.536564$,  $r_{s}^+ =1.21047M$, $r_\gamma^-=1.44708M$, and $r_\gamma^+ =1.55292M$.
Circular orbits exist with angular momentum
$L=L^{-}$ (gray curve) and energy  $E=E^{-}$ (black curve) in $r_{s}^{+}<r<r_\gamma^-$;
$L=0$ at $r=r_{s}^{+}$ (left plot);
$L=L_{\pm}$ in $r_\gamma^-\leq r<r_\gamma^+ $ (center plot); and
$L=L^{-}$ in $r\geq r_\gamma^+ $ (right  plot).
}}
\label{Caracoll}
\end{figure*}
\begin{figure*}
\centering
\begin{tabular}{lcl}
\includegraphics[width=0.3\hsize,clip]{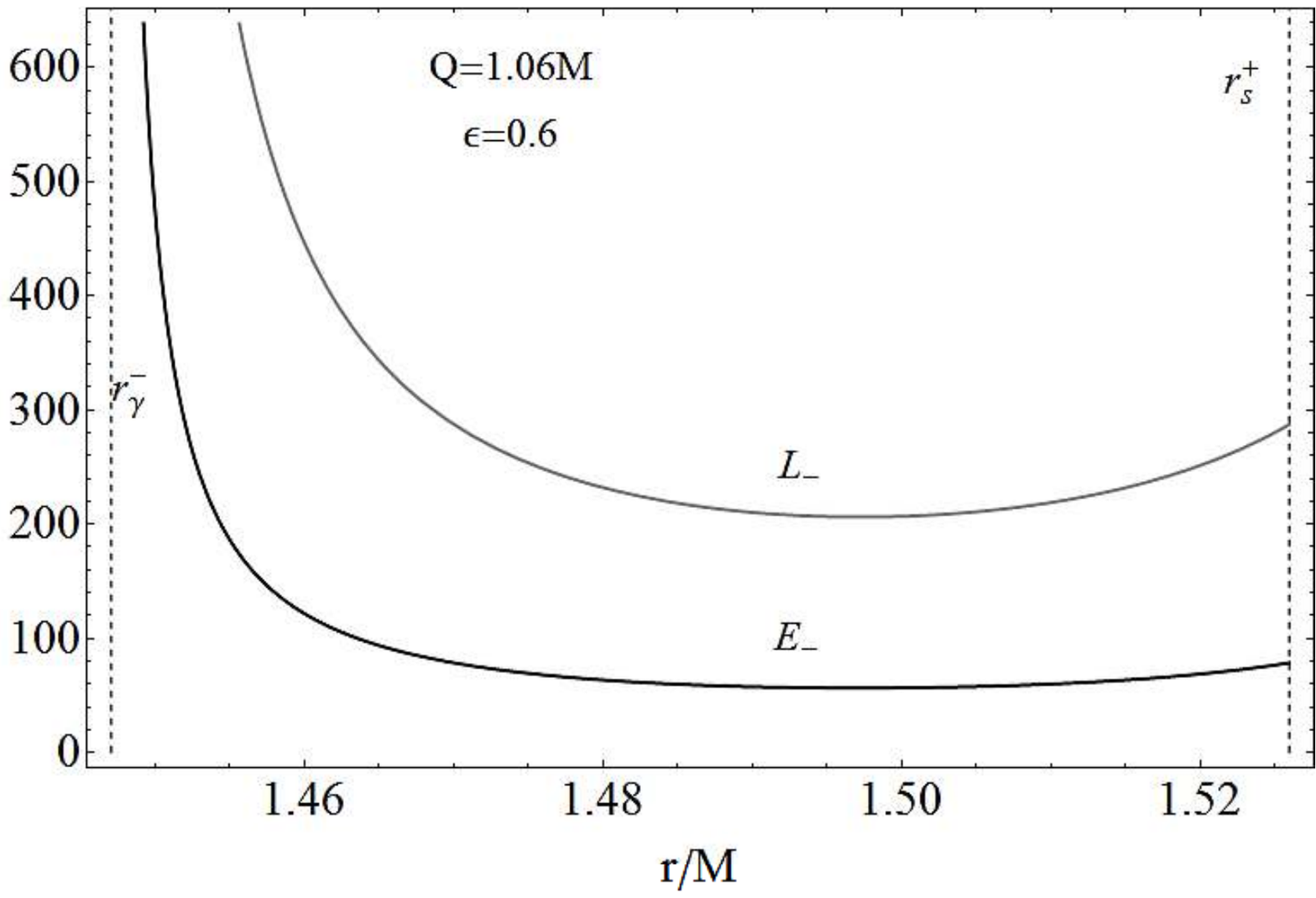}
\includegraphics[width=0.3\hsize,clip]{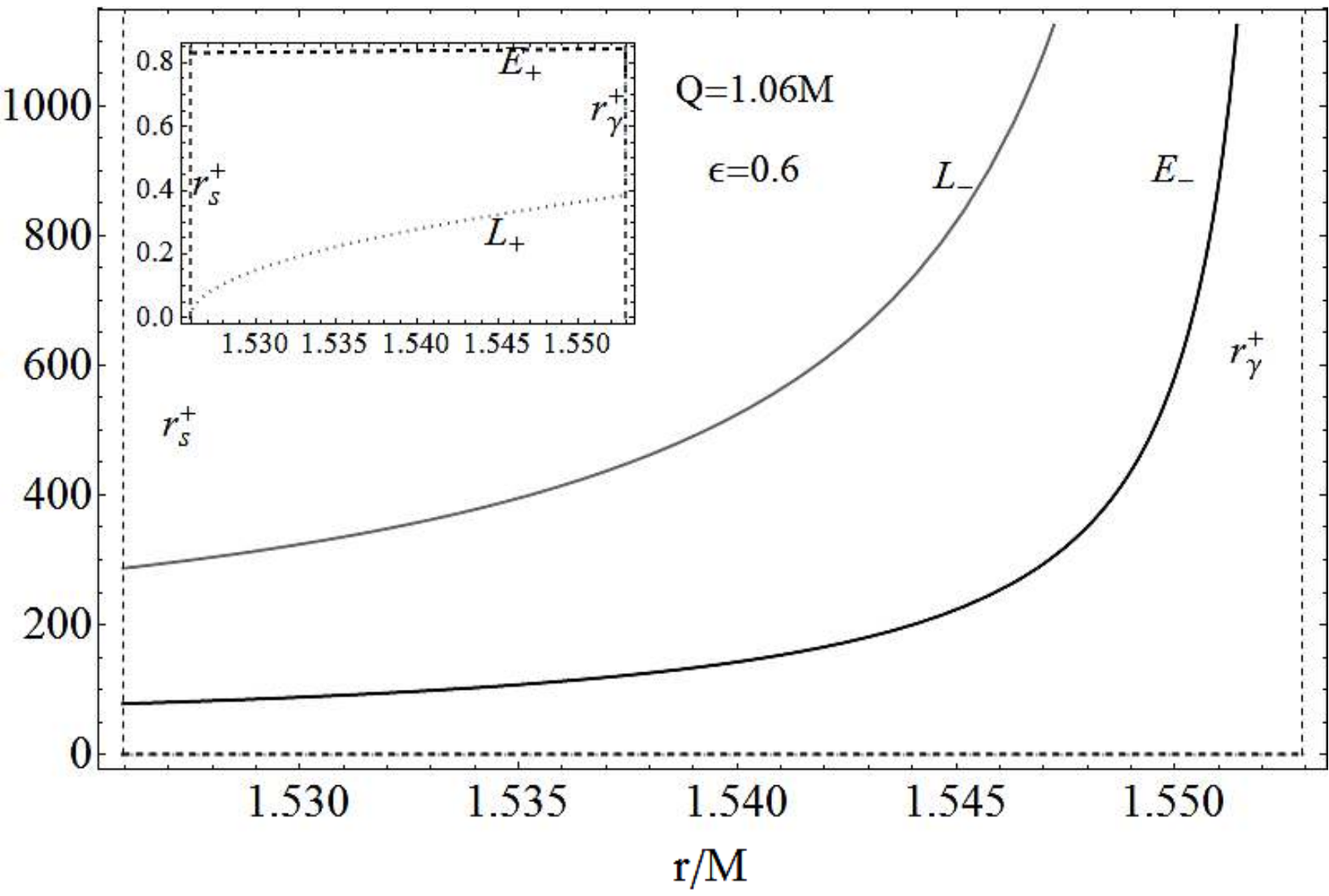}
\includegraphics[width=0.3\hsize,clip]{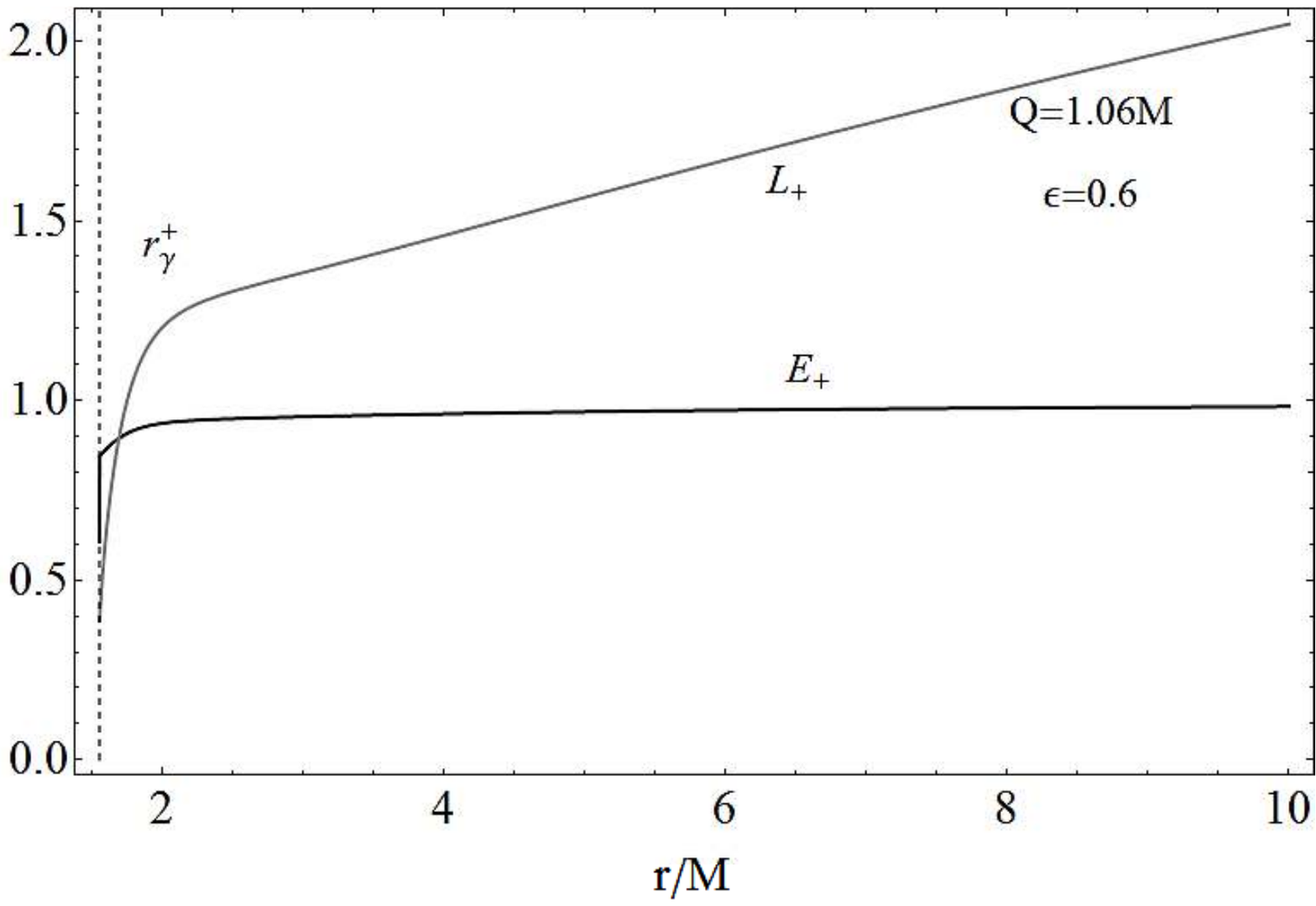}
\end{tabular}
\caption[font={footnotesize,it}]{\footnotesize{\textbf{Class}: $(3\sqrt{6}/7)M\leq Q\leq\sqrt{9/8}M$ and  $\widetilde{\widetilde{\epsilon}}_{-}<\epsilon<\widetilde{\widetilde{\epsilon}}_{+}$.
Parameter choice: $Q= 1.06M$ and $\epsilon =0.6$.
Then $\widetilde{\widetilde{\epsilon}}_{+}=0.618133$, $\widetilde{\widetilde{\epsilon}}_{-}=0.536564$, $r_{s}^+ =1.52596M$, $r_\gamma^-=1.44708M$, and $r_\gamma^+ =1.55292M$.
Circular orbits exist with angular momentum
$L=L_+$ (gray curve) and energy  $E=E_+$ (black curve) in $r_\gamma^-<r<r_{s}^{+}$ (left plot);
$L=L_{\pm}$ in $r_{s}^+ \leq r<r_\gamma^+ $ (center plot); and
$L=L^{-}$ in $r\geq r_\gamma^+ $ (right  plot).
}}
\label{Villart}
\end{figure*}
\begin{figure}
\centering
\begin{tabular}{cc}
\includegraphics[width=0.51\hsize,clip]{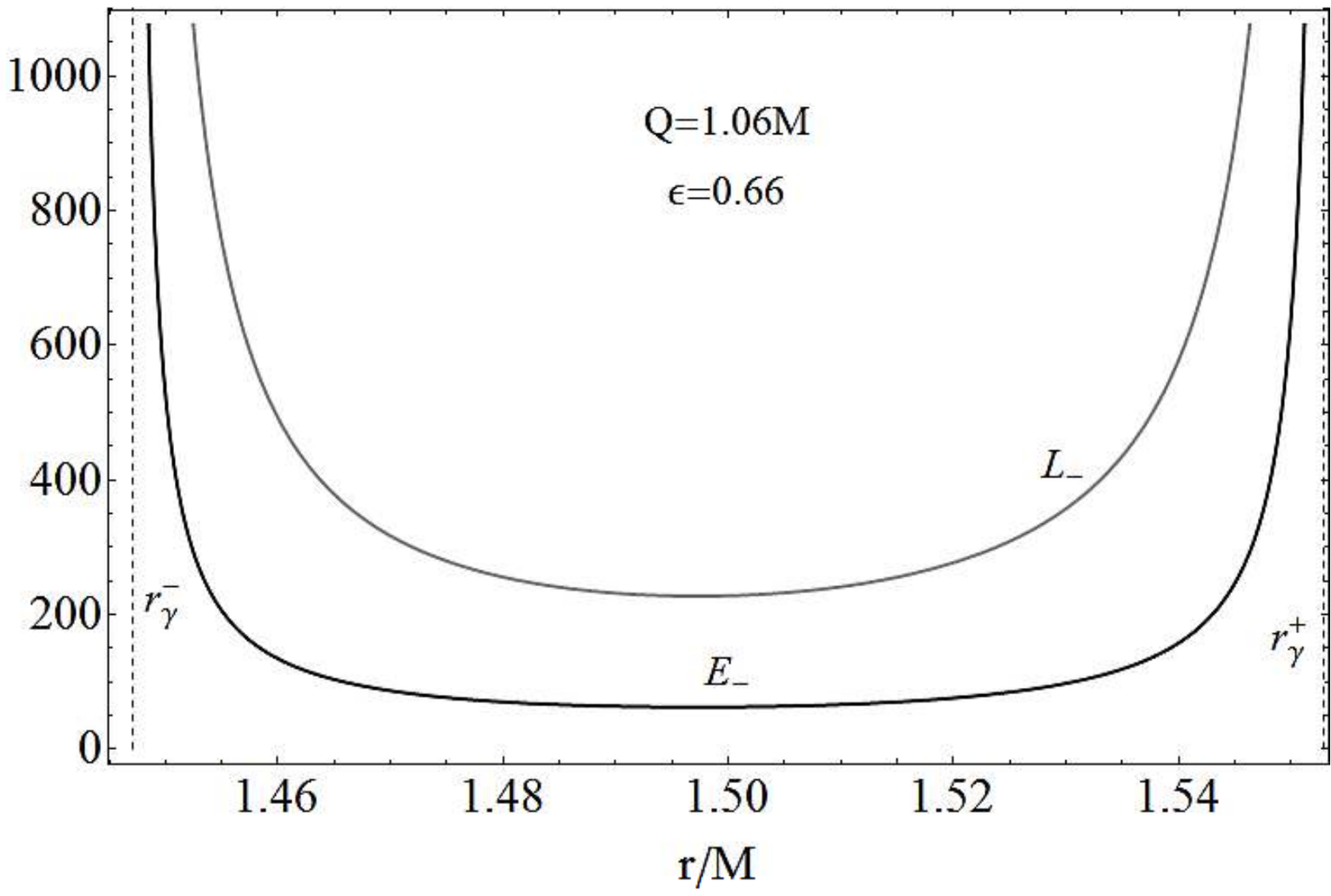}
\includegraphics[width=0.51\hsize,clip]{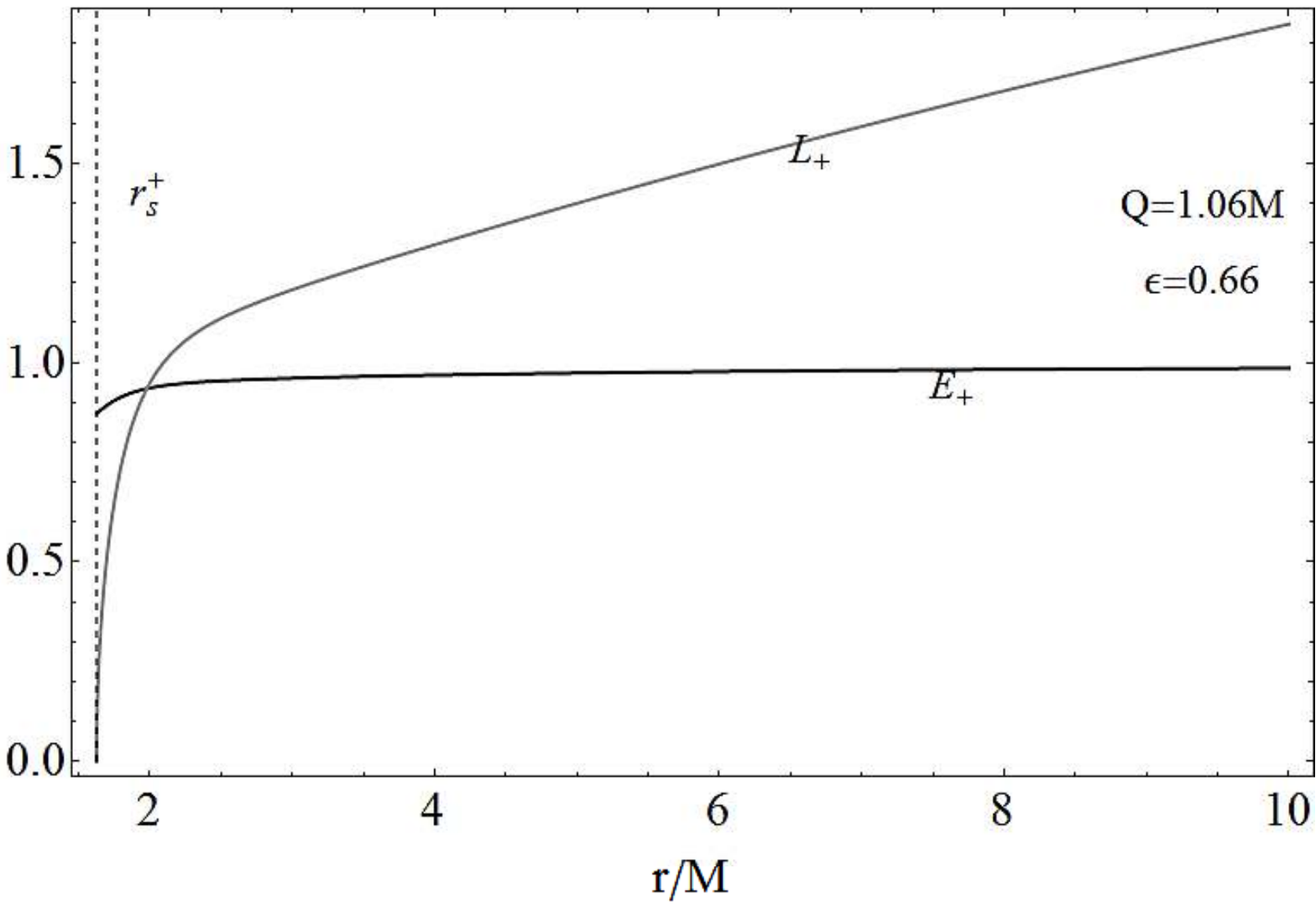}
\end{tabular}
\caption[font={footnotesize,it}]{\footnotesize{\textbf{Class}: $(3\sqrt{6}/7)M\leq Q\leq\sqrt{9/8}M$ and  $\widetilde{\widetilde{\epsilon}}_{+}\leq \epsilon<M/Q$.
 choice: $Q= 1.06M$ and $\epsilon =0.66$.
 Then $\widetilde{\widetilde{\epsilon}}_{+}=0.618133$, $M/Q=0.943396$  $r_{s}^+ =1.62572M$, $r_\gamma^-= 1.44708,M$, and $r_\gamma^+ =1.55292M$.
Circular orbits exist with angular momentum
$L=L_+$ (gray curve) and energy  $E=E_+$ (black curve) in $r_\gamma^-<r<r_\gamma^+$ (left plot) and
$L=L^{-}$ in $r>r_{s}^{+}$;
$L=0$ at $r=r_{s}^{+}$ (right  plot).
}}
\label{Fig:Catalani}
\end{figure}
\begin{figure}
\centering
\begin{tabular}{cc}
\includegraphics[width=0.71\hsize,clip]{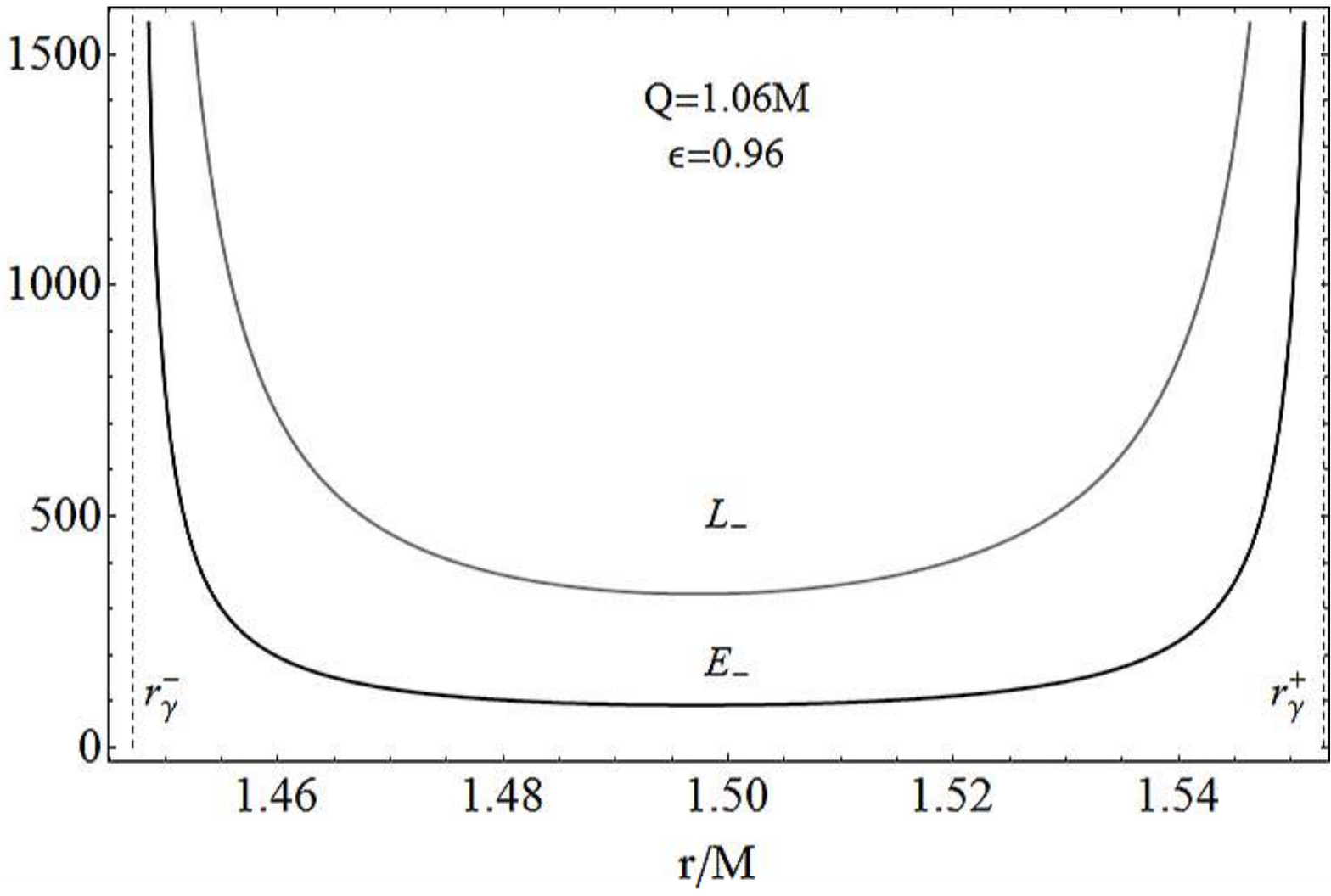}
\end{tabular}
\caption[font={footnotesize,it}]{\footnotesize{\textbf{Class}: $(3\sqrt{6}/7)M\leq Q\leq\sqrt{9/8}M$ and  $M/Q\le\epsilon <1$.
Parameter choice: $Q= 1.06M$ and $\epsilon =0.96$.
Then $M/Q=0.943396$, $r_\gamma^-=1.44708,M$ and $r_\gamma^+ =1.55292M$.
Circular orbits exist with angular momentum
$L=L_+$ (gray curve) and energy  $E=E_+$ (black curve) in $r_\gamma^-<r<r_\gamma^+$.
}}
\label{Epinay}
\end{figure}
%
\begin{figure}
\centering
\begin{tabular}{cc}
\includegraphics[width=0.71\hsize,clip]{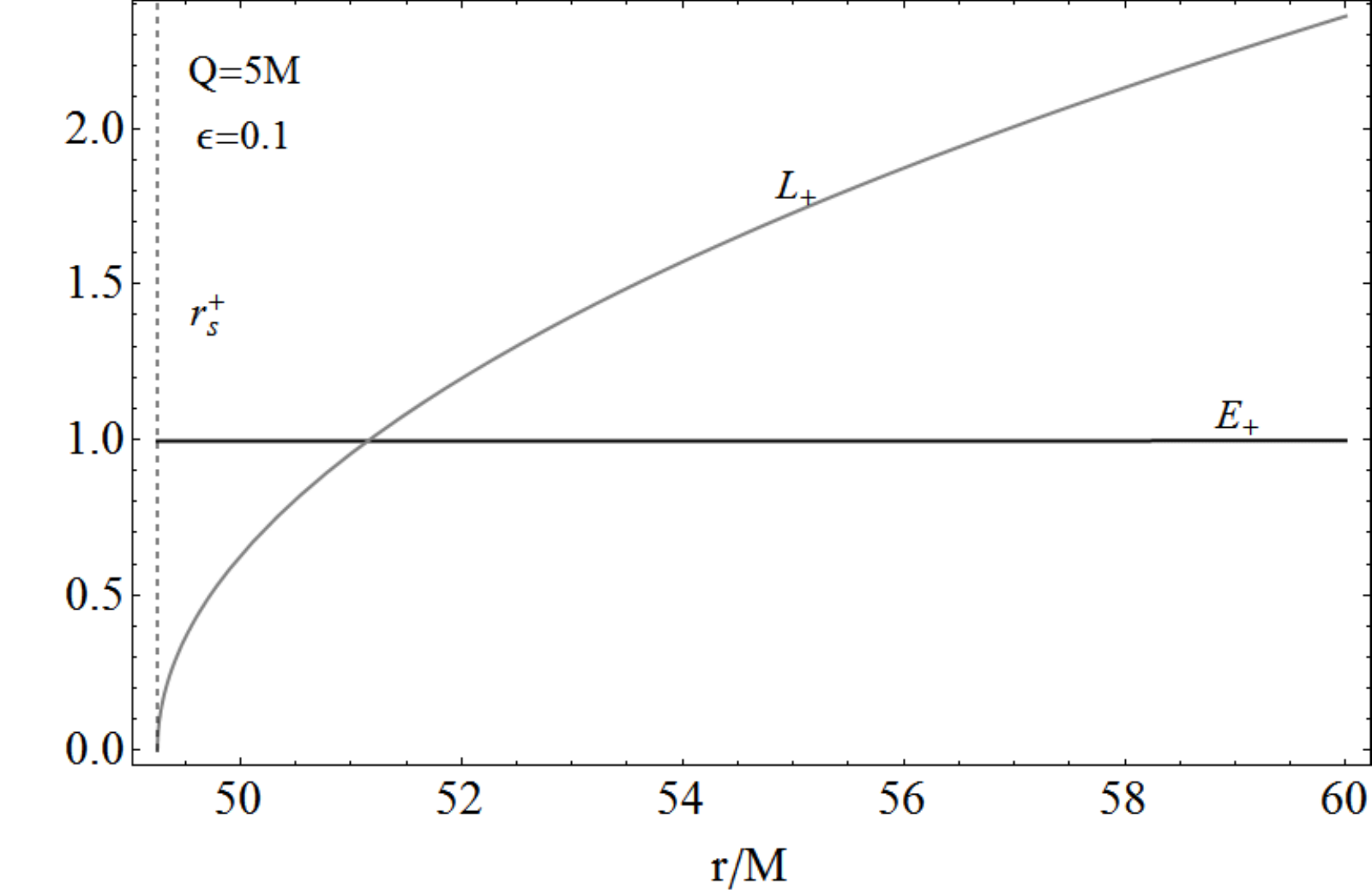}
\end{tabular}
\caption[font={footnotesize,it}]{\footnotesize{\textbf{Class}:$0<\epsilon<M/Q$ and $Q>\sqrt{9/8}M$.
Parameter choice: $Q= 5M$ and $\epsilon =0.1$.
Then $M/Q=0.2$ and $r_{s}^+ =49.2481M$.
Circular orbits exist with angular momentum
$L=L^{-}$ (gray curve) and energy  $E=E^{-}$ (black curve) in $r> r_{s}^+ $ and $L=0$ at $r= r_{s}^+ $.
}}
\label{Fig:Greca}
\end{figure}

\clearpage

\section{Behavior of the angular momentum and energy of negative test charges}
\label{ap:fign}
\begin{figure}
\centering
\begin{tabular}{cc}
\includegraphics[width=0.51\hsize,clip]{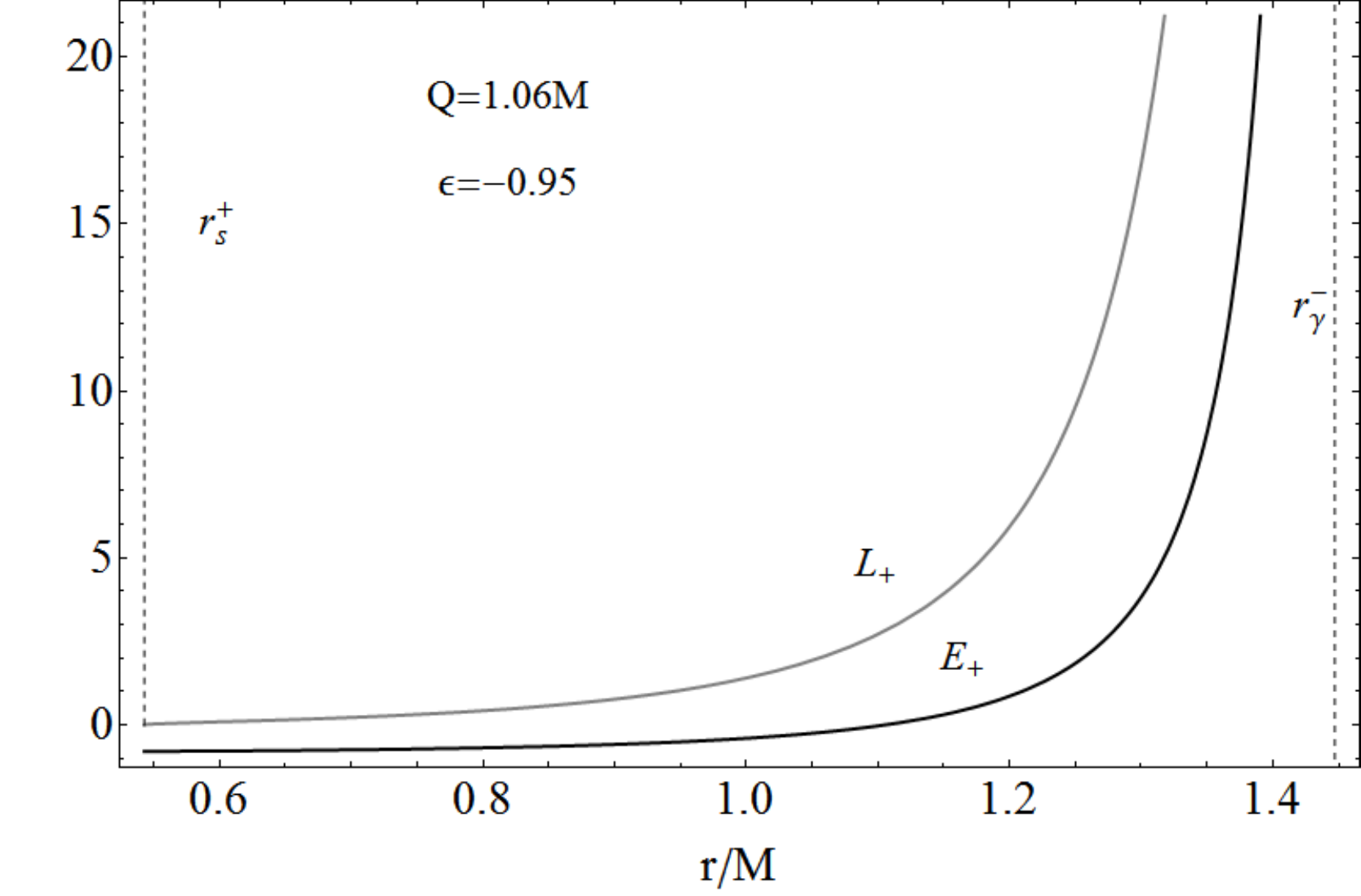}
\includegraphics[width=0.51\hsize,clip]{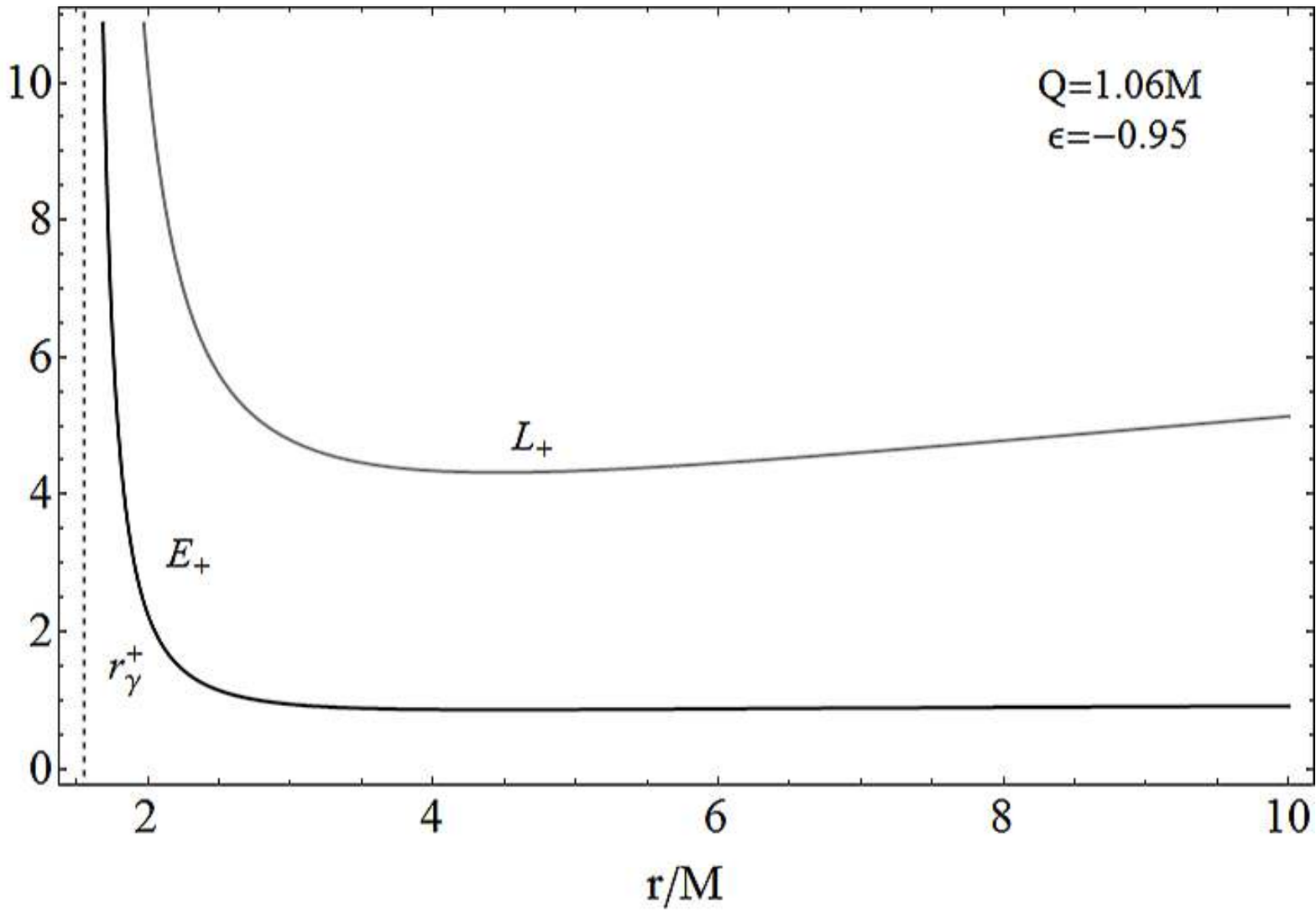}
\end{tabular}
\caption[font={footnotesize,it}]{\footnotesize{\textbf{Class}: $M<Q\leq\sqrt{9/8}M$ and  $-1<\epsilon<-M/Q$.
Parameter choice: $Q= 1.06M$ and  $\epsilon=-0.95$.
Then  $r_{s}^+ =0.542901M$, $r_\gamma^-=1.44708M$, and $r_\gamma^+ =1.55292M$.
Circular orbits exist with angular momentum
$L=L_+$ (gray curve) and energy  $E=E_+$ (black curve) in $r_{s}^+ < r<r_\gamma^-$ (left plot) and in $r>r_\gamma^+ $ for (right  plot).
For $r=r_{s}^+ $,  $L=0$.
 }}
\label{Reveng}
\end{figure}
\begin{figure}
\centering
\begin{tabular}{cc}
\includegraphics[width=0.51\hsize,clip]{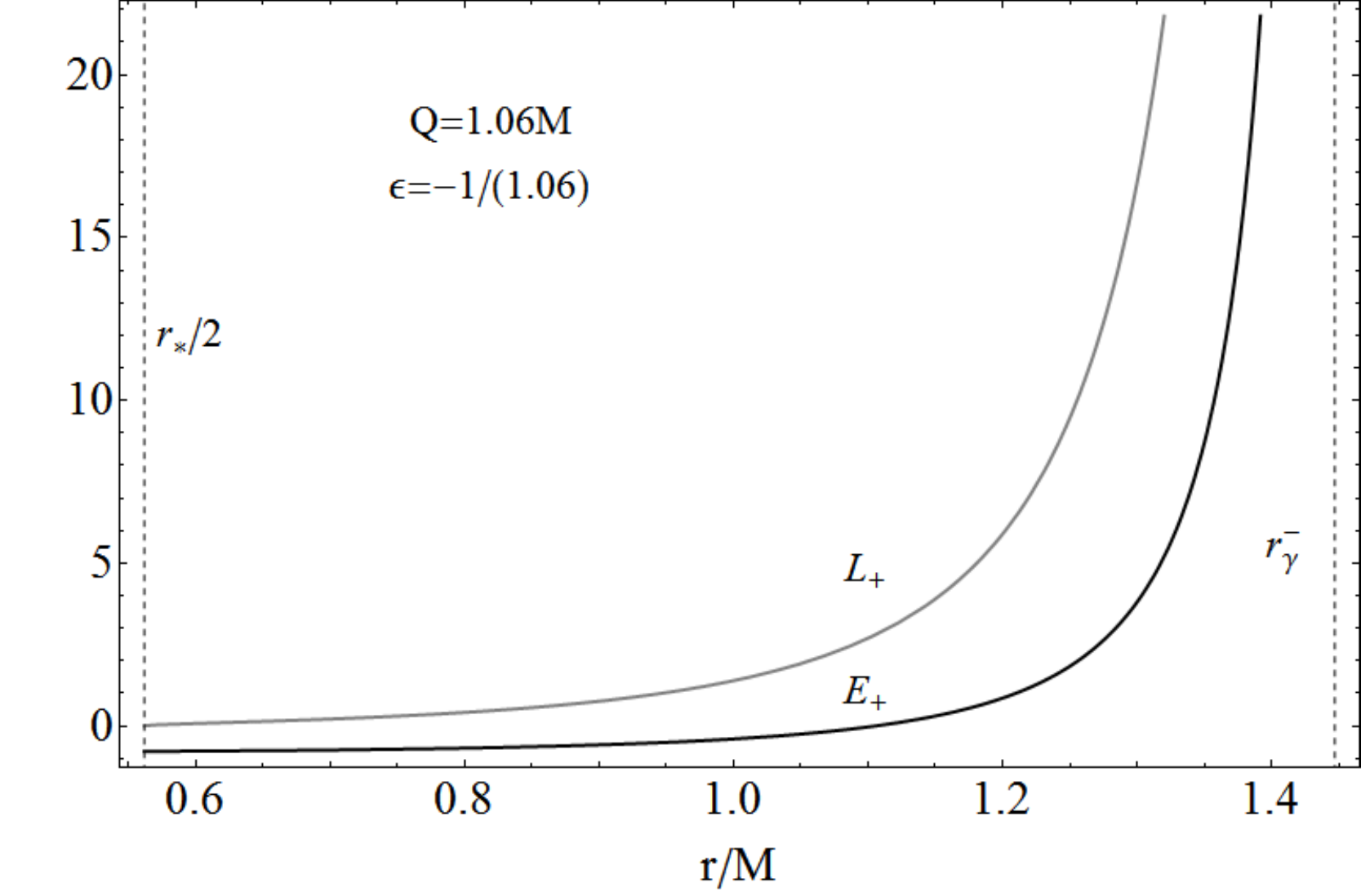}
\includegraphics[width=0.51\hsize,clip]{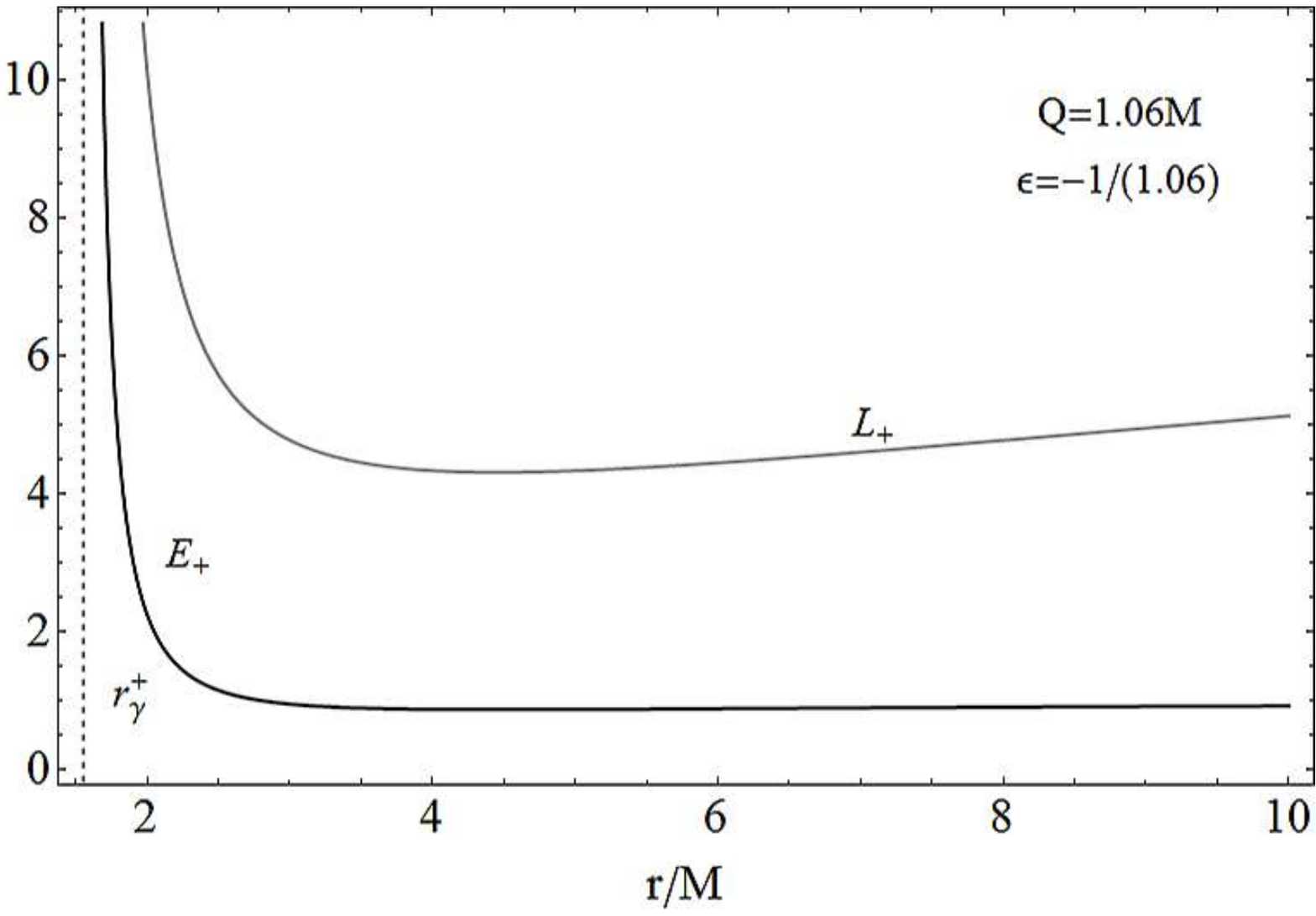}
\end{tabular}
\caption[font={footnotesize,it}]{\footnotesize{\textbf{Class}: $M<Q\leq\sqrt{9/8}M$ and  $\epsilon=-M/Q$.
Parameter choice: $Q= 1.06M$. Then $Q^2/(2M)=0.5618M$, $r_\gamma^-= 1.44708M$ and $r_\gamma^+ =1.55292M$.
Circular orbits exist with angular momentum
$L=L_+$ (gray curve) and energy  $E=E_+$ (black curve) in $Q^2/(2M)<r<r_\gamma^-$  (left plot) and in $r>r_\gamma^+ $(right  plot).
For $r=Q^2/(2M)$, $L=0$.
}}
\label{Oro}
\end{figure}
\begin{figure}
\centering
\begin{tabular}{cc}
\includegraphics[width=0.51\hsize,clip]{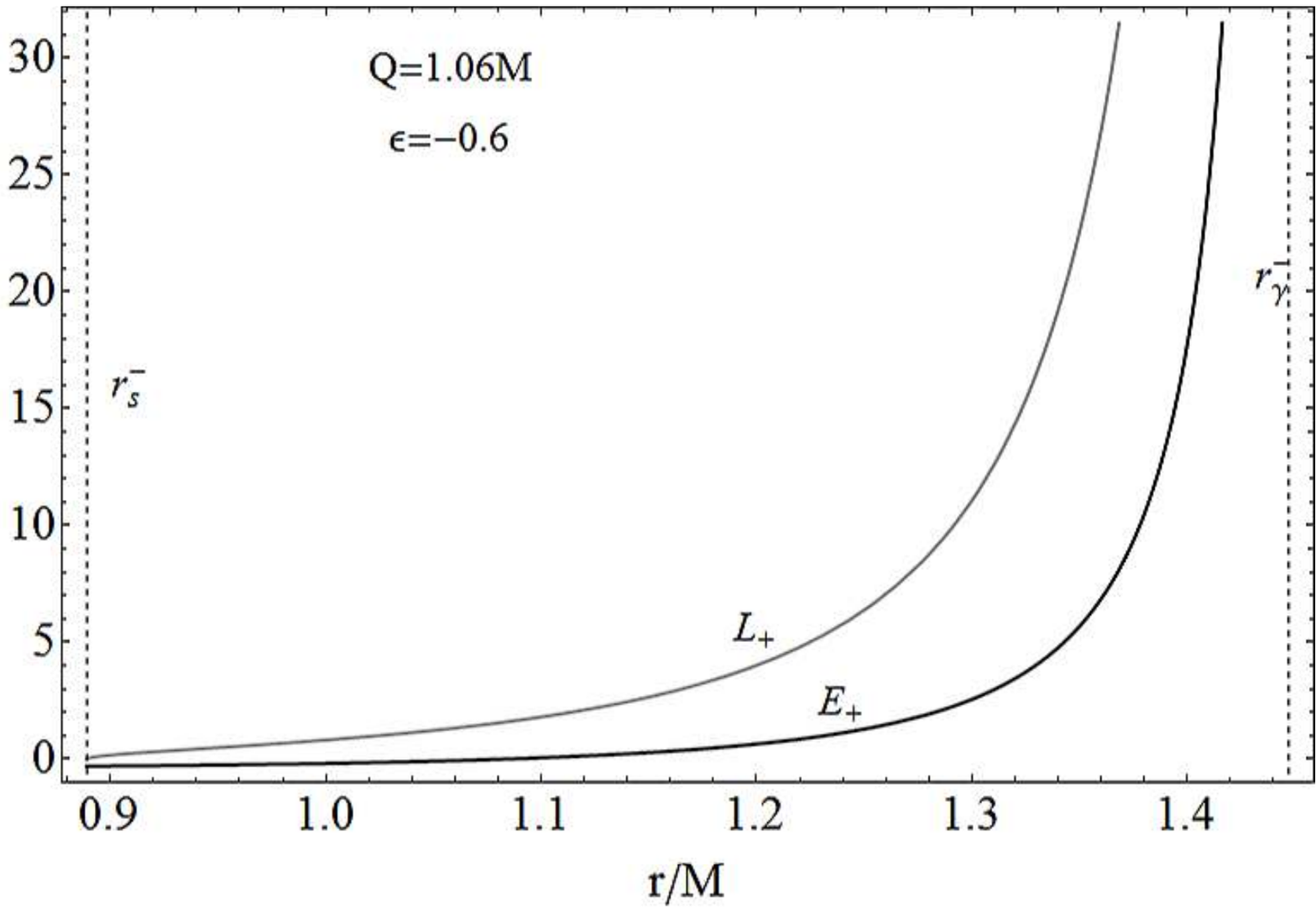}
\includegraphics[width=0.51\hsize,clip]{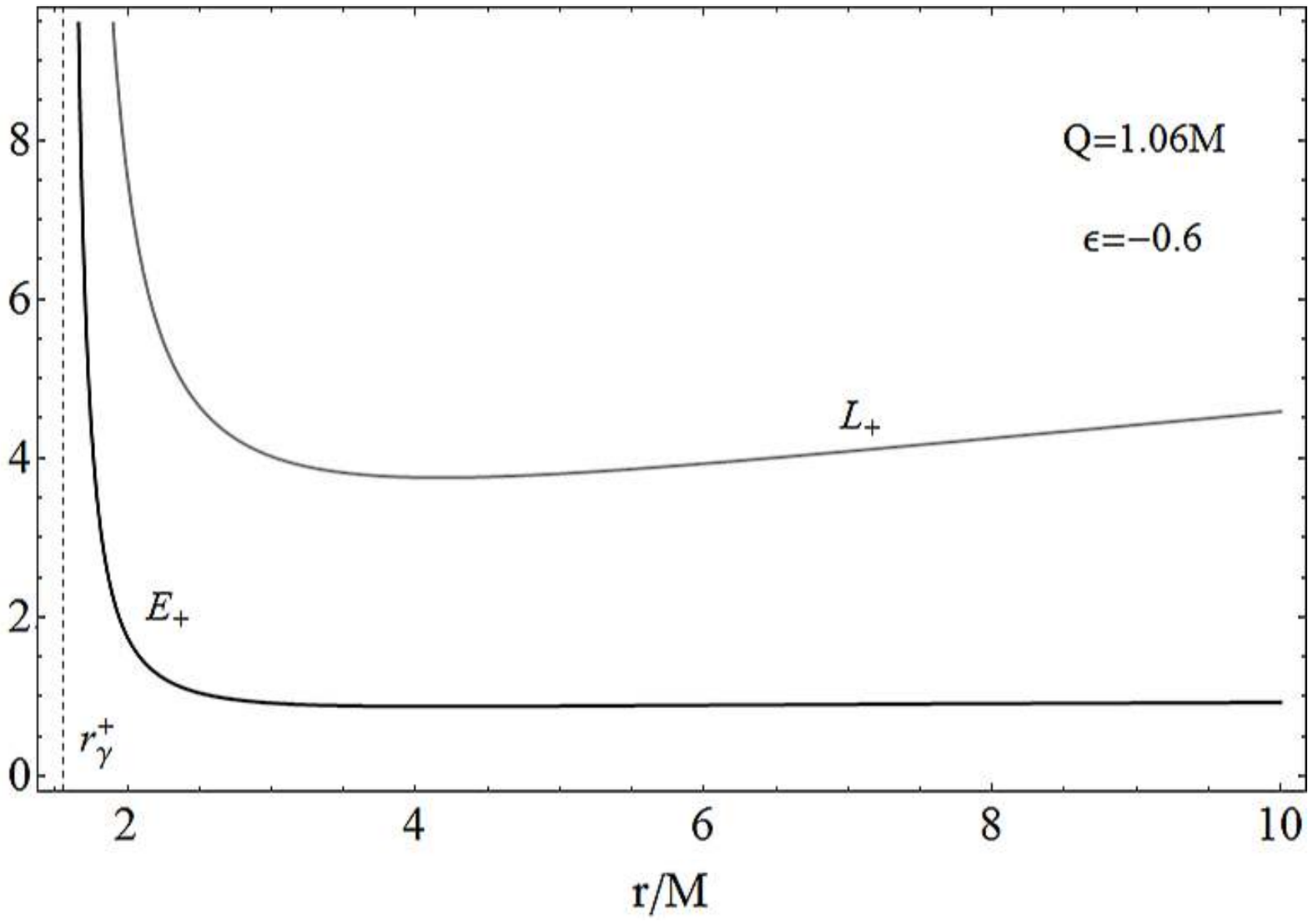}
\end{tabular}
\caption[font={footnotesize,it}]{\footnotesize{\textbf{Class}: $M<Q\leq\sqrt{9/8}M$ and  $-M/Q<\epsilon<0$.
Parameter choice: $Q= 1.06M$ and $\epsilon=-0.6$.
Then $M/Q=0.943396$,  $r_\gamma^-=1.44708M$, $r_\gamma^+ =1.55292M$, and $r_{s}^-=0.889152M$.
Circular orbits exist with angular momentum
$L=L_+$ (gray curve) and energy  $E=E_+$ (black curve) in $r_{s}^- <r<r_\gamma^-$   (left plot) and in $r>r_\gamma^+ $ (right  plot).
For $r=r_{s}^-$, $L=0$.
}}
\label{Grotta}
\end{figure}
\begin{figure}
\centering
\begin{tabular}{c}
\includegraphics[width=0.7\hsize,clip]{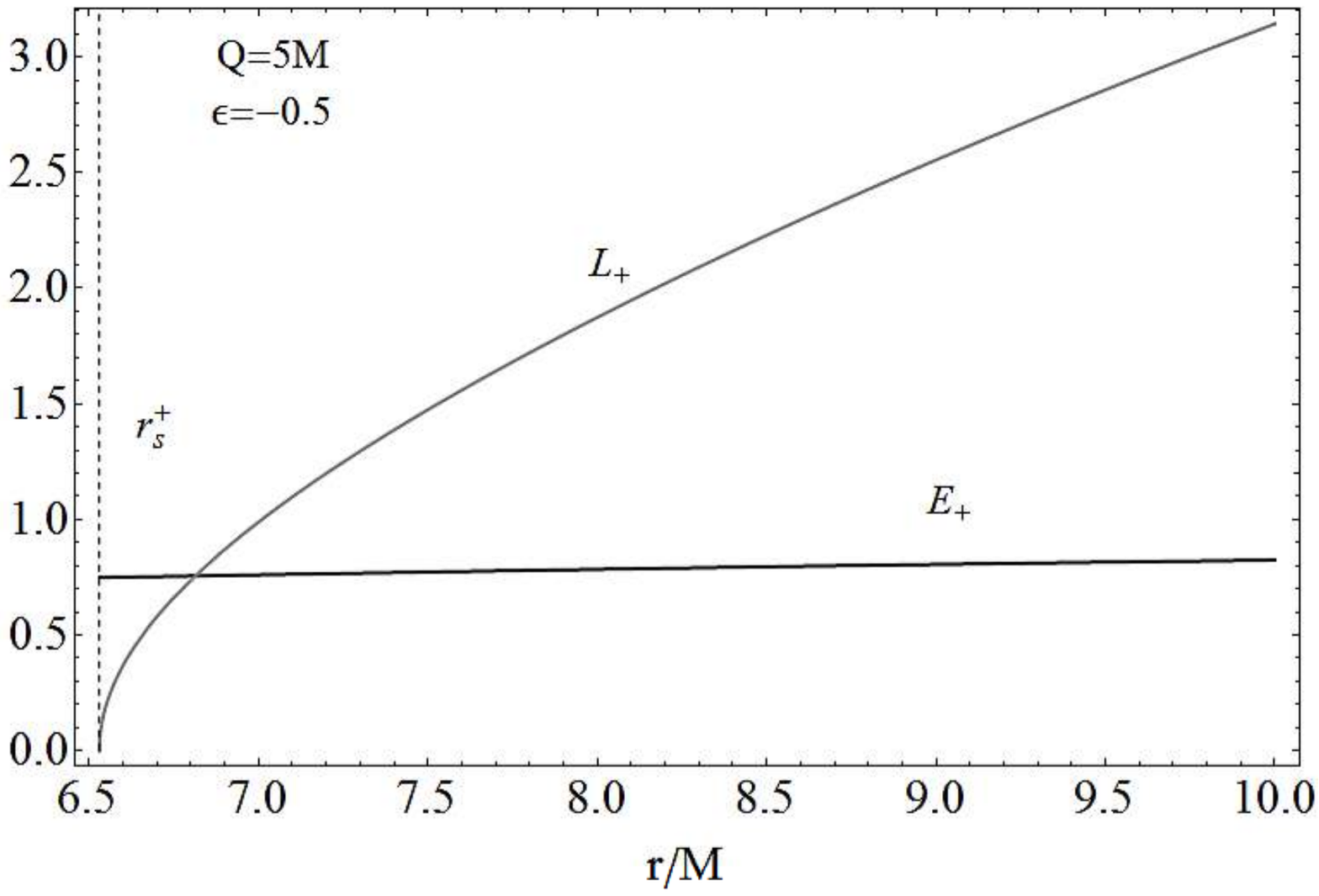}
\end{tabular}
\caption[font={footnotesize,it}]{\footnotesize{\textbf{Class}: $Q>\sqrt{9/8}M$ and  $-1<\epsilon<-M/Q$.
Parameter choice: $Q= 5M$ and $\epsilon =-0.5$.
Then $M/Q=0.2$, $r_{s}^{+}=6.5301M$.
Circular orbits exist with angular momentum
$L=L_+$ (gray curve) and energy  $E=E_+$ (black curve) in $r>r_{s}^{+}$.
For $r=r_{s}^{+}$, $L=0$.
}}
\label{Franz}
\end{figure}
\begin{figure}
\centering
\begin{tabular}{c}
\includegraphics[width=0.71\hsize,clip]{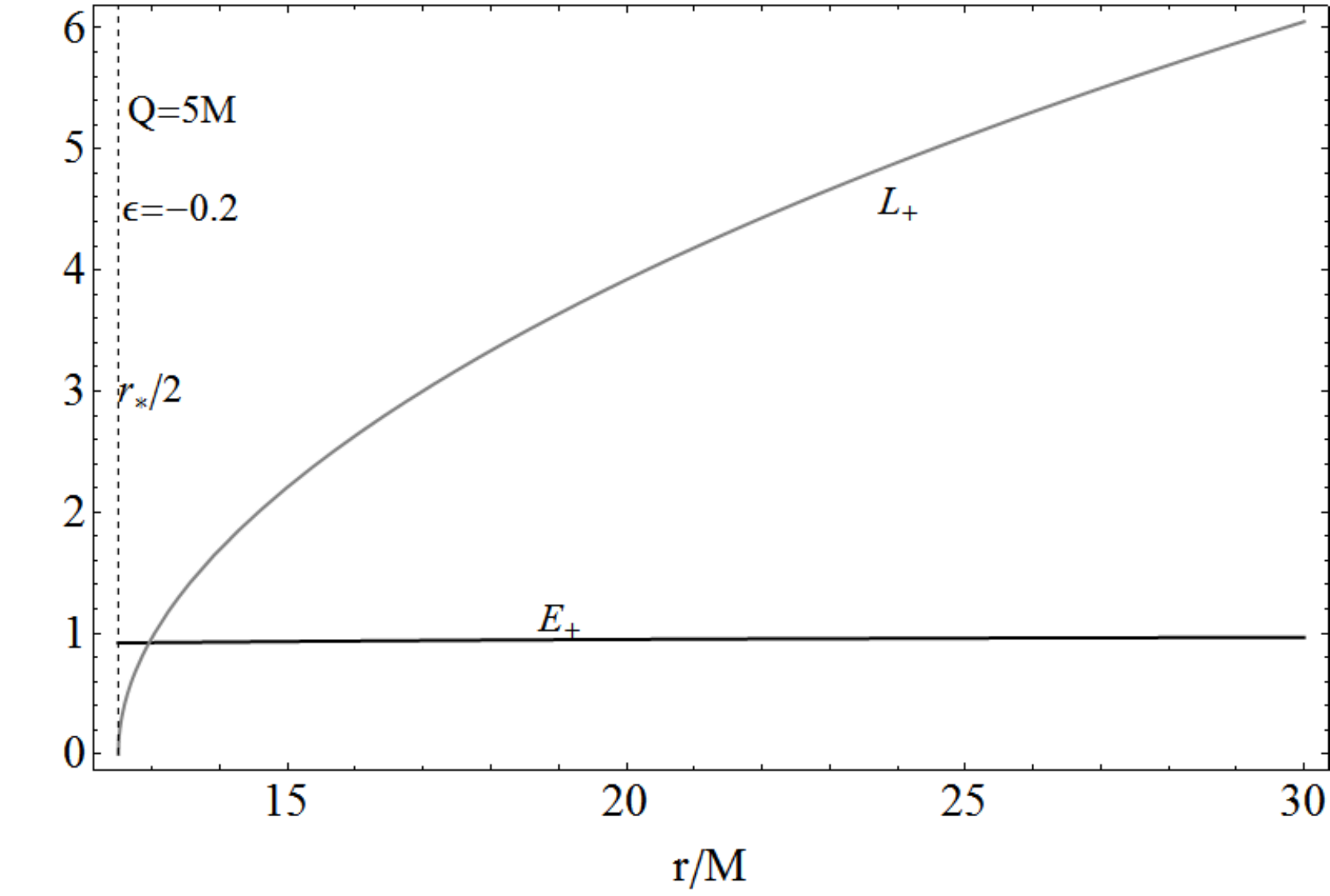}
\end{tabular}
\caption[font={footnotesize,it}]{\footnotesize{\textbf{Class}: $Q>\sqrt{9/8}M$ and  $\epsilon=-M/Q$.
Parameter choice: $Q= 5M$ and $\epsilon =-0.2$. Then $r_{s}^{+}=16M$.
Circular orbits exist with angular momentum
$L=L_+$ (gray curve) and energy  $E=E_+$ (black curve) in the region $r>Q^2/(2M)$.
For $r=Q^2/(2M)$, $L=0$.
}}
\label{Beuch}
\end{figure}
\begin{figure}
\centering
\begin{tabular}{c}
\includegraphics[width=0.81\hsize,clip]{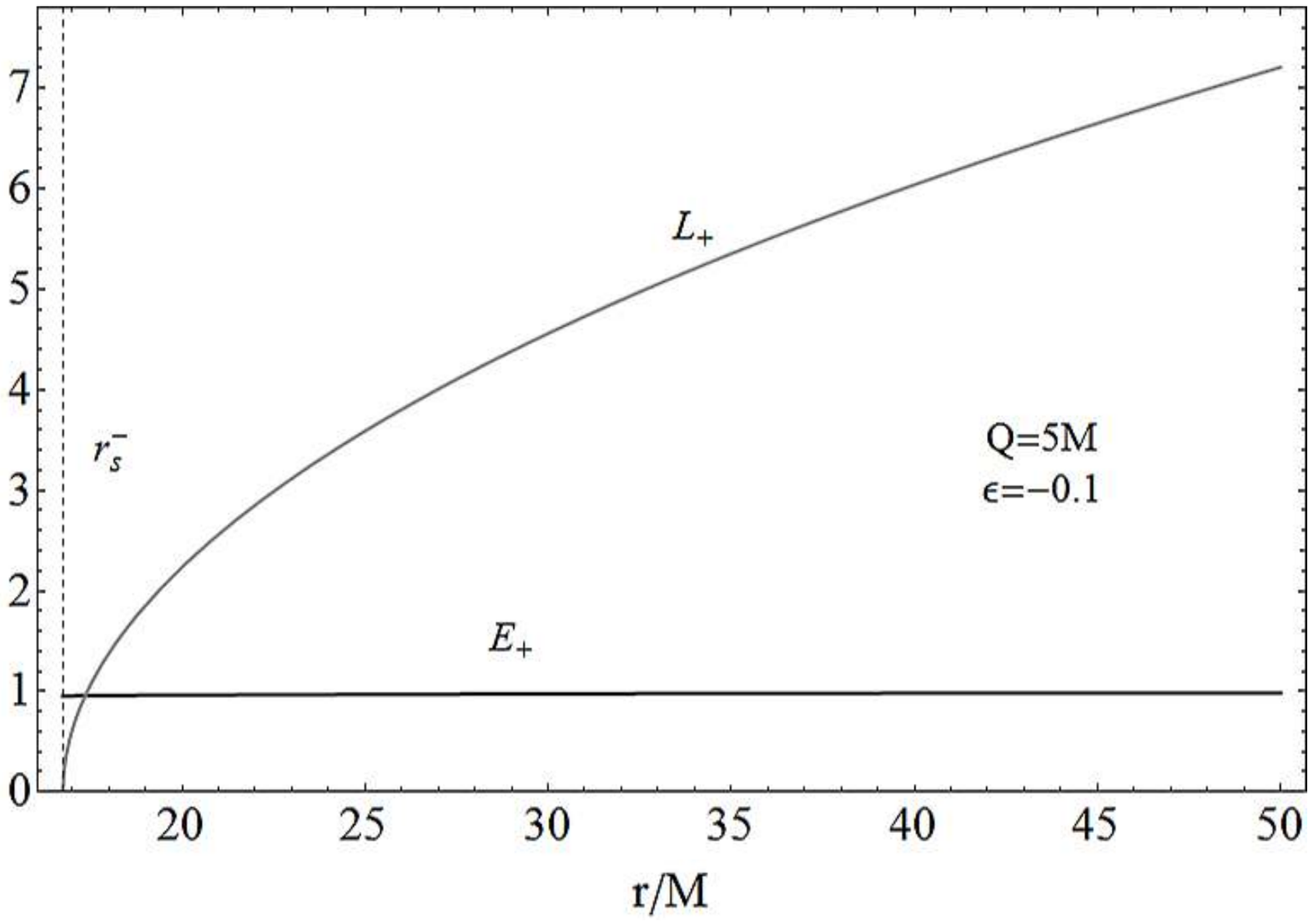}
\end{tabular}
\caption[font={footnotesize,it}]{\footnotesize{\textbf{Class}: $Q>\sqrt{9/8}M$ and  $-M/Q<\epsilon<0$.
Parameter choice: $Q=5M$ and  $\epsilon =-0.1$. Then $r_{s}^-=16.7519M$.
Circular orbits exist with angular momentum
$L=L_+$ (gray curve) and energy  $E=E_+$ (black curve)  in $r>r_{s}^- $.
For $r=r_{s}^-$, $L=0$.
}}
\label{Beuchnuovo}
\end{figure}


\end{document}